\documentclass[a4paper,11pt]{article}
\pdfoutput=1 % if your are submitting a pdflatex (i.e. if you have
\usepackage{jcappub} 
\usepackage[T1]{fontenc} 
\usepackage{graphicx}
\usepackage{epstopdf}
\usepackage{changepage}
\usepackage{xcolor}
\usepackage{bm}
\usepackage{mathrsfs}
\usepackage{hyperref}
\usepackage{orcidlink}
\usepackage{url}

\definecolor{myblue}{rgb}{0.2,0.3,0.7}
\definecolor{darkgreen}{rgb}{0,0.3,0}
\definecolor{midgreen}{rgb}{0,0.5,0}
\definecolor{mygreen}{rgb}{0,0.7,0}

\newcommand{\bdot}{{\displaystyle \cdot}}
\newcommand{\csdot}{{\left(c_{s(\mathrm{eff})}^2 \right)^{\!\bdot}}}
 \newcommand{\csddot}{{\left(c_{s(\mathrm{eff})}^2 \right)^{\! \bdot \bdot}}}

\newcommand{\CD}{{\mathcal{D}}}
\newcommand{\CP}{{\mathcal{P}}}
\newcommand{\CQ}{{\mathcal{Q}}}

\newcommand{\CV}{{\mathcal{V}}}
\newcommand{\CZ}{{\mathcal{Z}}}
\newcommand{\SP}{{\mathscr{P}}}
\newcommand{\QD}{{\mathcal{Q}_\mathcal{D}}}
\newcommand{\RD}{{\left\langle \mathcal{R} \right\rangle_\mathcal{D}}}
\newcommand{\VD}{{\mathcal{V}_\mathcal{D}}}
\newcommand{\WD}{{\mathcal{W}_\mathcal{D}}}
\newcommand{\sR}{{{}^{(3)} \! R}}

\newcommand{\eff}{^{(\mathrm{eff})}}
\newcommand{\baro}{_{\mathrm{bar}}}
\newcommand{\com}{_{\mathrm{com}}}

\newcommand{\avg}[1]{{\left\langle {#1} \right\rangle_\mathcal{D}}}

\newcommand{\FO}{\simeq}
\newcommand{\aref}{\bar a}
\newcommand{\rhoref}{\bar \varrho_d}
\newcommand{\tinit}{t_\mathbf{i}}
\newcommand{\Ii}{\mathrm{I}_\mathbf{i}}
\newcommand{\IIi}{\mathrm{II}_\mathbf{i}}
\newcommand{\IIIi}{\mathrm{III}_\mathbf{i}}
\newcommand{\rhoinit}{\rho_\mathbf{i}}
\newcommand{\ainit}{a_\mathbf{i}}
\newcommand{\arefi}{\bar a_\mathbf{i}}

\newcommand{\tinitb}{\bar t_\mathbf{i}}

\newcommand{\rhotot}{\rho_\mathrm{tot}}
\newcommand{\wtot}{w_\mathrm{tot}}
\newcommand{\rholoc}{\varrho_d}
\newcommand{\rholocinit}{\varrho_\mathbf{i}}

\title{Cosmological perturbations on an averaged background}

\author[a,1]{Marco Galoppo\orcidlink{0000-0003-2783-3603},\note{Corresponding author.}}
\author[a]{Pierre Mourier\orcidlink{0000-0001-8078-6901}}

\affiliation[a]{School of Physical \& Chemical Sciences, University of Canterbury, \\ Private Bag 4800, Christchurch 8140, New Zealand}
\emailAdd{Marco.Galoppo@canterbury.ac.nz}
\emailAdd{Pierre.Mourier@canterbury.ac.nz}

\abstract{In relativistic cosmology, the formation of nonlinear inhomogeneities can induce non-negligible backreaction on late-time expansion. Among the important consequences for precision cosmology is the potential impact on the linear growth of large-scale structures. We address this impact by combining covariant spatial averaging with covariant and gauge-invariant perturbation theory. We focus on irrotational dust model spacetimes. The effects of backreaction and nontrivial dynamical curvature on the average cosmological dynamics are formulated as the addition of an effective perfect fluid with pressure. We then introduce an effective background driven by both the averaged dust density and the emergent effective fluid, and derive the general evolution equations for linear perturbations of this system. The residual freedom in this framework amounts to specifying the properties of the effective-fluid perturbations as a closure condition. We analyse two physically motivated choices for this condition. In addition, we clarify the conditions under which the coupling between linear structure growth and perturbations of the effective fluid can be neglected. Finally, we apply this formalism to four examples of averaged cosmological models from the literature, three of which---intended as effective full descriptions of the largest scales---have been shown to provide a good fit to observational data. Our results highlight the importance of backreaction effects in shaping linear structure growth in such models. Neglecting these effects may thus lead to biased predictions for the development of large structures, even when the models provide a good description of the general background observables.
}

\begin{document}
\bibliographystyle{JHEP}

\maketitle
\flushbottom
\section{Introduction}
\label{sec:intro}

The growth of matter structures from nearly Gaussian primordial perturbations via gravitational instability is a central topic in cosmology. This process is highly sensitive to the cosmic expansion history, making the observed Large-Scale Structure (LSS) a powerful probe of cosmological evolution. Moreover, the LSS encodes valuable information about the large-scale spatial geometry of the Universe, which is ultimately reflected in the statistical properties of voids and clusters.

Within the standard $\Lambda$ -- Cold Dark Matter ($\Lambda$CDM) framework, small and intermediate-scale gravitational phenomena are effectively neglected in the large-scale evolution and geometry. The latter is described by introducing a spatially flat, isotropic large-scale metric, together with specific large-scale homogeneous matter components, namely dust, radiation, and a cosmological constant. The linear growth of LSS is then typically modeled using cosmological perturbation theory on this Friedmann--Lema\^itre--Robertson--Walker (FLRW) background~\cite{KodamaSasaki_1984,Mukhanov_1992}.

Remarkably, even with these simplifying assumptions, the $\Lambda$CDM framework has successfully described a variety of cosmological probes such as the Cosmic Microwave Background (CMB) power spectrum and Baryon Acoustic Oscillations (BAO) in the primordial plasma~\cite{Planck_2018}.

However, despite its empirical success, the $\Lambda$CDM paradigm exhibits several anomalies across a wide range of scales~\cite{Peebles_1967,BuchertEtAl_2016,Freedman_2017,Bullock_2017,Riess_2019,Perivolaropoulos_2022,Secrest_2022,Aluri_2023}. Furthermore, novel tensions with cosmological data emerge as evidence grows for a possible preference for Dynamical Dark Energy (DDE) over a cosmological constant $\Lambda$, as suggested by recent Type~Ia supernova and BAO measurements from DES~ \cite{Abbott_2024} and DESI~\cite{Abdul-Karim_2025,Adame_2025} when combined with CMB observations. 

Amidst the growing tensions between the $\Lambda$CDM model and cosmological data, it is therefore meaningful to re-examine some of the foundational assumptions underlying the construction of the standard cosmological model. Specifically, one may question the assumed negligible influence of the formation of small- and intermediate-scale nonlinear structures on the average large-scale dynamics, i.e., the \emph{backreaction} of inhomogeneities. Although in standard linear cosmological perturbation theory and within Newtonian gravity (even in the nonperturbative regime) this effect vanishes on average~\cite{BuchertEhlers_1997,BuchertRasanen_2012}, the result does not carry over to nonperturbative General Relativity (GR), where the geometry itself is dynamical. Hence, \emph{a priori}, the backreaction of inhomogeneities could have a significant impact on the average expansion rate, the evolution of spatial curvature, and light propagation in the Universe on large scales, thereby affecting cosmological inference. We refer the reader to, e.g.,~\cite{Clarkson_2011,BuchertRasanen_2012} for an in-depth discussion.

In cosmology, the issue of identifying a model that best describes the average behaviour of the real Universe with its complex structures---and thereby accounting for the backreaction of inhomogeneities---is known as the \emph{fitting problem}, or equivalently, the \emph{averaging problem}~\cite{Wald_1977,Ellis_1984,Ellis_1987,EllisBuchert_2005}. Indeed, tackling this issue necessarily requires the construction of an appropriately defined coarse-graining or averaging operation over the local spacetime description, which must account for the hierarchy of scales and nonlinear gravitational phenomena, in order to derive the effective dynamics on the largest scales.

Although numerous contributions have been made to this issue (see, e.g.,~\cite{Zalaletdinov_2008,Gasperini_2009,Korzynski_2010,Green_2011,SussmanBolejko_2012}), the most widely studied approach to cosmological averaging is the Buchert averaging scheme for scalar variables~\cite{Buchert_2000,Buchert_2001,BuchertMourierRoy_2020}, which we also adopt in this work. \textcolor{black}{While built upon integration over spatial hypersurfaces, this covariantly defined scheme is also especially suited for the interpretation of cosmological observations on the past light cone. Indeed, provided the hypersurfaces of averaging are statistically homogeneous and isotropic and structures evolve slowly compared to the light travel time across the statistical homogeneity scale~\cite{Koksbang_2019b,Koksbang_2020b}, the averaged scale factor and expansion prescribed by this scheme are closely connected with the distance--redshift relations obtained on the light cone, as suggested in~\cite{Rasanen_2010,Rasanen_2012} and successfully tested for various cosmological toy models~\cite{Lavinto_2013,Sikora_2017,Koksbang_2019a,Koksbang_2020a}. Within the Buchert averaging framework, backreaction from structure formation typically leads to a departure from pure FLRW dynamics, which emerge as a repeller in a dynamical-system analysis of model universes averaged in this way~\cite{Roy_etal_2011}.} A number of such \emph{averaged cosmological models} have been developed within this formalism and successfully tested against cosmological data~\cite{Wiltshire_2007a,Wiltshire_2007b,HeinesenBuchert_2020,Giani_2025a,Giani_2025b}. Remarkably, these models have been shown to outperform the standard $\Lambda$CDM model in fitting observations, helping to address tensions such as the Hubble and late-time structure growth-rate discrepancies, whilst also offering novel insights into the possible nature of dark energy and prospects for dynamical dark energy.

Interestingly, to our knowledge, the issue of structure growth has been theoretically investigated within the context of averaged cosmological models only in~\cite{Roy_2012}, where evolution equations for exact deviations of some scalar quantities with respect to their underlying averages were derived. Furthermore, in~\cite{Giani_2025a,Giani_2025b}, the authors adopted an alternative approach---based on an approximate application of conventional cosmological perturbation theory on an averaged background---to directly fit averaged cosmological models to LSS observational data for the first time in the literature.

In this work, we aim to provide a comprehensive treatment of the linear evolution of matter perturbations within averaged cosmological models, formulated in the framework of covariant and gauge-invariant perturbation theory~\cite{Ellis_1989a,Ellis_1989b,Ellis_1990,Bruni_1992a,Bruni_1992b,Dunsby_1992}. We thus bridge the important gap between covariant averaging~\cite{Buchert_2000,Buchert_2001,Gasperini_2010,HeinesenMourierBuchert_2019,BuchertMourierRoy_2020} and structure growth modelling in cosmological \mbox{perturbation theory}. 

Indeed, we must emphasise that without a well-defined notion of an averaging operation, any discussion of deviations from the mean becomes inconsistent. Hence, regardless of whether backreaction effects ultimately prove to be dynamically relevant on cosmological scales developing a meaningful framework in which large-scale perturbations can be evolved w.r.t.\ a consistent average remains a fundamental theoretical requirement. In the following, we construct such a framework, distinct from that developed in~\cite{Roy_2012} as we build upon well-established perturbation theory, thereby allowing for straightforward implementation, and enabling a natural extension to vector and tensor mode perturbations. On the other hand, we restrict our framework to first-order perturbations at this stage, in order to study linear growth, and to clearly separate the perturbations from the inhomogeneities accounted for in the background. This linear separation also implies that the perturbations we consider need not have a vanishing average value, and can have wavelengths exceeding the size of the averaging domain. Furthermore, we clarify the validity range of the approximation employed in~\cite{Giani_2025a,Giani_2025b} by elucidating the role of local linear perturbations in the dynamical curvature and backreaction.

The remainder of the paper is organised as follows. In Sec.~\ref{sec:Ellis&Bruni}, we recall the fundamentals of covariant and gauge-invariant perturbation theory as developed by Ellis and Bruni. In Sec.~\ref{sec:buchert}, we introduce and discuss Buchert's covariant averaging formalism. In Sec.~\ref{sec:Us}, we develop the framework for studying linear perturbations within averaged cosmological models and derive the corresponding evolution equations, focusing on the unconstrained degrees of freedom and their corresponding closure conditions. In Sec.~\ref{sec:examples}, we analyse the behaviour of perturbations across a range of averaged cosmological models, comparing the different scenarios. Finally, Sec.~\ref{sec:conc} presents our conclusions and outlines future perspectives.

Throughout this work, we set the vacuum speed of light $c$ to unity and use the metric signature $(-,+,+,+)$. We use Greek letters $\mu,\nu,\kappa$... to denote spacetime indices, running from $0$ to $3$, and Latin letters $a,b,c$..., $i,j,k$... to denote spatial indices or counters running from $1$ to $3$.

\section{Covariant and gauge-invariant perturbation theory}
\label{sec:Ellis&Bruni}

Perturbation theory plays a fundamental role in cosmology, providing a systematic framework to study the origin and early evolution of large-scale structure in the Universe. Conventionally, perturbation theory prescribes a splitting of all physical quantities into a homogeneous background and inhomogeneous perturbations. However, while GR is covariant---its equations do not depend on the choice of coordinates---this background/perturbation split is itself not a covariant operation. As a result, it inevitably introduces a gauge dependence, which reflects the ambiguity in identifying points of the fictitious background spacetime with those of the physical, inhomogeneous Universe. Hence, geometric and physical observables should be described by gauge-invariant quantities~\cite{Hawking_1966, Stewart_1974,Bardeen_1980, KodamaSasaki_1984, Ellis_1989a, Stewart_1990, Mukhanov_1992, Bruni_1997, Bruni_1999, Malik_2009}.

To address this issue, one can introduce Bardeen-type variables~\cite{Bardeen_1980}, which are specific linear combinations of non-covariant, gauge-dependent quantities, originally constructed to remain gauge-invariant at first order. While this approach can be systematically extended to higher orders \cite{Malik_2009}, working within the Bardeen formalism often complicates the interpretation of results, as the physical and geometrical meaning of these variables, which remain non-covariant, typically stays obscure unless a particular gauge is explicitly specified.

An alternative strategy is the covariant and gauge-invariant (CGI) approach originally developed by Ellis and Bruni \cite{Ellis_1989a,Ellis_1989b,Ellis_1990,Bruni_1992a,Bruni_1992b,Dunsby_1992}, and later expanded upon in \cite{Langlois_2005a,Langlois_2005b}. In this framework, one works directly with covariantly-defined spatial gradients of physical quantities, e.g.,  the energy density or the expansion scalar. The variables are defined exactly---without any approximation---and independently of a background. As a result, no non-covariant background splitting is required, and perturbations can be computed in a fully covariant manner, allowing the formalism to be applied to a variety of backgrounds.

\subsection{Covariant and gauge-invariant variables}

To illustrate this framework, let us consider an arbitrary spacetime with metric $g_{\mu\nu}$, filled with a mixture of perfect fluids\footnote{%
A direct generalisation to non-perfect fluids can be found in \cite{Dunsby_1992}.%
} characterised by $4-$velocities $u^{(i)}_\mu$ (such that $u^{(i)}_\mu u^{(i)\mu} = -1$), with proper energy densities $\rho^{(i)}$, isotropic pressures $p^{(i)}$, and proper kinematic expansion scalars $\Theta^{(i)}:=\nabla_\mu u^{(i)\mu}$. We also introduce a 4-velocity field $n^\mu$---tangent to the worldlines of some reference family of observers in the Universe---and the associated projection tensor $h_{\mu\nu}$ onto the local three-spaces orthogonal to $n^\mu$, namely, $h_{\mu\nu}:= g_{\mu\nu}+n_\mu n_\nu$. 

It is then possible to define, in a geometrical way, covariant quantities that can be interpreted as exact deviations with respect to an FLRW configuration, by introducing spatially projected gradients of the relevant physical quantities for the fluids. Defining the spatial gradient operator, $D_\mu := h^{\nu}_{\phantom{\mu}\mu}\nabla_\nu$ for a scalar variable (extendable to higher-rank, $n^\mu$-orthogonal tensors, with an additional $h_{\mu\nu}$ projection of each tensor index), we consider the covariant vector variables~\cite{Ellis_1989a,Ellis_1989b,Ellis_1990,Bruni_1992a,Bruni_1992b,Dunsby_1992}
\begin{align}
    X_\mu^{(i)} :=D_\mu \rho^{(i)} \quad ; \quad Y_\mu^{(i)} := D_\mu p^{(i)} \quad ; \quad
     Z_\mu^{(i)} := D_\mu \Theta^{(i)} \;\;  . \label{eq:FirstQuantities}
     \end{align}
These covariantly defined quantities identically vanish in a strictly FLRW spacetime. Following the Stewart–Walker lemma~\cite{Stewart_1974}, this guarantees their gauge invariance at first order in a perturbative scheme. These CGI quantities represent spatial gradients of the inhomogeneous physical variables with respect to the rest frame of the observers. As such, the CGI formalism enables direct calculations in the physical spacetime itself, without the strict need to introduce a fictitious background.

However, we note that the gradient variables in Eq.~\eqref{eq:FirstQuantities} represent absolute variations of the corresponding scalar quantities over a fixed distance scale. Conventionally, we wish to (i) compare relative changes with respect to a standard reference, and (ii) consider variations over a fixed comoving length scale. Therefore, following~\cite{Ellis_1989a,Ellis_1989b,Ellis_1990,Bruni_1992a,Bruni_1992b,Dunsby_1992}, we introduce the \emph{local scale factor} relative to the observer rest frame, $a$, and define the comoving fractional gradients: 
\begin{align}
    \mathcal{D}_\mu^{(i)} := \frac{a}{\rho^{(i)}}D_\mu \rho^{(i)} \quad ; \quad \mathcal{P}_\mu^{(i)} := \frac{a}{h^{(i)}} D_\mu p^{(i)} \quad ; \quad
     \mathcal{Z}_\mu^{(i)} := a \, D_\mu \Theta^{(i)} \;\; ,
\end{align}
where $h^{(i)}:= \rho^{(i)}+p^{(i)}$ is the enthalpy of each fluid\footnote{%
In contrast with previous works \cite{Ellis_1989a,Ellis_1989b,Ellis_1990,Bruni_1992a,Bruni_1992b,Dunsby_1992} in which the comoving fractional pressure had been normalised with respect to the pressure, here we employ an enthalpy-based normalisation.%
}. The local scale factor is defined by its evolution as $\dot{a}/a := \Theta /3$, where $\Theta := \nabla_\mu n^\mu$ is the kinematic expansion scalar of the observer congruence, and we have denoted by an overdot $\dot{\ }$ the observer-frame covariant evolution operator $n^\mu\nabla_\mu\,$; this is complemented by an arbitrary initialisation of $a$ to a homogeneous or near-homogeneous value on a given spatial slice. It is this set of CGI variables that can be then adopted as perturbation variables when analysing their evolution and constraint equations as linearised with respect to an FLRW background model. Although both the CGI approach and the conventional Bardeen-variables method ultimately lead to equivalent equations, we emphasise the fundamental difference in perspective between these two frameworks: in the CGI formalism, the starting point is an anisotropic, inhomogeneous spacetime that is kinematically close to FLRW, and the approximation consists in neglecting higher-order terms when the dynamical variables remain near their FLRW values. This contrasts with the standard method, where one begins with an exact FLRW background and subsequently introduces perturbations around it. 

Moreover, we note that a further set of CGI variables is given by considering the vorticity tensor ($\omega^{(i)}_{\mu\nu}$), shear tensor ($\sigma^{(i)}_{\mu\nu}$) and $4-$acceleration ($a^{(i)}_\mu$) of each fluid component,the corresponding quantities $\omega_{\mu \nu}$, $\sigma_{\mu \nu}$, $a_\mu$ for the observer congruence $n^\mu$, and the electric and magnetic parts of the Weyl tensor \cite{Ellis_1989a,Ellis_1989b,Ellis_1990,Bruni_1992a,Bruni_1992b,Dunsby_1992}. All of these tensors vanish in an FLRW background. Furthermore, within a multi-fluid scenario, to each fluid component is associated a tilt velocity, $V_\mu^{(i)}$, describing the velocity of the $i$th fluid with respect to the observer frame via the orthogonal decomposition $u_{\mu}^{(i)} =: (-n^\nu u_{\nu}^{(i)}) \, (n_\mu +V_\mu^{(i)})$, with $n^\mu V_\mu^{(i)} = 0$. By assuming the same $4-$velocity for every fluid in the FLRW background, the tilt velocity of each fluid component represents another CGI perturbation variable of this framework. 

Finally, the above set of CGI variables can be complemented by introducing their corresponding  quantities for the combination of all fluid sources. We define 
\begin{equation}
    \rho:= \sum_{(i)}\rho^{(i)} \quad ; \quad p = \sum_{(i)}p^{(i)} \quad ; \quad h = \sum_{(i)}h^{(i)} = \rho + p \;\; ,
\end{equation}
and introduce the combined-fluid CGI variables within the observer frame:
\begin{equation}
    \mathcal{D}_\mu := \sum_{i}\frac{\rho^{(i)}}{\rho}\mathcal{D}_\mu^{(i)} \quad ; \quad \mathcal{P}_\mu:= \sum_{i}\frac{h^{(i)}}{h}\mathcal{P}_\mu^{(i)} \quad ; \quad 
    \mathcal{Z}_\mu := a \, D_\mu\Theta \quad ; \quad 
    V_\mu := \sum_{i}\frac{h^{(i)}}{h}V_\mu^{(i)} \;\; . \label{eq:tot_fluid_CGI}
\end{equation}
The combined-fluid variables introduced in Eq.~\eqref{eq:tot_fluid_CGI} are dependent variables with respect to the set of single-fluid CGI quantities. Indeed, here we note that---within a first-order perturbative scheme---the single-fluid component expansion scalars and that of the observer congruence are related by\footnote{%
Hereafter we employ the $\FO$ symbol to identify equivalence up to first order in the perturbative scheme.%
}:
\begin{equation}
    \Theta \FO \Theta^{(i)} + \nabla^{\mu}V_\mu^{(i)} \; ,
    \label{eq:ThetaThetai}
\end{equation}
so that $\mathcal{Z}_\mu^{(i)}$ for a given fluid component $(i)$ is directly related to $\mathcal{Z}_\mu$ and $V_\mu^{(i)}$. In this work, we will ultimately employ the set of CGI variables $\{ \mathcal{Z}_\mu \, ; \, \mathcal{D}_\mu^{(i)}, \mathcal{P}^{(i)}_\mu, V^{(i)}_\mu\}$.

\subsection{Fundamental equations}

We are considering a $4-$dimensional, Lorentzian spacetime manifold with energy-momentum tensor $T_{\mu\nu} = \sum_{(i)}T^{(i)}_{\mu\nu}$, where the energy-momentum tensor of each component is given by
\begin{align}
    T^{(i)}_{\mu\nu}  := {} & \left(\rho^{(i)}+p^{(i)} \right) \,  u^{(i)}_\mu u^{(i)}_\nu + p^{(i)} \, g_{\mu\nu} \nonumber\\ 
    {} \FO  {} & \left(\rho^{(i)}+p^{(i)} \right) \,  n_\mu n_\nu + p^{(i)} \, g_{\mu\nu} + q^{(i)}_{\mu} \, n_{\nu } + q^{(i)}_{\nu} \, n_{\mu } \; . 
    \label{eq:Tmunu_i}
\end{align}
In the second line, we have decomposed the energy-momentum tensor of the $i$th fluid component with respect to the observer frames, introducing $q^{(i)}_{\mu} := h^{(i)} \, V^{(i)}_\mu$ as the linearised energy flux of the $i$th component. As in this work we consider only non-interacting fluids, we impose a direct conservation of the energy-momentum tensor of each component, i.e., for each $i$,
\begin{equation}
    \nabla^\nu T^{(i)}_{\mu\nu} = 0 \; . \label{eq:cons_ithfluid}
\end{equation}
Projecting the energy–momentum conservation relations \eqref{eq:cons_ithfluid} along $n_\mu$, and $h_{\mu\nu}$, using the $n^\mu$-frame decomposition of $T_{\mu\nu}^{(i)}$ in Eq.~\eqref{eq:Tmunu_i}, we obtain the continuity equation of each perfect fluid component, and an equation relating the observer-frame acceleration to perturbation variables of any given fluid component $i$, up to linear order in the perturbation variables:
\begin{align}
   & \dot{\rho}^{(i)} \FO - h^{(i)}\,\Theta - h^{(i)}D^\mu V_\mu^{(i)}  \, ; \label{eq:continuity} \\
   & a_\mu \FO -\frac{1}{a} \CP^{(i)}_\mu - F^{(i)}_\mu\, . \label{eq:acceleration}
\end{align}
Here, following the notation of \cite{Bruni_1992a,Dunsby_1992}, we have defined $F^{(i)}_\mu := \dot{V}^{(i)}_\mu  + \big((1/3) - c_{s(i)}^2 \big) \, \Theta {V}^{(i)}_\mu$, with the squared sound speed of the $i$th fluid component given by $c_{s(i)}^2:=\dot{p}^{(i)}/\dot{\rho}^{(i)}$, here evaluated for the background (at zeroth order).

We can supplement these equations with those for the combined-fluid CGI variables by projecting the total energy-momentum tensor conservation equation with respect to $n_\mu$, and to $h_{\mu\nu}$, to derive the linearised combined-fluid continuity and acceleration equations, respectively:
\begin{align}
    &\dot{\rho} \FO -h \, \Theta  - h \, D^\mu V_\mu \; , \label{eq:continuity_tot}  \\
    & a_\mu \FO -\frac{1}{a}\mathcal{P}_\mu - F_\mu \; , \label{eq:acceleration_tot}
\end{align}
where $F_\mu:=  \dot{V}_\mu  + \big((1/3) - c_{s}^2 \big) \, \Theta {V}_\mu$, with the effective combined-fluid squared sound speed defined with respect to the combined-fluid variables as $c_{s}^2:=\dot{p}/\dot{\rho} = \sum_{(i)} \, (h^{(i)}/h) \, c^2_{s(i)}$, again evaluated at background level.

The above continuity and acceleration equations are then coupled with the evolution equation for $\Theta$, given by the Raychaudhuri equation---the fundamental equation for gravitational attraction---which reads to first order and in the observer frame:
\begin{equation}
    \dot{\Theta} + \frac{1}{3}\Theta^2 - \nabla_\mu a^\mu + 4\pi G (\rho + 3p) - 2 \Lambda \FO 0\, , \label{eq:Raychaudhuri}
\end{equation}
where we have introduced a cosmological constant $\Lambda$, and neglected the contributions from the shear and vorticity scalars, which are second-order perturbations. Interestingly, as the Raychaudhuri equation arises from a projection of the Ricci tensor and thus of the full energy momentum-tensor ($T_{\mu\nu} = \sum_{(i)}T^{(i)}_{\mu\nu}$), to first order (where $T^{\mu\nu} u_\mu^{(i)} u_\nu^{(i)} \FO T^{\mu\nu} n_\mu n_\nu \; \forall i$), the Raychaudhuri equations obtained from the $4-$velocities of the individual fluid components are automatically satisfied once we impose the validity of Eq.~\eqref{eq:Raychaudhuri} --- i.e., they do not constitute additional independent equations.

Finally, we can also write the Hamiltonian constraint relating the spatial Ricci scalar $\sR$ of the hypersurfaces orthogonal to the observer congruence,\footnote{%
We remark that, strictly speaking, the direct interpretation of $\sR$ as a hypersurface curvature requires the observer congruence $n^\mu$ to be hypersurface-forming, i.e. its vorticity $\omega_{\mu\nu}$ to vanish. This is in particular the case in the FLRW background, and in practice we only need to apply the Hamiltonian constraint~\eqref{eq:Hamilton} at the background level due to the local $\sR$ multiplying first-order terms. However, we note that this scalar can still be defined in the general case from $n^\mu$ as a local $3-$Ricci scalar, satisfying the Hamiltonian constraint by construction~\cite{Ellis_1990}.%
} $n^\mu$, with its expansion scalar and energy density, and the cosmological constant, i.e., to first order,
\begin{equation}
    \sR \FO -\frac{2}{3}\Theta^2 + 16\pi G \rho - 2\Lambda \, . \label{eq:Hamilton}
\end{equation}

\subsection{Evolution equations of the first-order perturbations}

To derive the evolution equation for the density perturbation of each fluid component, we take the spatial gradient of the respective continuity equation, Eq.~\eqref{eq:continuity}, using Eq.~\eqref{eq:acceleration} to reexpress the $4-$acceleration of $n^\mu$ which appears upon commuting the observer-frame temporal and spatial derivatives. We find \cite{Dunsby_1992} 
\begin{equation}
   \dot{\mathcal{D}}_\mu^{(i)} - w_{(i)}\Theta \:\! \mathcal{D}_\mu^{(i)}  + \left(1+w_{(i)}\right)\mathcal{Z}_\mu - a\,\Theta\left(1+w_{(i)}\right)F_\mu^{(i)} + a \left(1+w_{(i)}\right) D_\mu D^\alpha V_\alpha^{(i)} \FO 0 \; ,
    \label{eq:grad_cont_comp}
\end{equation}
where we have introduced the equation of state (EoS) parameter of the $i$th fluid component $w^{(i)}:= p^{(i)}/\rho^{(i)}$. Furthermore, the evolution equation for the expansion perturbation can be derived from the spatial gradient of the Raychaudhuri equation, Eq.~\eqref{eq:Raychaudhuri}, where the $4-$acceleration $a_\mu$ is rewritten via Eq.~\eqref{eq:acceleration_tot}, and Eq.~\eqref{eq:continuity_tot} is used to express the time derivative of the total energy density. We obtain \cite{Dunsby_1992}
\begin{align}
    & \dot{\mathcal{Z}}_\mu + \frac{2}{3}\Theta \mathcal{Z}_\mu + 4\pi G\rho \, \mathcal{D}_\mu + \left[ {}^{(3)}\!\Delta + \frac{ {}^{(3)}\!R}{2}\right]\mathcal{P}_\mu -2 \;\! a \, \Theta \, c_s^2 \, D_\alpha \omega_{\mu}^{\phantom{\mu}\alpha} \nonumber \\
    & \qquad +\frac{1}{2} \:\! a \;\! {}^{(3)}\!R \, F_\mu - 12\pi G \:\! h \:\! a \;\! F_\mu + a \, D_\mu D^\alpha F_\alpha \FO 0 \, , \label{eq:grad_ray}
\end{align}
where ${}^{(3)}\!\Delta := D_\mu D^\mu$ is the spatial three-Laplacian (i.e., the Laplace--Beltrami operator of the spatial metric on the hypersurfaces) arising upon commuting the observer-frame temporal and spatial derivatives. We note that, in Eq.~\eqref{eq:grad_cont_comp}, the density perturbations of the individual fluid components couple directly to the total expansion perturbation $\mathcal{Z}_\mu$, whose evolution is, in turn, determined by the total density and pressure perturbations in Eq.~\eqref{eq:grad_ray}. Such a coupling is expected, since in a multi-fluid scenario the perturbations of the different components still interact gravitationally with one another, notwithstanding the assumed decoupling of the individual energy-momentum contributions.

The presence of a vorticity term in Eq.~\eqref{eq:grad_ray} is also of interest, showing the linear relation between the divergence of the vorticity and the density gradients in almost-FLRW models (see~\cite{Ellis_1990} for details).
Due to this term, we also need in general the evolution equation for the vorticity tensor of the observer congruence, $\omega_{\mu\nu}$, up to first order, namely~\cite{Hawking_1966,Ellis_1990}: 
\begin{equation}
    \dot{\omega}_{\mu\nu} \FO -\frac{2}{3} \Theta \, \omega_{\mu\nu} + D_{[\mu}a_{\nu]} \FO \left( c_s^2 - \frac{2}{3} \right) \Theta \, \omega_{\mu\nu} - D_{[\mu} F_{\nu]} \; , \label{eq:vorticity}
\end{equation}
where, in this work, we are using an opposite sign convention in the definition of the vorticity tensor, i.e., $\omega_{\mu\nu} := h_{[\mu}^\alpha h_{\nu]}^\beta \:\! \nabla_{\alpha} n_\beta$, compared to~\cite{Hawking_1966,Ellis_1990,Bruni_1992a,Dunsby_1992}.

The evolution equations \eqref{eq:grad_cont_comp}--\eqref{eq:vorticity} are complemented by taking the difference between the two alternative expressions for the $4-$acceleration of the observer frames, Eqs.~\eqref{eq:acceleration} and \eqref{eq:acceleration_tot}, generating first-order evolution equations for relative tilt velocities of the fluid components~\cite{Dunsby_1992}:
\begin{equation}
    \dot{V}^{(i)}_\mu  - \dot V_\mu + \left(\frac{1}{3} - c_{s(i)}^2 \right) \Theta  V^{(i)}_\mu - \left(\frac{1}{3} - c_{s}^2 \right) \Theta  V_\mu  +\frac{1}{a} \left( \mathcal{P}_\mu^{(i)} - \CP_\mu \right) \FO 0\, . \label{eq:acc_comp}
\end{equation}

The system of evolution equations \eqref{eq:grad_cont_comp}--\eqref{eq:acc_comp} only closes once we additionally specify (i) $V_\mu$, or the tilt velocity $V_\mu^{(i)}$ for one fluid component, or some combination of the tilt velocities; and (ii) the pressure perturbations $\CP_\mu^{(i)}$ for all fluid components. These correspond to two choices which need to be made depending on the physical system considered. Setting (i) corresponds to explicitly defining the observer frame $n^\mu$, which we have kept unspecified in the above derivation. For instance, one can define $n^\mu$ as the total energy frame, defined by setting $V_\mu := 0$; or one may use the energy frame of one of the individual perfect-fluid components $i_0$, $n_\mu := u_\mu^{(i_0)}$, which sets $V_\mu^{(i_0)} := 0$, as we will do in later sections. Specifying (ii) amounts to introducing an EoS for each of the fluid components, $p^{(i)}(\rho^{(i)}, s^{(i)})$, where $s^{(i)}$ is the single-component entropy density. Indeed, the pressure perturbations $\mathcal{P}_\mu^{(i)}$ can be written in general as
\begin{equation}
     (1+w_{(i)})\mathcal{P}_\mu^{(i)} = c_{s(i)}^2\mathcal{D}_\mu^{(i)} + w_{(i)}\mathcal{E}_\mu^{(i)} \; ,
     \label{eq:entropy_perturb}
 \end{equation}
where $\mathcal{E}_\mu^{(i)}  := (a/p^{(i)})(\partial p^{(i)}/\partial s^{(i)}) \, D_\mu s^{(i)}$ represent the entropy perturbations, whose evolution equations can be derived by taking the spatial gradient of the conservation equation for the specific entropy of each fluid component, i.e., $u^{(i)}_\mu\nabla^\mu s^{(i)}\FO 0\,$. Alternatively, effective closure conditions for $\mathcal{P}^{(i)}_\mu$---which do not directly state any EoS---can  also be employed to close the system of evolution equations. 

Note that, up to this point,  we have considered all perturbation variables as vector fields (plus the vorticity tensor field) in a nearly-FLRW spacetime. However, when dealing with cosmological perturbations, it is standard to introduce a decomposition into scalar, vector, and tensor modes through a nonlocal procedure \cite{KodamaSasaki_1984,Stewart_1990}, the scalar perturbations being then solely responsible for the clustering of matter in the Universe at the linear level. 

Instead, within the CGI approach, scalar and tensor perturbation fields are built via a \emph{local} decomposition of the covariant derivative of the above vector variables. For instance, for the density–gradient perturbation $\mathcal{D}^\mu$, we can introduce the decomposition
\begin{equation}
a \, D_\mu \mathcal{D}_\nu =: \Delta_{\mu\nu} = \frac{1}{3} h_{\mu\nu} \, \Delta + \Sigma_{\mu\nu} + W_{\mu\nu} \; ,
\end{equation}
where $\Sigma_{\mu\nu} := \Delta_{(\mu\nu)} - (1/3) \, h_{\mu\nu} \, \Delta$ and $W_{\mu\nu} := \Delta_{[\mu\nu]} \FO - a^2 (1 + w) \Theta \, \omega_{\mu\nu}$ (with $w := p / \rho$) are the symmetric trace-free, and skew-symmetric parts of $\Delta_{\mu\nu}$, respectively, whilst the spatial divergence $\Delta := \Delta^\mu{}_\mu \FO (a^2 / \rho) \, {}^{(3)}\!\Delta \rho$ is the scalar CGI variable measuring the aggregation of matter \cite{Ellis_1990,Bruni_1992a,Bruni_1992b,Dunsby_1992}.
It is therefore convenient, within a cosmological setting, to introduce this local splitting for all perturbative variables of interest. Hence, we define the following set of scalar CGI variables for the fluid components:
\begin{equation}
\begin{gathered}
\Delta^{(i)} := a \, D^\mu \mathcal{D}_\mu^{(i)} 
\quad ; \quad  \mathcal{P}^{(i)} := a \, D^\mu \mathcal{P}_\mu^{(i)} \quad ; \quad V^{(i)} := a \, D^\mu V_\mu^{(i)} \;\; ; \\
F^{(i)} := a \, D^\mu F_\mu^{(i)} \FO \dot V^{(i)} + \left( \frac{1}{3} - c^2_{s(i)} \right) \:\! \Theta V^{(i)} \;\; ;  \label{eq:scalar_part_fluid}
\end{gathered}
\end{equation}
and the combined multi-fluid scalar CGI variables:
\begin{equation}
\begin{gathered}
\Delta = a \, D^\mu \mathcal{D}_\mu \quad ; \quad  \mathcal{P} := a \, D^\mu \mathcal{P}_\mu \quad ; \quad  \mathcal{Z} := a \, D^\mu \mathcal{Z}_\mu \quad ; \quad V := a \, D^\mu V_\mu \;\; ; \\
F := a \, D^\mu F_\mu \FO \dot V + \left( \frac{1}{3} - c_{s(i)}^2 \right) \:\! \Theta V\; . \label{eq:scalar_part_tot}
\end{gathered} 
\end{equation}
We note that, like $\Delta$ above, the first-order variables $\CZ$, $\CP$, $\Delta^{(i)}$ and $\CP^{(i)}$ are all proportional to $3-$Laplacians of local source fields ($\Theta$, $p$, $\rho^{(i)}$ and $p^{(i)}$ respectively), normalised to the local comoving length scale via $a^2$. 

We can then derive the evolution equations for the scalar CGI variables  by taking another spatial gradient $D^\mu$ of Eqs.~\eqref{eq:grad_cont_comp}--\eqref{eq:acc_comp}. We find the system of ordinary differential equations (ODEs) which we will employ in this work~\cite{Dunsby_1992}: 
\begin{align}
    & \dot{\Delta}^{(i)} - w_{(i)}\Theta\Delta^{(i)}  + \left(1+w_{(i)}\right)\mathcal{Z} - a \, \Theta\left(1+w_{(i)}\right)F^{(i)} + a\left(1+w_{(i)}\right) {}^{(3)}\!\Delta V^{(i)} \FO 0 \; ,\label{eq:1} \\
    &     \dot{\mathcal{Z}} + \frac{2}{3}\Theta \mathcal{Z} + 4\pi G\rho \;\! \Delta + \left[ {}^{(3)}\!\Delta + \frac{ {}^{(3)}\!R}{2}\right]\mathcal{P} +\frac{1}{2} a \,  {}^{(3)}\!R \, F - 12\pi G \:\! h \:\! a \;\! F + a \, {}^{(3)}\!\Delta F \FO 0 \; , \label{eq:2}\\
    &  \dot{V}^{(i)} - \dot V + \Theta \left(\frac{1}{3} - c_{s(i)}^2  \right) V^{(i)} - \Theta \left(\frac{1}{3} - c_{s}^2  \right) V  +\frac{1}{a}\mathcal{P}^{(i)} -\frac{1}{a}\mathcal{P} \FO 0 \; . \label{eq:3}
\end{align}
Interestingly, in the scalar equations~\eqref{eq:1}--\eqref{eq:3}, the coupling between individual and combined fluid contributions remains, as expected, but the vorticity plays no role, thus reducing the number of variables within the system compared to Eqs.~\eqref{eq:grad_cont_comp}--\eqref{eq:acc_comp}. Hence, the system of evolution equations~\eqref{eq:1}--\eqref{eq:3} is directly closed by specifying the rest frame of the observers, and the EoSs of the individual fluid components.

In Sec.~\ref{sec:Us}, we will be specialising the set of evolution equations \eqref{eq:1}-\eqref{eq:3} to a two-fluid system composed of irrotational dust and an effective perfect fluid, and derive the relevant evolution equations for the dust perturbations. The second, effective fluid source arises from a covariant scalar averaging procedure applied to an inhomogeneous Universe, to which we turn now.

\section{Spatial averaging scheme}
\label{sec:buchert}

In a generic globally hyperbolic spacetime with fluid sources, the dynamics of a given ensemble of fluid elements will not in general match the evolution of a homogeneous-isotropic model universe with the same (initial) average energy densities, due to the presence of dynamical, spatially inhomogeneous structures in the energy distribution.

The contributions of local inhomogeneities to the overall dynamics of an extended region of fluid---as a flow tube---\textcolor{black}{can} be described in a 3+1 foliation of spacetime into spacelike hypersurfaces, by explicitly averaging local variables over the region considered within each spatial slice.
\textcolor{black}{Such a covariant} relativistic averaging scheme---for scalar variables---and the resulting average dynamics for a spatial domain comoving with the fluid flow, were introduced by Buchert for an irrotational dust or perfect-fluid source in its rest frames in~\cite{Buchert_2000,Buchert_2001}, and later generalised to arbitrary fluid contents and spatial foliations in~\cite{BuchertMourierRoy_2018,BuchertMourierRoy_2020}. In this work, we are interested in effective dynamics in the matter-dominated universe, in presence of structures, considering scales large enough that effects of velocity dispersion and vorticity can be neglected. We can thus restrict ourselves to the irrotational dust averaging framework of~\cite{Buchert_2000}. We now summarize its construction and main features relevant for the perturbed models studied in this work.

\subsection{Defining spatial averages for scalars}

The $4-$dimensional, Lorentzian spacetime manifold is assumed to be globally hyperbolic, and to obey the Einstein equations sourced by a single, irrotational pressureless (dust) fluid source with energy density $\rholoc$ and timelike $4-$velocity $u^\mu$, \emph{i.e.}, with energy-momentum tensor $T_{\mu\nu} = \rholoc \;\! u_\mu u_\nu$. The metric $\bm{g} = g_{\mu\nu} \, \mathrm{d}x^\mu \mathrm{d}x^\nu$ defines a local projector orthogonal to the $4-$velocity, with components $b_{\mu \nu} := g_{\mu \nu} + u_\mu u_\nu$. The $4$-velocity field $\bm u$ is further characterised by its expansion scalar $\Theta$ and by its shear tensor $\sigma_{\mu \nu}$ and scalar $\sigma^2 := (1/2) \, \sigma^{\mu \nu} \sigma_{\mu\nu}$. These quantities arise from the kinematic decomposition of the fluid's expansion tensor,
\begin{equation}
    \Theta_{\mu \nu} := b^\kappa_{\,(\mu} b^\lambda_{\,\nu)}\nabla_\kappa u_\lambda = \nabla_\mu u_\nu = \frac{1}{3} \Theta \, b_{\mu \nu} + \sigma_{\mu\nu} \;\; ; \;\; \Theta := \Theta^\mu_{\,\mu} = b^{\mu \nu} \nabla_\mu u_\nu \;\; ; \;\; \sigma_{\mu \nu} := \Theta_{\mu \nu} - \frac{1}{3} \Theta \, b_{\mu \nu} \; ,
\end{equation}
with the $4-$acceleration $u^\nu \nabla_\nu u^\mu$ and vorticity $b^\kappa_{\,[\mu} b^\lambda_{\,\nu]} \nabla_\kappa u_\lambda$ of the fluid vanishing by the irrotational dust assumption. The shear tensor is symmetric, traceless and fluid-orthogonal by construction, $\sigma_{[\mu \nu]} = 0$, $\sigma^{\mu}_{\,\mu} = 0$, $u^\mu \sigma_{\mu \nu} = 0$.

By Frobenius' theorem, the irrotational velocity field $\bm u$ defines a spacetime foliation into hypersurfaces everywhere orthogonal to $\bm u$. These hypersurfaces are thus spacelike, and correspond to global rest frames for the fluid.
For convenience, we introduce a coordinate system $(x^\mu) = (t, X^i)$ adapted to the fluid flow and the foliation, with $t$ constant on each spatial slice, normalised to be a proper time of the fluid (the lapse is set to $1$), and with the spatial coordinates $X^i$ comoving with the fluid flow (the shift is set to $0$). In these coordinates, the line element reads:
\begin{equation}
  \label{eq:lineelem}
    \mathrm{d}s^2 = - \mathrm{d}t^2 + g_{ij} \, \mathrm{d}X^i \mathrm{d}X^j = - \mathrm{d}t^2 + b_{ij} \, \mathrm{d}X^i \mathrm{d}X^j \; .
\end{equation}
We will denote as an overdot ${}^\bdot$ the evolution operator along $\bm u$, $u^\mu \nabla_\mu$, and, consistently, the time derivative of any scalar variable $\varphi(t)$ only depending on time (such as a spatially integrated or averaged quantity), $\dot\varphi := \mathrm{d}\varphi(t) / \mathrm{d}t$.
Note that the averaging scheme remains covariantly defined, \emph{via} the above characterisation of the coordinate set. Manifestly covariant expressions of this formalism and its generalisation to other fluid sources and foliations can be found in~\cite{Gasperini_2009,Gasperini_2010,HeinesenMourierBuchert_2019}.

An arbitrary compact domain $\CD$ on the spatial slices can then be selected as the region of interest over which averaged properties will be computed. The propagation of the domain between slices is defined to be comoving with the fluid flow, so that it encompasses a given set of fluid elements throughout their evolution. In other words, a given flow tube of the fluid is chosen, and the domain $\CD$ considered on each hypersurface is the intersection of the flow tube and that slice.
The volume of the averaging domain $\CD$ is then defined as its Riemannian volume from the induced spatial metric on the hypersurfaces:
\begin{equation}
   \label{eq:defVD}
    \VD(t) = \int_\CD \, \sqrt{b} \; \mathrm{d}^3X \; , \; \text{with} \;\; b := \det(b_{ij}) \; .
\end{equation}
In the set of fluid-comoving spatial coordinates $X^i$, the boundaries of the integration domain are time-independent.
One can then define the corresponding volume-weighted spatial average $\avg{\psi}$ of any scalar variable $\psi$ over the domain of interest:
\begin{equation}
    \avg{\psi}(t) := \frac{1}{\VD(t)} \int_\CD \; \sqrt{g} \; \psi \; \mathrm{d}^3 X \; .
\end{equation}
The fluid-comoving propagation of the domain implies that the evolution rate of its volume is directly given by the averaged expansion scalar: $\dot{\mathcal{V}_\CD} / \VD = \avg{\Theta}$. More generally, from this assumption the following evolution equation is derived for the average of any scalar variable $\psi$, here expressed as a \emph{commutation rule} between the spatial averaging and time evolution operators:
\begin{equation}
    \label{eq:comm_rule}
    \left\langle \vphantom{\dot\psi} \psi \right\rangle_\CD^{\displaystyle .} - \avg{\dot\psi} = \avg{ \vphantom{\dot\psi} \Theta \, \psi} - \avg{\vphantom{\dot\psi} \Theta} \, \avg{ \vphantom{\dot\psi} \psi} \; .
\end{equation}

\subsection{Averaged evolution equations for a fluid region and effective Friedmann form}

From the above definition of the volume of the averaging domain, an \emph{effective scale factor} $a_\CD$ can be defined at any given time for the domain to track the evolution of its characteristic size: $a_\CD(t) \propto \VD(t)^{1/3}$. In the above references, it is normalised to unity at a given initial time; in this work, the normalisation of $a_\CD(t)$ is instead specific to each of the models considered in Sec.~\ref{sec:examples}---and is rather of order unity at present time. The evolution rate of this scale factor defines an effective Hubble parameter $H_\CD(t)$ for the comoving domain,
\begin{equation}
    H_\CD(t) := \frac{\dot a_\CD}{a_\CD} = \frac{1}{3} \frac{\dot{\mathcal{V}}_\CD}{\VD} = \frac{1}{3} \avg{\Theta} \; .
\end{equation}
The dynamical evolution equations for the effective scale factor---the Buchert equations---are obtained from the spatial averages of the Hamilton constraint and of the Raychaudhuri equation, using the commutation rule~\eqref{eq:comm_rule}. They involve the average of the local scalar intrinsic 3-curvature on the hypersurfaces $\mathcal{R}$, as well as a term specifically measuring the inhomogeneity and anisotropy in the fluid's expansion --- and thus absent from the Friedmann equations in FLRW spacetimes: the \emph{kinematical backreaction} term,\footnote{%
When vorticity is included (either in Newtonian gravity~\cite{BuchertEhlers_1997} or by considering a foliation no longer orthogonal to the fluid $4-$velocity~\cite{BuchertMourierRoy_2018}), it contributes an additional, positive-definite term to the kinematical backreaction.%
}
\begin{equation}
   \label{eq:defQD}
    \QD = \frac{2}{3} \left( \, \avg{\Theta^2} - \big\langle{\Theta} \big\rangle^{\,2}_\CD \, \right) - 2 \, \avg{\sigma^2} \; .
\end{equation}
The effective scale factor evolution equations then read:
\begin{align}
   \label{eq:avg_Hamilton}
    3 \frac{\dot a_\CD^2}{a_\CD^2} & =  8 \pi G \;\! \big\langle \rholoc \big\rangle_\CD + \Lambda - \frac{1}{2} \RD - \frac{1}{2} \QD \; ; \\
   \label{eq:avg_Raych}
    3 \frac{\ddot a_\CD}{a_\CD} & = - 4 \pi G \;\! \big\langle \rholoc \big\rangle_\CD + \Lambda + \QD \; .
\end{align}
This system is complemented by the averaged energy conservation equation:
\begin{equation}
 \label{eq:avg_energyconserv}
    \big\langle \rholoc \big\rangle_\CD^{\bdot} \, + \, 3 \, \frac{\dot a_\CD}{a_\CD} \;\! \big\langle \rholoc \big\rangle_\CD \, = 0 \;\; ;
\end{equation}
that is, like its local counterpart, the average energy density is inversely proportional to volume. This directly expresses the conservation of the total fluid rest mass $\avg{\rholoc} \VD = \int_\CD \, \rholoc \;\! \sqrt{b} \; \mathrm{d}^3 X$ within the domain comoving with the fluid flow --- although it can also be derived from averaging the local rest mass conservation equation $\dot\rholoc + \Theta \, \rholoc = 0$ and applying the commutation rule~\eqref{eq:comm_rule}.

Unlike their FLRW counterparts, Eqs.~\eqref{eq:avg_Hamilton}--\eqref{eq:avg_energyconserv} are independent, due to the additional presence of the backreaction term and the \emph{a priori} unknown evolution of the averaged curvature. Combining the time derivative of Eq.~\eqref{eq:avg_Hamilton} with Eqs.~\eqref{eq:avg_Raych} and \eqref{eq:avg_energyconserv} gives an additional, dependent \emph{integrability condition} which couples the evolutions of these two terms:
\begin{equation}
  \label{eq:intcond}
    a_\CD^{-6} \left(a_\CD^6 \, {\mathcal{Q}}_\CD  \right)^\bdot+ a_\CD^{-2} \left( a_\CD^2 \, \RD \right)^\bdot = 0 \; . 
\end{equation}
In this equation, the averaged curvature $\RD$ may equivalently be replaced by a curvature deviation variable, $\WD := \RD - 6 K_\CD /a_\CD^{2}$, expressing the discrepancy with respect to the scaling of spatial curvature in FLRW spacetimes, where $K_\CD$ is a constant which may \emph{a priori} be freely chosen for a given averaging domain---a natural choice being to equate $6 K_\CD$ to $\RD \, a_\CD^2$ at some initial time.

The two coupled contributions $\QD$ and $\WD$ encompass the deviations of the effective scale factor's evolution to that of a homogeneous $\Lambda$CDM model (which here may even have nonzero curvature). They can be combined into an effective additional perfect fluid source~\cite{Buchert_2001,BuchertMourierRoy_2018}, with effective (still domain- and thus scale-dependent) energy density and pressure $\rho_\CD^\mathrm{eff}(t)$, $p_\CD^\mathrm{eff}(t)$, given by:
\begin{align}
    \rho_\CD^\mathrm{eff}(t) & := - \frac{1}{16 \pi G} \QD - \frac{1}{16 \pi G} \WD \quad ; \label{eq:def_rhoeff} \\
    p_\CD^\mathrm{eff}(t) & := - \frac{1}{16 \pi G} \QD + \frac{1}{48 \pi G} \WD \; \; .   \label{eq:def_peff}
\end{align}
The evolution equations \eqref{eq:avg_Hamilton}--\eqref{eq:avg_Raych} for the effective scale factor can accordingly be rewritten as (domain-dependent) effective Friedmann equations including this additional source,
\begin{align}
   \label{eq:avg_Hamilton_eff}
     3 \frac{\dot a_\CD^2}{a_\CD^2} & =  8 \pi G \left( \big\langle \rholoc \big\rangle_\CD + \rho_\CD^\mathrm{eff} \right)  - 3 \frac{K_\CD}{a_\CD^2}  + \Lambda \; ; \\
   \label{eq:avg_Raych_eff}
    3 \frac{\ddot a_\CD}{a_\CD} & = - 4 \pi G \left( \big\langle \rholoc \big\rangle_\CD + \rho_\CD^\mathrm{eff} + 3 \, p_\CD^\mathrm{eff} \right) + \Lambda  \; .
\end{align}
In this system, the average dust energy density scales like its homogeneous counterpart,
\begin{equation}
    \big\langle \rholoc \big\rangle_\CD(t) \propto a_\CD^{-3}(t) \; ,
\end{equation}
from the rest mass conservation equation~\eqref{eq:avg_energyconserv}, while the additional, effective fluid source consistently obeys the expected energy conservation equation for a homogeneous perfect fluid decoupled from the dust---as a rewriting of the integrability condition~\eqref{eq:intcond}:
\begin{equation}
   \label{eq:intcond_eff}
    \dot\rho_\CD^\mathrm{eff} + 3 \frac{\dot a_\CD}{a_\CD} \left( \rho_\CD^\mathrm{eff} + p_\CD^\mathrm{eff} \right) = 0 \; .
\end{equation}
For this effective source, one can also define an effective equation of state parameter $w_{\mathrm{eff}\!\!\;,\CD}(t) \! := p_\CD^\mathrm{eff}(t) / \rho_\CD^\mathrm{eff}(t)$, whenever $\rho_\CD^\mathrm{eff}(t) \neq 0$, and an effective squared speed of sound parameter $c_{s,\mathrm{eff},\CD}^2 (t) := \dot p_\CD^\mathrm{eff}(t) / \dot\rho_\CD^\mathrm{eff}(t)$. Both parameters are time- and domain-dependent. Note that $c_{s,\mathrm{eff},\CD}^2$ may diverge at times where $\dot\rho_\CD^\mathrm{eff}(t) = 0$; from Eq.~\eqref{eq:intcond_eff}, assuming $\rho_\CD^\mathrm{eff}(t) \neq 0$, this corresponds to the effective source crossing a phantom equation of state, $w_{\mathrm{eff},\CD}(t) = -1$. We stress that this is not pathological behaviour, as these parameters emerge from the dynamics of nonlinear structures within the domain and do not correspond to the energy contents of a fundamental fluid or scalar field. We will only see that this may lead to formal issues with the definition of a perturbation scheme around such averaged dynamics as a background, specifically when relying on a barotropic parametrisation of the perturbations of the effective fluid sources (see Sec.~\ref{subsec:barotropic}, and the examples of sections~\ref{subsec:GMC} and \ref{subsec:GMP}).

The above effective fluid source terms may also be cast into an effective minimally coupled scalar field source $\Phi_\CD(t)$, still dependent on time and on the averaging domain: the \emph{morphon} field~\cite{Buchert_2006}, characterised by its domain-dependent potential and kinetic energy defined from $(\rho^\mathrm{eff}_\CD(t), p^\mathrm{eff}_\CD(t))$.
The integrability condition~\eqref{eq:intcond}, seen as the conservation equation~\eqref{eq:intcond_eff} for the effective fluid source, then becomes a minimally coupled Klein-Gordon equation for the morphon.

The system~\eqref{eq:avg_Hamilton}--\eqref{eq:intcond}, or its alternative effective-source form~\eqref{eq:avg_Hamilton_eff}--\eqref{eq:intcond_eff}---each consisting of three independent equations---is not closed in general. This is a consequence of local information being lost in spatial averaging and in considering only scalar parts of the Einstein equations. As for FLRW models, an equation of state relating $\rho_\CD^\mathrm{eff}$ and $p_\CD^\mathrm{eff}$---yet as an \emph{effective} relation in the average case---would be required as a closure condition. This could be prescribed directly as a phenomenological ansatz, or arise from the knowledge of the local dynamics within a given model. Nevertheless, even without an explicit closure condition, the above system may at least be viewed as a set of balance equations to quantify the contributions and the coupling of curvature and inhomogeneities in the dynamics of a given model, globally or for a given region and scale. In this respect, it can be relevant for instance to rewrite the averaged Hamilton constraint (e.g. under the form of Eq.~\eqref{eq:avg_Hamilton}) as a sum of time-dependent $\Omega$ parameters, as for the standard Friedmann equations---although here \emph{a priori} dependent on the averaging domain $\CD$---, via the normalisation of the equation by\footnote{%
Note that these definitions would not hold at a time where $\dot a_\CD(t) = 0$, which can occur if the region of fluid considered corresponds to a collapsing overdense region.%
} $3 H_\CD^2$:
\begin{gather}
     \Omega^\CD_d + \Omega^\CD_\Lambda + \Omega^\CD_\mathcal{R} + \Omega^\CD_\mathcal{Q} = 1 \quad ;  \nonumber \\
   \Omega^\CD_d := \frac{8 \pi G \;\! \big\langle \rholoc \big\rangle_\CD}{3 H_\CD^2}  \quad ; \quad \Omega^\CD_\Lambda := \frac{\Lambda}{3 H_\CD^2} \quad ; \quad \Omega^\CD_\mathcal{R} := - \frac{\RD}{6 H_\CD^2} \quad ; \quad \Omega^\CD_\mathcal{Q} := - \frac{\QD}{6 H_\CD^2} \quad . \label{eq:omegas} 
\end{gather}
The corresponding present-day value (at time $t=t_0$) of a given $\Omega^\CD_i$ variable for the domain $\CD$ will be denoted as $\Omega_i^{\CD,0}$.

In Sec.~\ref{sec:examples}, we will be considering several examples of averaged dynamics for inhomogeneous irrotational dust spacetimes described by these equations, arising from various models of the local dynamics, or from phenomenological ans\"atze for the backreaction and curvature contributions as an effective fluid source.

\section{Linear perturbations in an averaged spacetime}
\label{sec:Us}

We shall now introduce perturbations, described within the CGI formalism presented in Sec.~\ref{sec:Ellis&Bruni}, on top of the evolution of quantities averaged over a certain comoving region. As in the previous section, our model universe is sourced solely by one irrotational dust fluid, otherwise fully general, of inhomogeneous density $\rholoc$ and $4-$velocity field $u_\mu^{(d)}$, defining a flow-orthogonal spacetime foliation and adapted dust proper-time coordinate $t$. We select a certain averaging domain $\CD$ within the spatial hypersurfaces, comoving with the dust fluid flow, and define its effective scale factor $a_\CD(t)$ as above, which characterises its size evolution. The averaged dynamics of the domain is then determined by the Buchert equations, which can be written as effective Friedmann equations for $a_\CD(t)$ (Eqs.~\eqref{eq:avg_Hamilton_eff}--\eqref{eq:intcond_eff}) with two fluid sources: the average dust density $\avg{\rholoc}(t)$ which scales like a homogeneous dust density, and the effective perfect fluid of energy density $\rho^\mathrm{(eff)}_\CD(t)$ and isotropic pressure $p^\mathrm{(eff)}_\CD(t)$. This consistently defines a two-fluid FLRW model for the given $\CD$ characterising its average behaviour, with its spatial curvature set by the choice of $K_\CD$. Nontrivial dynamics of the averaged spatial curvature within the domain are fully encompassed in the effective fluid source.
The latter also incorporates all other contributions from local, linear and nonlinear, inhomogeneity and anisotropy---the backreaction effect---which alter the regional dynamics compared to a strictly homogeneous dust model universe with the same average density.

We can now isolate a small inhomogeneity in the matter density, e.g., at a fixed wavelength, within the domain of interest, and examine how the linear growth of this structure is influenced by a regional expansion that deviates from that of a strictly homogeneous and isotropic dust fluid. For this purpose, we map the system onto a simpler model spacetime consisting of a two-fluid almost-FLRW Universe, at all times close to the above effective FLRW model describing the average expansion of the inhomogeneous domain $\CD$, serving as a background, and with linear perturbations encompassing the additional matter structure we are focusing on. We note that the domain's average properties defining our background already include, by construction, all inhomogeneities within $\CD$ (notably contrasting with \emph{a priori} prescriptions of a homogeneous dust background). The particular small-amplitude density fluctuation that we are tracking as a linear perturbation on that background can be thought of either as one small contribution split apart and removed from the background, or as a perturbation added to the original model spacetime in which the averaged properties of $\CD$ were computed. In either case, by definition, any modification this perturbation may incur on the average properties of the domain will be at linear order, and within a first-order scheme the growth of the perturbation would remain unaffected by such a first-order shift of the background definition---even if the linear density fluctuation is not strictly required to have a vanishing spatial average.

Within this framework, the effective backreaction/curvature fluid present in the background must also be perturbed for consistency, as will be apparent below; this represents the linear local dynamical coupling of the dust perturbation to the expansion and spatial curvature fluctuations. However, the (physical) dust fluid does remain separately conserved, and we can accordingly describe this setup within the CGI perturbation framework introduced in Sec.~\ref{sec:Ellis&Bruni} as a set of two non-interacting perfect fluids, one being pressureless.

\subsection{General derivation}
\label{subsec:Us-general}

The scalar linear CGI perturbations in our two-fluid model are described by the system of equations~\eqref{eq:1}-\eqref{eq:3}, specialised to an irrotational dust component $(d)$ and an effective perfect-fluid component $(\mathrm{eff})$. We find 
\allowdisplaybreaks
\begin{adjustwidth}{-0.2cm}{-0.2cm}
\vspace{-\abovedisplayskip} 
\begin{align}
    & \dot{\Delta}^{(d)} +\mathcal{Z}  - a \, \Theta F^{(d)} + a {}^{(3)}\!\Delta V^{(d)}   \FO 0 \; ,\label{eq:system1_dust} \\
    & \dot{\Delta}^{(\mathrm{eff})} -w_\mathrm{(eff)}\Theta\Delta^{(\mathrm{eff})} + (1+w_\mathrm{(eff)})\mathcal{Z}  - a\Theta (1+w_\mathrm{(eff)})F^{(\mathrm{eff})}+(1+w_\mathrm{(eff)}) a {}^{(3)}\!\Delta V^{(\mathrm{eff})}\FO 0 \; , \label{eq:system1_eff}\\
   &     \dot{\mathcal{Z}} + \frac{2}{3}\Theta \mathcal{Z} + 4\pi G\rho \;\! \Delta + \left[ {}^{(3)}\!\Delta + \frac{ {\!}^{(3)}\!R}{2}\right]\mathcal{P} +\frac{ {\!}^{(3)}\!R}{2} \;\! a \;\! F - 12\pi G \:\! h \:\! a \;\! F + a \, {}^{(3)}\!\Delta F \FO 0 \; , \label{eq:system1_Z}\\
    &  \dot{V}^{(d)} - \dot{V}^{(\mathrm{eff})} + \frac{1}{3}\Theta V^{(d)} - \Theta \left(\frac{1}{3} - c_{s,\mathrm{(eff)}}^2  \right) V^{(\mathrm{eff})}  - \frac{1}{a}\mathcal{P}^{(\mathrm{eff})} \FO 0 \; , \label{eq:system1_V}
\end{align}
\end{adjustwidth}
where we used $w_{(d)} = c^2_{s(d)}=0$, $\mathcal{P}^{(d)} = 0$, and derived Eq.~\eqref{eq:system1_V} as the difference between the two realisations of Eq.~\eqref{eq:3} for the dust fluid and the effective fluid. We note that the variables $\rho^{(d)}$; $\rho\eff$; $p\eff$ (through the total $\rho = \rho^{(d)}+\rho\eff$ and $h = \rho + p\eff$); $w_{(\mathrm{eff})}$; $c^2_{s(\mathrm{eff})}$; $a$; and $\Theta$, appearing in this system, all multiply first-order quantities. Hence, they can as well be reduced to their respective time-dependent background values, i.e.: $\avg{\rholoc}$; $\rho\eff_\CD$; $p\eff_\CD$; $w_{\mathrm{eff},\CD}$; $c^2_{s,\mathrm{eff},\CD}$; $a_\CD$; and $3 \;\! \dot a_\CD / a_\CD$.

The same holds for the $3-$Laplacian ${}^{(3)}\!\Delta$, which is applied to first-order variables and thus can be reduced to the spatial Laplace-Beltrami operator of the FLRW background; and it also holds for the scalar $3-$curvature term $\sR$, which can be reduced to its background value $- 6 K_\CD / a_\CD^2$. However, we stress that this background $3-$curvature is in general distinct from the averaged spatial curvature over the domain $\CD$ within the original fully inhomogeneous spacetime. Indeed, the evolution of the averaged curvature is coupled to the kinematical backreaction and therefore does not need to scale as ${} \propto a_\CD^{-2}(t)$. This difference in scaling underlines that the effects of a nontrivial spacetime geometry are directly captured in our approach at the \emph{dynamical} level, through the background effective fluid source (here, specifically through the averaged Hamiltonian constraint), and do not necessarily have a \emph{geometric} counterpart in the effective FLRW that serves as background for the CGI perturbation variables. Additionally, the choice of $K_\CD$ does not even \emph{a priori} need to be such that the background and average curvature coincide at an initial time. The residual ambiguity in choosing this background curvature parameter can naturally be mitigated by requiring the topology of the almost-FLRW model to match that of the original inhomogeneous Universe, thereby constraining the sign of $K_\CD$. Moreover, the choice of background curvature appears to be ultimately meaningful only in relation to the closure conditions for the full set of perturbation evolution equations (see Eqs.~\eqref{eq:system2_dust}--\eqref{eq:system2_V} and related discussion below, and, e.g., Sec.~\ref{subsec:comoving}). We therefore develop our analysis of the linear CGI perturbations regardless of the ultimate choice of background curvature.

As discussed in Sec.~\ref{sec:Ellis&Bruni}, to close the system of evolution equations for the scalar CGI perturbation variables we need to (i) specify the observer rest-frame---thus introducing a further relation between the tilt velocities---and (ii) introduce an ansatz for the EoS of the effective fluid. Thus, we now turn to the selection of a physically relevant observer rest-frame. We argue that the dust energy frame, namely $V^{(d)} = 0$, represents the natural choice in our framework. We wish to ultimately interpret the CGI density perturbations in terms of gradients in the \emph{observed} density field of the matter-dominated Universe. We should then select the observer rest-frame to coincide with that of real potential observers (and light sources), which, on cosmological scales, would typically be comoving with the dust fluid. Moreover, as we employ the fluid-orthogonal averaging scheme of~\cite{Buchert_2000} to introduce the regional averaged dynamics and the corresponding effective background, it is meaningful to select the same foliation slices employed in the averaging procedure as the observer rest frames\footnote{%
The Buchert averaging scheme recalled above in Sec.~\ref{sec:buchert} is specifically built within the global rest frames of the irrotational dust. Extending it to a generic choice of foliation introduces a few additional terms in the averaged dynamics~\cite{BuchertMourierRoy_2018}. Furthermore, while this general formalism is both covariant and foliation-invariant, individual average terms such as the effective sources are defined up to a time-dependent factor set by the choice of lapse function~\cite{BuchertMourierRoy_2020}, and have a first-order dependence in the tilt velocity selecting the foliation~\cite{Mourier_2024}. Hence, for an irrotational dust spacetime, it is convenient to restrict the analysis and interpretation of CGI perturbations to within the same foliation employed for the averaging procedure.%
}.

We therefore cast Eqs.~\eqref{eq:system1_dust}--\eqref{eq:system1_V} in terms of the reference frame of an observer comoving with the dust, and obtain:
\begin{adjustwidth}{-0.2cm}{-0.2cm}
\vspace{-\abovedisplayskip} 
\begin{align}
    & \dot{\Delta}^{(d)} +\mathcal{Z}\FO 0 \; ,\label{eq:system2_dust} \\
    & \dot{\Delta}^{(\mathrm{eff})} -w_\mathrm{(eff)}\Theta\Delta^{(\mathrm{eff})} + (1+w_\mathrm{(eff)})\mathcal{Z} + \Theta (1+w_\mathrm{(eff)})\mathcal{P}^{(\mathrm{eff})}+a \:\!(1+w_\mathrm{(eff)}) \:\! {}^{(3)}\!\Delta V^{(\mathrm{eff})}\FO 0 \; , \label{eq:system2_eff}\\
   &     \dot{\mathcal{Z}} + \frac{2}{3}\Theta \mathcal{Z} + 4\pi G\left(\rho^{(d)}\Delta^{(d)}+\rho^{(\mathrm{eff})}\Delta^{(\mathrm{eff})}\right)+12\pi G h^{(\mathrm{eff})}\mathcal{P}^{(\mathrm{eff})} \FO 0 \; , \label{eq:system2_Z}\\
    &   \dot{V}^{(\mathrm{eff})} + \Theta \left(\frac{1}{3} - c_{s,\mathrm{(eff)}}^2  \right) V^{(\mathrm{eff})}  + \frac{1}{a}\mathcal{P}^{(\mathrm{eff})} \FO 0 \; , \label{eq:system2_V}
\end{align}
\vspace{-\belowdisplayskip} 
\end{adjustwidth}
where we have used $V^{(d)} = F^{(d)} = 0$, re-expressed $\Delta$ in terms of $\Delta^{(d)}$ and $\Delta^{(\mathrm{eff})}$, and employed Eq.~\eqref{eq:system2_V} to replace $F^{(\mathrm{eff})}$ in Eq.~\eqref{eq:system2_eff}. In Eq.~\eqref{eq:system2_Z}, we have also used the relation $\mathcal{P} + a \;\! F = 0$, which arises from the vanishing of the $4-$acceleration in the reference dust frame, expressed through Eq.~\eqref{eq:acceleration_tot}. In this way, adopting the dust geodesic reference frame removes the \emph{direct} influence of the background $3-$curvature on the evolution of the CGI perturbations (\emph{via} the $\sR$ terms in Eq.~\eqref{eq:system1_Z}), which only still plays a role through the spectrum of the $3-$Laplacian operator in Eq.~\eqref{eq:system1_Z} for a nonzero $V\eff$.

The evolution equation for $V\eff$, Eq.~\eqref{eq:system2_V}, explicitly contains the background sound speed of the effective fluid. However, the effective fluid, as such, may present phantom crossings in its dynamical background evolution, where the sound speed formally diverges and the above system of ODEs becomes ill-defined. It is then convenient to introduce the change of variable $\tilde{V}^{(\mathrm{eff})} :=(1+w_{(\mathrm{eff})}) V^{(\mathrm{eff})}$, which eliminates this explicit dependence on $c^2_{s,\mathrm{(eff)}}$ (see Eq.~\eqref{eq:system4_V} below). Likewise, we introduce the rescaled variable $\tilde{\mathcal{P}}^{(\mathrm{eff)}}:=(1+w_{(\mathrm{eff})}) \mathcal{P}^{(\mathrm{eff})}$  to further simplify the system.
Moreover, as we are dealing with linear perturbations relative to an effective homogeneous background, we can further simplify the study of the system dynamics by moving to spatial frequency space\footnote{%
Given the ambiguity in the definition of the background spatial curvature, hereafter we denote by $k^2/a^2$ the eigenvalues of the (background) $3-$Laplacian operator regardless of the topology assigned to the $3-$space. The spatial frequency space is the space of the corresponding eigenfunctions, corresponding to Fourier space in the case of a spatially flat background. %
}. We find 
\begin{align}
    & \dot{\Delta}^{(d)} +\mathcal{Z}\FO 0 \; ,\label{eq:system4_dust} \\
    & \dot{\Delta}^{(\mathrm{eff})} -w_\mathrm{(eff)}\Theta\Delta^{(\mathrm{eff})} + (1+w_\mathrm{(eff)})\mathcal{Z} +\Theta \tilde{\mathcal{P}}^{(\mathrm{eff)}}-\frac{k^2}{a}\tilde{V}^{(\mathrm{eff})}\FO 0 \; , \label{eq:system4_eff}\\
   &     \dot{\mathcal{Z}} + \frac{2}{3}\Theta \mathcal{Z} + 4\pi G\left(\rho^{(d)}\Delta^{(d)}+\rho^{(\mathrm{eff})}\Delta^{(\mathrm{eff})}\right)+12\pi G \rho^{(\mathrm{eff})}\tilde{\mathcal{P}}^{(\mathrm{eff)}} \FO 0 \; , \label{eq:system4_Z}\\
    &   \dot{\tilde{V}}^{(\mathrm{eff})} + \Theta \left(\frac{1}{3} - w_{\mathrm{(eff)}}  \right) \tilde{V}^{(\mathrm{eff})}  +\frac{1}{a}\tilde{\mathcal{P}}^{(\mathrm{eff})} \FO 0 \; . \label{eq:system4_V}
\end{align}
The system of ordinary evolution equations \eqref{eq:system4_dust}--\eqref{eq:system4_V} is the one we will employ hereafter.

An additional relation is still required as a closure condition of the system, which amounts to specifying one more constraint on the effective fluid's perturbation variables. Indeed, the energy-momentum contribution of the effective fluid source is only uniquely defined at the background level (as $\rho\eff_\CD(t)$, $p\eff_\CD(t)$), through the averaged properties of the original domain $\CD$. In a perturbed setting, further assumptions must be made in order to characterise the spatial gradients of its energy density and pressure, and its velocity relative to the dust flow, besides considering, as above, that it remains a perfect fluid. As mentioned above, this fluid component cannot simply be assumed to remain strictly homogeneous, since it is coupled, gravitationally, to the dust fluctuations, e.g. through the total density perturbation appearing in the perturbed Raychaudhuri equation~\eqref{eq:system4_Z}. Indeed, if the perturbations in the effective fluid's density and pressure, $\Delta\eff$, $\tilde \CP\eff$, are both assumed to vanish identically, it can be shown that the system of four ODEs \eqref{eq:system4_dust}--\eqref{eq:system4_V} reduced to the remaining three perturbation variables $\{\Delta^{(d)}, \, \CZ , \, \tilde V\eff \}$ is indeed overdetermined, and forces all perturbations to zero. A less restrictive, single additional relation between the three perturbation variables of the effective fluid should thus be imposed, for which there is no unique choice. In the following subsections, we turn to discussing a series of possible physically motivated ans\"atze for closure conditions or approximations for the system of evolution equations. A further example of possible closure condition and the resulting closed system of perturbation evolution equations are presented in Appendix~\ref{appendix:mc}.

\subsection{Barotropic effective fluid}
\label{subsec:barotropic}

A common assumption for non-interacting perfect fluids is to assume that they retain a barotropic relation $p^{(i)} = p^{(i)}(\rho^{(i)})$ throughout their evolution, i.e., to neglect the entropy perturbation $\mathcal{E}^{(i)}_\mu$ in Eq.~\eqref{eq:entropy_perturb}. In terms of scalar perturbations and to first order, for the corresponding fluid $(i)$, this implies $\tilde \CP^{(i)} := (1 + w_{(i)}) \CP^{(i)} = c^2_{s(i)} \;\! \Delta^{(i)}$ (in any observer frame $n^\mu$), where $c_{s(i)}^2$ is evaluated in the background, equivalently as $\partial p^{(i)} / \partial \rho^{(i)}$ or as $\dot p^{(i)} / \dot \rho^{(i)}$. Similarly, we may impose that the perturbed effective fluid retains its background pressure to density relation $p\eff_\CD(t) = p\eff_\CD \big[\rho\eff_\CD (t) \big]$ up to first order:
\begin{equation}
    \tilde \CP\eff = c^2_{s(\mathrm{eff})}  \, \Delta\eff \; ,
    \label{eq:barotropic_condition}
\end{equation}
where $c_{s(\mathrm{eff})}^2$ is understood as  the background effective squared speed of sound $c^2_{s,\mathrm{eff},\CD}(t) = \dot p^\mathrm{eff}_\CD / \dot\rho^\mathrm{eff}_\CD$ within the first-order perturbative scheme. 

With this closure relation used to rewrite $\tilde \CP\eff$, the system of evolution equations for the first-order scalar CGI variables~\eqref{eq:system4_dust}-\eqref{eq:system4_V} becomes:
\begin{align}
    & \dot{\Delta}^{(d)}  + \CZ  \FO 0 \; ;  \label{eq:dotdeltad_bar} \\
    & \dot{\Delta}^{(\mathrm{eff})}  -\left(w_{(\mathrm{eff})}+c^2_{s(\mathrm{eff})}\right)\Theta\Delta^{(\mathrm{eff})}  + (1+w_{(\mathrm{eff})}) \CZ  -\frac{k^2}{a}\tilde{V}^{(\mathrm{eff})}  \FO 0 \; ; \label{eq:dotdeltaeff_bar} \\
    & \dot{\CZ}  + \frac{2}{3}\Theta \CZ  + 4\pi G\left[\rho^{(d)} \:\! \Delta^{(d)}  +\rho\eff \left(3 \;\! c^2_{s(\mathrm{eff})}+1\right)\Delta^{(\mathrm{eff})}  \right] \FO 0 \; ; \label{eq:dotZ_bar} \\
    & \dot{\tilde{V}}^{(\mathrm{eff})}  + \Theta \left(\frac{1}{3} -  w_{(\mathrm{eff})} \right)\tilde{V}^{(\mathrm{eff})}  +\frac{1}{a} c^2_{s(\mathrm{eff})} \, \Delta^{(\mathrm{eff})}  \FO 0 \; , \label{eq:dotV_bar} 
\end{align}

One may then eliminate $\CZ$ straightforwardly through Eq.~\eqref{eq:dotdeltad_bar}, as well as $\tilde V\eff$ via Eq.~\eqref{eq:dotdeltaeff_bar}, to obtain the two coupled second-order (in derivatives) ODEs satisfied by the density perturbations:
\begin{align}
    & \ddot{\Delta}^{(d)} +\frac{2}{3}\Theta \dot{\Delta}^{(d)}  -4\pi G\rho^{(d)} \Delta^{(d)}  \FO 4\pi G \rho\eff \left(1+ 3 \;\! c^2_{s(\mathrm{eff})}\right)\Delta^{(\mathrm{eff})}  \; ;\label{eq:ddotdeltad_bar} \\
    & \ddot\Delta^{(\mathrm{eff})}  + \left(\frac{2}{3} - 2 \:\! w_{(\mathrm{eff})} - c_{s(\mathrm{eff})}^2 \right) \Theta \dot\Delta^{(\mathrm{eff})}  + \left[ \frac{2}{3} \Theta^2 \left( - 2 \:\! w_{(\mathrm{eff})} + c_{s(\mathrm{eff})}^2 (1 + 3 \:\! w_{(\mathrm{eff})}) \right)  - \Theta \csdot \right. \nonumber \\
    & \quad + \left( 4 \pi G \rho^{(d)}  - \Lambda \right) \left(w_{(\mathrm{eff})} + c_{s(\mathrm{eff})}^2 \right)  
     - 4 \pi G \rho\eff \left(1 - 3 w_{(\mathrm{eff})}^2 + 2 c_{s(\mathrm{eff})}^2 \right) + c_{s(\mathrm{eff})}^2 \frac{k^2}{a^2}  \bigg]  \Delta^{(\mathrm{eff})}  \nonumber \\
     & \quad \qquad \FO 4 \pi G \rho^{(d)} (1 + w_{(\mathrm{eff})}) \Delta^{(d)}  - c_{s(\mathrm{eff})}^2 (1 + w_{(\mathrm{eff})}) \, \Theta \dot\Delta^{(d)}  \; , \label{eq:ddotdeltaeff_bar}
\end{align}
where the Raychaudhuri equation~\eqref{eq:Raychaudhuri} (with $a^\mu = 0$) has been used at the background level to eliminate $\dot\Theta$. The evolution equation for the dust perturbation itself, Eq.~\eqref{eq:ddotdeltad_bar}, would reduce to the familiar linear perturbation equation of a pure-dust FLRW background (possibly with $\Lambda$ and spatial curvature) if its right-hand side vanished. However, as discussed above, the effective fluid density perturbation $\Delta\eff $ cannot be assumed to identically vanish (which would also set $\tilde\CP\eff  = 0$ in this case), as this would also enforce a vanishing dust density perturbation.
A general source-free evolution equation for the dust perturbation alone may however still be derived for this barotropic case by combining the above two equations. The resulting fourth-derivative-order equation is provided for completeness in App.~\ref{appendix:4thorder}.

We emphasize that the barotropic condition~\eqref{eq:barotropic_condition} is only one particular ansatz for the effective density and pressure perturbations. The effective fluid source, as such, does not have a fundamental EoS or well-defined entropy variations. The barotropic property of its defining homogeneous background can, in fact, be violated at a certain point in time $t_1$ if $\dot \rho^\mathrm{eff}_\CD(t_1)$ vanishes at that time while $\dot p^\mathrm{eff}_\CD(t_1)$ does not. As noted above in Sec.~\ref{sec:buchert}, this would correspond to this background source crossing a phantom effective EoS, $w_{(\mathrm{eff})}(t_1) = -1$, and to a diverging $c_{s(\mathrm{eff})}^2$ at that point. While this is no fundamental issue with the dynamics of the averages which define the background, the barotropic condition makes the background effective sound speed appear in all forms of the perturbation evolution equations above, and thus the perturbative framework would generally cease to be valid at such a time within this ansatz. This situation will indeed occur in some of the examples we consider below in Sec.~\ref{sec:examples}.

\subsection{Comoving effective fluid}\label{subsec:comoving}

The background effective fluid is defined from average properties over a spatial domain within the hypersurfaces orthogonal to the dust $4-$velocity, the domain propagating along that same dust flow between slices. Accordingly, another natural possible constraint for the perturbed model spacetime would be to define the effective fluid $4-$velocity as comoving with the dust flow, as would be the case for an unperturbed, FLRW two-fluid model. This also amounts to tying the effective fluid to the $4-$velocity $n^\mu$ of the observers, which we have already picked as coinciding with the dust frame.

We may thus set the comoving condition,
\begin{equation}
    u_\mu\eff = u_\mu^{(d)} = n_\mu \quad : \quad V_\mu\eff = 0 \;\; .
\end{equation}
In terms of the scalar CGI variables, this implies $V\eff = \tilde V\eff = 0$, and Eq.~\eqref{eq:system4_V} reduces to also setting the effective first-order pressure perturbation to zero, $\tilde\CP\eff \FO 0$. This is consistent with the effective fluid now being also geodesic. The other three evolution equations~\eqref{eq:system4_dust}--\eqref{eq:system4_Z} then couple the remaining three perturbation variables $\Delta^{(d)}$, $\Delta\eff$ and $\CZ$, and read in this case:
\begin{align}
    & \dot{\Delta}^{(d)} +\mathcal{Z}\FO 0 \; ,\label{eq:dotdeltad_com} \\
    & \dot{\Delta}^{(\mathrm{eff})} -w_\mathrm{(eff)}\Theta\Delta^{(\mathrm{eff})} + (1+w_\mathrm{(eff)})\mathcal{Z} \FO 0 \; , \label{eq:dotdeltaeff_com}\\
   &     \dot{\mathcal{Z}} + \frac{2}{3}\Theta \mathcal{Z} + 4\pi G\left(\rho^{(d)}\Delta^{(d)}+\rho^{(\mathrm{eff})}\Delta^{(\mathrm{eff})}\right)  \FO 0 \; , \label{eq:dotZ_com}
\end{align}
Whilst this system is significantly simpler overall than what we obtained above for the barotropic ansatz (Eqs.~\eqref{eq:dotdeltad_bar}--\eqref{eq:dotZ_bar}), its most remarkable feature is its reduction to ordinary differential equations even when not written in spatial frequency space. Hence, in the comoving ansatz, the dynamics of the perturbations are independent of their scale. This is due to the only coupling between different observers (or the only dependence in the wavenumber $k$) in the general system appearing via the $3-$Laplacian of $V\eff$ in Eq.~\eqref{eq:system2_eff} (or via the $(k^2/a) \;\! \tilde V\eff$ term in Eq.~\eqref{eq:system4_eff}), and thus disappearing in the comoving case. Remarkably, the disappearance of the spatial Laplacian operator also implies that there is no more dependence of the linear evolution on the spatial curvature choice for the background in this picture, besides the direct effect on the time evolution of the background coefficients $\Theta$, $w_{(\mathrm{eff})}$, $\rho^{(d)}$, $\rho\eff$. 

As for the barotropic ansatz above, Eq.~\eqref{eq:dotdeltad_com} can be inserted into Eq.~\eqref{eq:dotZ_com}, which then becomes a second-order differential equation on the dust perturbation $\Delta^{(d)}$, sourced by the effective density perturbation $\Delta\eff$,
\begin{equation}
    \ddot\Delta^{(d)} + \frac{2}{3}\Theta \dot\Delta^{(d)} - 4\pi G \rho^{(d)}\Delta^{(d)} \FO 4 \pi G \rho^{(\mathrm{eff})}\Delta^{(\mathrm{eff})} \; . \label{eq:ddotdeltad_com}
\end{equation}
The evolution of $\Delta\eff$, in turn, can be given in this case directly by Eq.~\eqref{eq:dotdeltaeff_com} above, as a first-order differential equation sourced by $- (1 + w_{(\mathrm{eff})} ) \;\! \CZ = (1 + w_{(\mathrm{eff})}) \, \dot\Delta^{(d)}$. Alternatively, this equation may be combined with its derivative, with Eq.~\eqref{eq:dotZ_com}, and with the Raychaudhuri equation~\eqref{eq:Raychaudhuri} evaluated for the background, to give instead a second-order differential equation for $\Delta\eff$ sourced directly by $\Delta^{(d)}$, and thus coupled to the above Eq.~\eqref{eq:ddotdeltad_com} on $\Delta^{(d)}$:
\begin{align}
        & \ddot{\Delta}^{(\mathrm{eff})}+ \left(\frac{2}{3} + c_{s(\mathrm{eff})}^2 - 2 \:\! w_{(\mathrm{eff})} \right) \Theta \dot{\Delta}^{(\mathrm{eff})} +\left[\left(c_{s(\mathrm{eff})}^2 - \frac{4}{3} w_{(\mathrm{eff})} \right) \Theta^2 - \Lambda w_{(\mathrm{eff})} \right. \nonumber \\
        & \quad \left. \vphantom{\frac{3}{3}} + 4\pi G \rho^{(d)} \:\! w_{(\mathrm{eff})} - 4\pi G \rho\eff \left(1- 3 \:\! w_{(\mathrm{eff})}^2 \right) \right] \Delta^{(\mathrm{eff})}  \FO 4\pi G \rho^{(d)} (1+w_{(\mathrm{eff})}) \Delta^{(d)} \; .
        \label{eq:ddotdeltaeff_com}
\end{align}
This rewriting is done at the expense of recovering contributions from the background squared speed of sound $c^2_{s,\mathrm{eff},\CD}(t)$, and thus possible formal divergences in the perturbative equations upon crossing a phantom effective EoS, $w^{\mathrm{eff}}_\CD = -1$. 

Finally, the effective-fluid perturbation may be eliminated from Eq.~\eqref{eq:dotZ_com} and its derivative via Eq.~\eqref{eq:dotdeltaeff_com}. Using again the background Raychaudhuri equation to rewrite $\dot\Theta$, and Eq.~\eqref{eq:dotdeltad_com}, this results in a second-order decoupled evolution equation for $\CZ$:
\begin{equation}
    \ddot \CZ + \frac{5}{3} \Theta \dot \CZ  + \left[ \frac{4}{9} \Theta^2 - \frac{20 \pi G}{3} \rho^{(d)} - \frac{4 \pi G}{3} \rho\eff (5 + 9 \:\! w_{(\mathrm{eff})}) + \frac{2}{3} \Lambda \right] \CZ \FO 0 \; .
    \label{eq:ddotZ_comoving}
\end{equation}
With $\CZ = -\dot{\Delta}^{(d)}$, the above relation also represents a third-order, source-free, and scale-independent differential equation that governs the evolution of the dust density perturbation $\Delta^{(d)}$, interestingly containing no terms proportional to $\Delta^{(d)}$ itself.

The trivial $\CZ = 0$ solution to the above source-free equation implies that, at any given point in space or wavelength, a constant $\Delta^{(d)}$ is a particular solution of the perturbation dynamics in this setup. Through Eq.~\eqref{eq:dotZ_com}, this corresponds (at that point or wavelength) to a vanishing total density perturbation, $\Delta \FO 0$, i.e., to $\Delta\eff(t) \FO - (\avg{\rholoc} (t) / \rho_\CD^\mathrm{eff}(t) ) \, \Delta^{(d)}$.

\subsection{Effective M\'esz\'aros approximation}
\label{subsec:meszaros}

Alternatively to prescribing a closure condition for Eqs.~\eqref{eq:system4_dust}--\eqref{eq:system4_V}, it is possible, under certain assumptions, to employ an effective M\'esz\'aros approximation within the perturbative scheme (see~\cite{Giani_2025a,Giani_2025b} for a recent implementation). The M\'esz\'aros approximation~\cite{Meszaros_1974, KodamaSasaki_1984}, originally formulated to describe the growth of sub-horizon dust perturbations in a dust--radiation multi-fluid background, effectively neglects the coupling between the perturbations of the two fluids. Instead, it encodes the influence of the non-dust fluid (conventionally radiation) solely at the background level; namely, it prescribes as the evolution equation for the dust perturbations (using CGI variables),
\begin{equation}
\ddot{\Delta}^{(d)} + \frac{2}{3}{\Theta} \dot{\Delta}^{(d)} - 4\pi G{\rho^{(d)}}\Delta^{(d)} \FO 0 \; , \label{eq:original_Meszaros}
\end{equation}
where ${\Theta}$ is the background expansion scalar driven by both fluid components, and $\rho^{(d)}$ (which can be reduced to its background value $\avg{\rholoc}$) evolves according to the full background scale factor. Physically, the M\'esz\'aros approximation is applicable whenever the pressure of the non-dust component suppresses its own clustering on small scales, as in the dust--radiation case, so that the non-dust perturbation remains very small and the dust one evolves primarily under the influence of the two-fluid background. Therefore, the validity of the M\'esz\'aros approximation for such a two-fluid system ultimately rests on the positivity of the equation of state parameter and squared sound speed of the non-dust fluid.

To illustrate this point, let us consider Eqs.~\eqref{eq:system4_dust}--\eqref{eq:system4_eff} for an effective fluid such that $c^2_{s(\mathrm{eff})} = w_{(\mathrm{eff})} = \mathrm{const}$ (as is the case for physical fluids with linear EoSs, such as radiation). We also assume adiabatic perturbations for the effective fluid, i.e., a barotropic picture, with $\tilde{\mathcal{{P}}}^{(\mathrm{eff)}} = w_{(\mathrm{eff})} \, \Delta^{(\mathrm{eff)}}$, and write the second-order evolution equations for both the dust and effective fluid linear perturbations. We find
\begin{align}
     &\ddot{\Delta}^{(d)} + \frac{2}{3}\Theta\dot{\Delta}^{(d)} - 4\pi G\rho^{(d)}\Delta^{(d)} \FO 4\pi G \rho^{(\mathrm{eff})}(1+3w_{(\mathrm{eff})})\Delta^{(\mathrm{eff)}} \; ; \label{eq:MZ1}\\    
     &\ddot{\Delta}^{(\mathrm{eff})} + \frac{2}{3}\Theta\dot{\Delta}^{(\mathrm{eff})} + \frac{k^2}{a^2}w_{\mathrm{(eff)}}\left(\Delta^{\mathrm{(eff)}}-a\Theta\tilde{V}\eff \right)
     \FO {} \nonumber \\
     & \qquad \qquad 4\pi G(1+w_{(\mathrm{eff})})\left[ (1+3w_{(\mathrm{eff})})\rho^{(\mathrm{eff})}\Delta^{(\mathrm{eff)}} +  \rho^{(d)}\Delta^{(d)}\right] \; . \label{eq:MZ2}
\end{align}
Note that we have kept a term involving $\tilde V\eff$, namely $\Delta^{\mathrm{(eff)}} - a\Theta \tilde{V}\eff$, which represents the energy density perturbations of the effective fluid in its own rest frame. On sufficiently small scales however, the tilt contribution in that term can be safely neglected. Indeed, from Eq.~\eqref{eq:scalar_part_fluid} we see that, in spatial frequency space, $\tilde{V}\eff$ scales linearly with $k$, while $\Delta^{\mathrm{(eff)}}$ exhibits a quadratic scaling. Then, considering perturbations much below the horizon scale ($k/a \gg \Theta^{-1}$), we observe from Eq.~\eqref{eq:MZ2} that $\Delta^{\mathrm{(eff)}}$ exhibits an oscillatory, non-growing behaviour whenever $w_{\mathrm{(eff)}} > 0$. That is, in that case, the pressure of the effective fluid counteracts the growth of its perturbations on small scales. Consequently, for an initially very small $|\Delta\eff|$, the right-hand side of Eq.~\eqref{eq:MZ1} can be neglected in the evolution of the dust perturbations, thereby recovering the M\'esz\'aros approximation given in Eq.~\eqref{eq:original_Meszaros}. Interestingly, in the special case of a constant ``curvature-like'' EoS, $w_{(\mathrm{eff})} = c^2_{s(\mathrm{eff})} = \mathrm{const} = -1/3$, the right-hand side of Eq.~\eqref{eq:MZ1} vanishes exactly, hence also leading to the above evolution equation~\eqref{eq:original_Meszaros} despite a possibly growing $\Delta\eff$---this time at all scales, and without approximation beyond the linearisation of perturbations.

We must stress that the M\'esz\'aros approximation cannot be applied \emph{a priori} to an arbitrary \{dust + secondary fluid\} system. In particular, this limitation arises when the secondary fluid---emerging as an effective source from a covariant averaging procedure applied to an inhomogeneous universe---is not guaranteed to have either a positive equation of state or a well-defined sound speed. Indeed, for the illustrative subcase discussed in Eqs.~\eqref{eq:MZ1} and \eqref{eq:MZ2}, a negative EoS would instead lead to an unstable growth regime at small scales for the perturbations of that fluid. Consequently, the applicability of the approximation must be carefully assessed within each model.

In Sec.~\ref{sec:examples}, for each of the averaged cosmological models we consider, we discuss the applicability of the M\'esz\'aros approximation and use it as a reference against which to compare the results obtained with the previously introduced closure conditions.

\section{Structure growth examples}
\label{sec:examples}

We now turn to several examples of structure growth within cosmological models that employ averaged dynamics of inhomogeneous, irrotational dust spacetimes to describe the evolution of the Universe or a finite region thereof. Specifically, we consider the timescape cosmology~\cite{Wiltshire_2007a,Wiltshire_2007b,Wiltshire_2009a,Wiltshire_2009b,Wiltshire_2014}, the Giani--von Marttens--Camilleri (GMC) and Giani--von Marttens--Piattella (GMP) models~\cite{Giani_2025a,Giani_2025b}, as well as a description of a local patch of the Universe based on the Relativistic Zel’dovich Approximation (RZA)~\cite{BuchertRZA_2012,BuchertRZA_2013,AllesRZA_2015,AlRoumiRZA_2017,LiRZA_2018,GasparBuchertRZA_2021}.

These cosmological models enable us to investigate the role of backreaction in shaping linear structure growth across different phenomenological scenarios. In the timescape framework, backreaction effects (including dynamical curvature) fully account for the inferred acceleration of the Universe, replacing conventional dark energy. In contrast, the GMC and GMP models incorporate cosmic-web backreaction primarily to induce a dynamical evolution within the dark energy sector, which nevertheless remains predominantly sourced by a cosmological constant. These three models are intended as potential full effective descriptions of the large-scale cosmological dynamics, with an averaging domain well above the scale of statistical homogeneity. Finally, our fourth, RZA-based example provides an effective description of how deviations of a local cosmic average from the global evolution influence the development of local perturbations. By analysing these complementary phenomenologies, we aim to investigate the coupling between backreaction from nonlinear structure formation and the linear growth of matter perturbations in a unified manner.

In the following subsections, we examine each model in turn. After outlining their averaged dynamics and the background properties of the associated effective fluid, we focus on the growth of dust perturbations under the two alternative closure conditions introduced in Sec.~\ref{sec:Us}. Note that, since we focus on structure growth in the matter-dominated era, in each model considered we neglect the sub-dominant radiation component in both the background and perturbations. We use identical initial conditions for the perturbations for each model, thereby enabling a direct and quantitative comparison between their respective predictions. To solve the systems of ODEs that determine the evolution of the perturbations, we numerically integrate the equations using numerical solvers from the \texttt{scipy} Python library~\cite{Virtanen_2020}. Our code, for each of the models and closure conditions, is available at~\cite{myrepo}.

For all of these examples, we will choose a spatially flat model for our two-fluid FLRW background ($K_\CD = 0$) for simplicity. For the perturbations in the barotropic representation, whose growth does depend on their scale, this implies that the space of eigenfunctions of the background spatial Laplacian is the ordinary Fourier space. Henceforth, when describing the scale of a given barotropic-framework perturbation as `horizon' or `sub-horizon', we will be referring to its status at present time. That is, we compare the comoving wavelength of the perturbations on this flat space with the value of $c/H_0$ in the $\Lambda$CDM model (following the best-fit of~\cite{Planck_2018}).

\subsection{The timescape model}
\label{subsec:ts}

The timescape cosmology~\cite{Wiltshire_2007a,Wiltshire_2007b,Wiltshire_2009a,Wiltshire_2009b,Wiltshire_2014} represents the first example in the literature of an averaged cosmological model applied to observational data~\cite{Smale_2011,Dam_2017,Lane_2025,Seifert_2024}. In particular, it has been shown to provide an excellent fit to Type~Ia supernova observations~\cite{Smale_2011,Dam_2017}, even outperforming the standard $\Lambda$CDM model in this context with recent observations~\cite{Lane_2025,Seifert_2024}. Interestingly, the model provides an alternative perspective on the origin of dark energy, proposing that it arises purely as a geometric effect, resulting from backreaction phenomena coupled to calibration effects of the varying clock rates among cosmological observers.

\subsubsection{Timescape: theoretical framework}
Within the timescape framework, the Universe is partitioned into two distinct domain classes, walls and voids~\cite{Wiltshire_2007a,Wiltshire_2007b,Wiltshire_2009a,Wiltshire_2009b,Wiltshire_2014}. Physically, walls correspond to marginally bound regions of the Universe, containing overdense matter structures, whereas voids are identified with underdense domains undergoing faster-than-average expansion.

Each domain is simplified and modelled by a dust-sourced FLRW spacetime, with no cosmological constant, and with a curvature depending on the domain type: spatially flat for walls\footnote{%
We note that, given the physical identification of walls as marginally bound regions and their mathematical description via an FLRW geometry, the curvature within such idealised wall regions is not a free choice within the framework, but instead must be vanishing by construction.%
} and negatively curved for voids. In addition, a third domain type with zero volume measure is formally introduced as a boundary to ensure the correct general-relativistic matching of walls and voids within a single spacetime. However, this domain remains ultimately unspecified, as the timescape cosmological model arises as a volume-weighted average, constructed according to Buchert's covariant averaging scheme~\cite{Buchert_2000} (described hereabove in Sec.~\ref{sec:buchert}), over an observable-Universe--sized domain $\CD$. The volume of $\CD$ is the sum of those of its two nonzero-measure subdomains, i.e., $v$ (union of all void regions within $\CD$) and $w$ (union of all wall regions within $\CD$). Thus, the Universe's volume fractions occupied by voids, $f_v(t) := \CV_v / \VD$, and by walls, $f_w(t) := \CV_w / \VD$, satisfy $f_v + f_w = 1$. In terms of the effective scale factors $a_\CD$, $a_v$, $a_w$---each defined from their corresponding (sub)domain volume and normalised at a given initial time $\tinitb$, i.e., $a_Y(t) := (\CV_Y(t) / \CV_Y(\tinitb))^{1/3}$, $Y = \CD,v,w$---this reads~\cite{Wiltshire_2007a,Wiltshire_2007b,Wiltshire_2009a,Wiltshire_2009b,Wiltshire_2014}:
\begin{equation}
    a_\CD^3 = f_v(\tinitb) \, a_v^3 + f_w(\tinitb) \, a_w^3\; . \label{ts:avg}
\end{equation}
Furthermore, the domain-averaged expansion rate $H_\CD = \avg{\Theta} /3$ can be expressed in an analogous two-scale decomposition in terms of the expansion rates within voids ($H_v$) and walls ($H_w$) at any given time, namely~\cite{Wiltshire_2007a,Wiltshire_2007b,Wiltshire_2009a,Wiltshire_2009b,Wiltshire_2014}:
\begin{equation}
H_\CD = f_v H_v + (1 - f_v) H_w \; .
\end{equation}

Additionally, as a first closure condition for the Buchert equations determining the evolution of the domain's effective scale factor, Eqs.~\eqref{eq:avg_Hamilton}--\eqref{eq:intcond}, the timescape model prescribes the \emph{uniform quasilocal Hubble expansion condition}. That is, it requires an exact compensation between the different clock rates of wall and void observers and the locally measured expansion rates~\cite{Wiltshire_2007a,Wiltshire_2007b,Wiltshire_2009a,Wiltshire_2009b,Wiltshire_2014}:
\begin{equation}
H_\CD = H_v^{\tau_v} = H_w^{\tau_w} \quad , \quad H_v^{\tau_v} := \frac{\mathrm{d}t}{\mathrm{d}\tau_v} H_v \quad , \quad H_w^{\tau_w} := \frac{\mathrm{d}t}{\mathrm{d}\tau_w} H_w \;\; , 
\label{eq:ts_rates}
\end{equation}
where $\tau_v$ ($\tau_w$) denotes the proper time of an observer within a void (wall)---while the cosmic time $t$ is associated with the idealised coarse-grained dust flow---, and $H_v^{\tau_v}$ ($H_w^{\tau_w}$) is the proper expansion rate of a void (wall) as measured by the corresponding observer.

Finally, a second closure condition is imposed by requiring that radial null geodesics match between the wall regions (where we reside as an observer in this framework) and the averaged large-scale geometry. This introduces a direct link between the scale factors of the two geometries, namely
\begin{equation}
    \label{eq:ts_aw}
    a_w = \gamma_w^{-1} a_\CD \, ,
\end{equation}
where $\gamma_w := \mathrm{d}t/\mathrm{d}\tau_w$ represents the relative lapse function between the average cosmic time and the proper time of a wall observer. 

It is precisely these two closure conditions, relating the clocks and rulers of different observers, that ultimately produce an apparent acceleration for wall observers---including us as observers---thus mimicking the effects of DDE in cosmological data. 

The timescape cosmology\footnote{%
Hereafter, we use ``timescape'' to refer specifically to the so-called \emph{tracker solution} of timescape cosmology, which indeed emerges as the unique physical solution upon restricting our attention to the matter-dominated era~\cite{Wiltshire_2007a,Wiltshire_2007b}. As such, it also represents the primary realisation of timescape cosmology that has been tested against observations~\cite{Smale_2011,Dam_2017,Lane_2025,Seifert_2024}. Nervertheless, in principle the timescape framework allows for a richer phenomenology; see e.g.,~\cite{Duley_2013} for an approximate model incorporating the radiation component within this framework.%
}, instead of being characterised by a matter density parameter relative to the critical density as in $\Lambda$CDM, is naturally described by the void fraction $f_v(t)$. Its present value, $f_{v,0}$, together with the present-time averaged cosmic expansion rate, $H_{\mathcal{D},0}$, constitute the two free parameters of the model. Moreover, as timescape represents a closed solution to Eqs.~\eqref{eq:avg_Hamilton}--\eqref{eq:intcond}, the domain-averaged scale factor can be expressed as an explicit function, i.e.,~\cite{Wiltshire_2007a,Wiltshire_2007b,Wiltshire_2009a,Wiltshire_2009b,Wiltshire_2014}
\begin{equation}
    a_\CD(t) = \frac{3 \:\! a_{\CD,0} \, H_{\CD,0} \,t}{2+f_{v,0}} \left(\frac{f_{v,0}}{f_v(t)}\right)^{1/3} \; ,\label{eq:ts_scalefactor}
\end{equation}
whilst the void fraction is expressed as~\cite{Wiltshire_2007a,Wiltshire_2007b,Wiltshire_2009a,Wiltshire_2009b,Wiltshire_2014}:
\begin{equation}
    f_v(t) = \frac{3f_{v,0} \;\! H_{\CD,0} \, t}{3f_{v,0} \;\! H_{\CD,0} \, t + (2+f_{v,0})(1-f_{v,0})} \; .\label{eq:ts_fv}
\end{equation}
One also finds a closed form for the averaged curvature and backreaction terms as~\cite{Wiltshire_2007a,Wiltshire_2007b,Wiltshire_2009a,Wiltshire_2009b,Wiltshire_2014}:
\begin{equation}
    \RD (t) = -6 \:\! f_{v,0}^{-2/3} \;\! \frac{f_v(t)}{t^2}\quad ; \quad \QD(t)= \frac{2 \dot{f_v}(t)^2}{3f_v(t)\;\![1-f_v(t)]} \;\; , \label{eq:ts_RandQ}
\end{equation}
whilst the averaged dust density follows the standard $a_\CD^{-3}$ scaling, with the normalisation $\avg{\rholoc}(t_0) = 12 \:\! H_{\CD,0}^2(1-f_{v,0})/[8\pi G \;\! (2+f_{v,0})^2]$. In addition, we note that due to the different expansion in wall and void regions, the average (\lq\lq bare\rq\rq) and observed (\lq\lq dressed\rq\rq) expansion rates do not coincide. Indeed, one gets at present time the relation $H_{\CD,0} = 2 \:\! H_0\, (2+f_{v,0})/(4 + f_{v,0} + 4 \;\! f_{v,0}^2)$, where $H_0$ is the \lq\lq dressed\rq\rq\, Hubble parameter directly extracted by observations~\cite{Wiltshire_2007a,Wiltshire_2007b,Wiltshire_2009a,Wiltshire_2009b,Wiltshire_2014}. Finally, in the timescape model, following Eqs.~\eqref{eq:ts_rates} and~\eqref{eq:ts_aw}, the observed cosmological redshift is defined by $1 + z= a_{w,0}/a_w = \gamma_w \:\! (a_\CD / a_{w, 0})^{-1}$, where one obtains $\gamma_w(t) = 1 +f_v(t)/2$~\cite{Wiltshire_2007a,Wiltshire_2007b,Wiltshire_2009a,Wiltshire_2009b,Wiltshire_2014}. Here, the initial time $\tinitb$ where the scale factors are normalised is set at the surface of last scattering.

\begin{figure}[htb!]
\begin{adjustwidth}{-2cm}{-1.5cm}
\centering
    \begin{minipage}{0.495\linewidth}
        \centering
        \includegraphics[width=\linewidth]{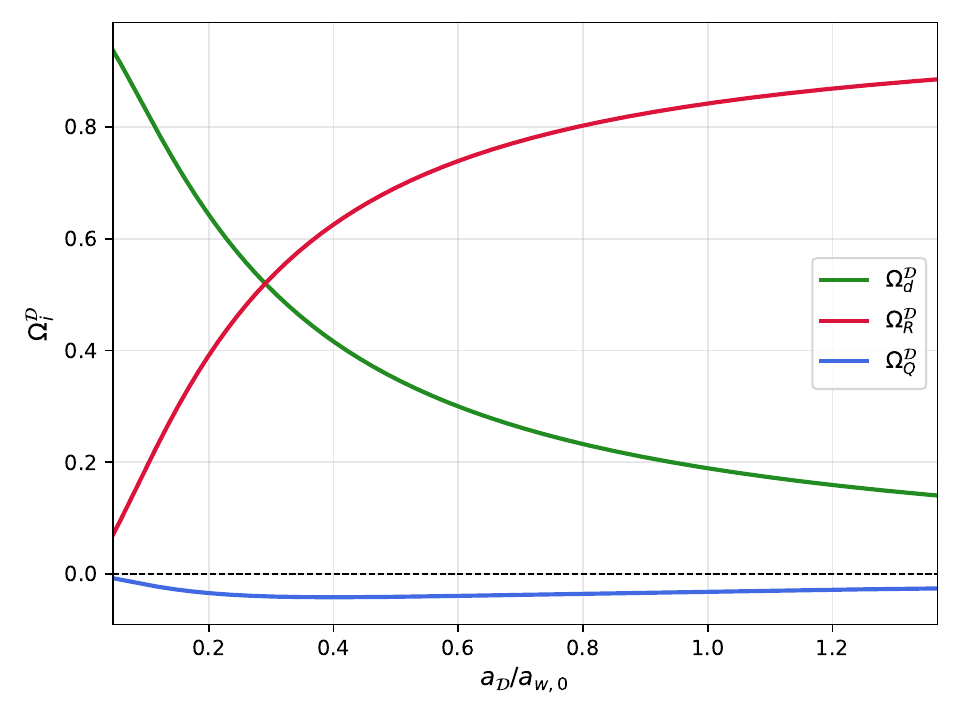}
        \centering
    \end{minipage}\hfill
    \begin{minipage}{0.495\linewidth}
        \centering        \includegraphics[width=\linewidth]{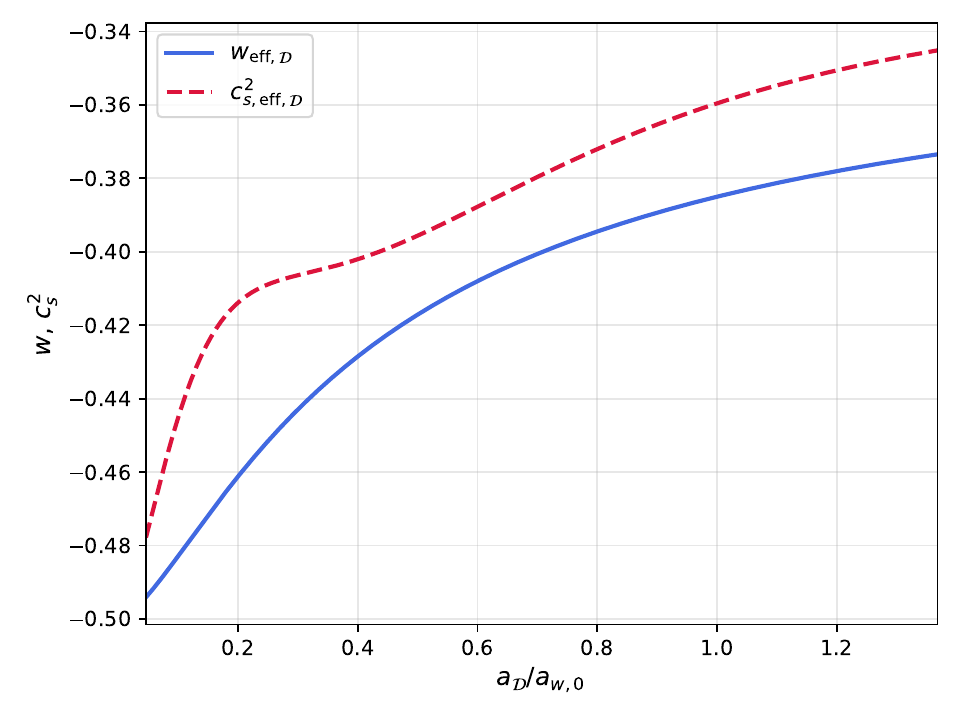}
        \centering
    \end{minipage}
\end{adjustwidth}
\caption{The $\Omega$ variables (left panel) and the effective fluid EoS parameter and squared sound speed (right panel) in the timescape model, as functions of the rescaled volume-average scale factor $a_\CD/a_{w,0}$, with $f_{v,0} = 0.737$, and $H_0 = 73.03$~km/(s$\cdot$Mpc). The model has no cosmological constant.}
\label{fig:ts_1}
\end{figure}

Employing Eqs.~\eqref{eq:ts_scalefactor}--\eqref{eq:ts_RandQ}, we can thus extract the $\Omega$ variables defining the cosmological energy balance in timescape (see Eq.~\eqref{eq:omegas}, with $\Lambda = 0$ in this case) as well as the energy density, pressure, EoS and squared sound speed of the effective fluid discussed in sections~\ref{sec:buchert}--\ref{sec:Us} (using Eqs.~\eqref{eq:def_rhoeff}--\eqref{eq:def_peff} in particular, with $K_\CD = 0$). In Fig.~\ref{fig:ts_1} we show, starting from redshift $z = 20$, the $\Omega$ variables (left panel), as well as the EoS parameter and squared sound speed of the effective fluid (right panel), against the average scale factor $a_\CD$ normalised by $a_{w,0}$ ($\approx 1090$) to offset the initial normalisation of $a_\CD = a_w = 1$ at $\tinitb$. We have set the model's parameters to the best-fit results from~\cite{Seifert_2024}, i.e., $f_{v,0} = 0.737$ and $H_0 = 73.03$~km/(s$\cdot$Mpc), corresponding to $H_{\CD,0} \approx 58.01$~km/(s$\cdot$Mpc), and, for the present-day volume-average scale factor, $a_{\CD,0} / a_{w,0} = 1 + f_{v,0}/2 \approx 1.37$.

From the left panel of Fig.~\ref{fig:ts_1}, we find that the dominant contribution to the averaged energy balance at late times arises from the curvature term---as opposed to the cosmological constant playing this role in the $\Lambda$CDM model. Meanwhile, the kinematical backreaction, although strictly positive ($\Omega^\CD_\CQ < 0$), remains subdominant compared to the other components. This behaviour highlights a common---and indeed expected~\cite{Buchert_2000,Buchert_2001}---feature of averaged cosmological models (see also the following subsections): they generally allow for a non-negligible and dynamically relevant curvature parameter, while the kinematical backreaction, consisting of two partially compensating terms, typically provides only a minor contribution.

The right panel of Fig.~\ref{fig:ts_1} shows that both the EoS parameter and the squared sound speed of the effective fluid remain negative throughout the evolution of the timescape model. Indeed, within averaged cosmological models, where the effective fluid ultimately arises as a combination of curvature and kinematical backreaction (see Eqs.~\eqref{eq:def_rhoeff}--\eqref{eq:def_peff}), negative values of $w_{\mathrm{eff},\CD}$ and $c^2_{s,\mathrm{eff},\CD}$ can generally be expected for at least part of the evolution. In particular, here, we note that the equation of state parameter satisfies $w_{\mathrm{eff},\CD} < -1/3$ at all times, thereby producing accelerated expansion within the model. In addition, following our discussion in Sec.~\ref{subsec:meszaros}, we conclude that the M\'esz\'aros approximation should not be applicable in the timescape framework. Indeed, since $w_{\mathrm{eff},\CD} < 0$ and $ c^2_{s,\mathrm{eff},\CD} < 0$, in a barotropic perturbation picture the effective fluid does not suppress structure formation on small scales; rather, it leads to the development of instabilities. Nonetheless, we can still \emph{formally} consider the M\'esz\'aros approximation in order to compare its predictions (from Eq.~\eqref{eq:original_Meszaros}) with those obtained from the other closure conditions discussed in Sec.~\ref{sec:Us}. In particular, since this approximation relies on a barotropic description of each fluid, it is not expected to hold either for other parametrisations of the effective fluid's perturbations, such as the comoving picture of Sec.~\ref{subsec:comoving}. Nonetheless, within such \emph{non}-barotropic descriptions, there is also no more need for any tight correlation between the sign of the effective parameters $w_{\mathrm{eff},\CD}$ and $ c^2_{s,\mathrm{eff},\CD}$, and the extent to which the dust perturbation growth deviates from the M\'esz\'aros approximation predictions.

\subsubsection{Timescape: CGI linear structure growth}
\label{subsubsec:ts_CGI}
\begin{figure}[htb!]
    \centering
    \includegraphics[width=0.7\textwidth]{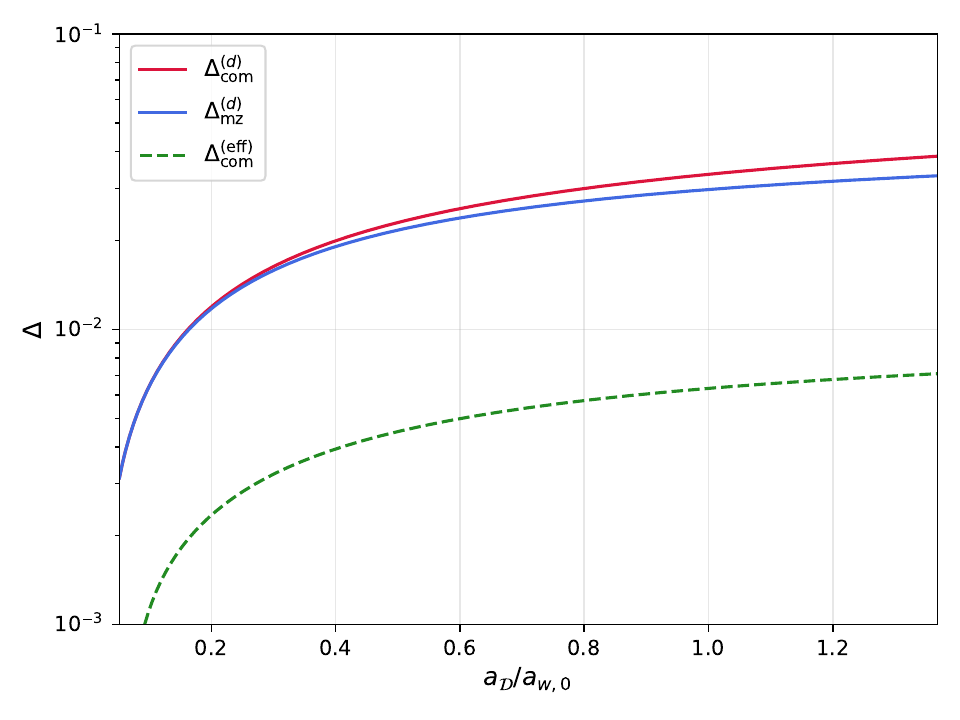} 
    \caption{Dust density perturbation growth within the comoving effective fluid framework, $\Delta^{(d)}_\mathrm{com}$, and the M\'esz\'aros approximation, $\Delta^{(d)}_\mathrm{mz}$, for the timescape model, plotted as functions of the rescaled volume-average scale factor $a_\CD/a_{w,0}$. For the first scenario we also show the effective fluid density perturbations, $\Delta^{(\mathrm{eff})}_\mathrm{com}$. The ICs are set at $z = 20$, seeded using a $\Lambda$CDM-based approximation for $1100 \geq z > 20$ for both the dust density and expansion perturbations. The IC for the effective fluid perturbation at $z=20$ is instead set to zero.}
    \label{fig:ts_2}
\end{figure}

In Fig.~\ref{fig:ts_2}, we show the growth of a linear matter perturbation as predicted by the M\'esz\'aros approximation via Eq.~\eqref{eq:original_Meszaros}, and the one arising from the comoving effective fluid closure condition, as well as the first-order perturbation in the effective fluid energy density for the latter scenario.

The Initial Conditions (ICs) for the perturbations are set at a time $\tinit$ corresponding to a redshift $z = 20$, with $\Delta^{(\mathrm{eff})}(\tinit) = 0$, while $\Delta^{(d)}(\tinit)$ is chosen to represent a typical horizon-scale dust density perturbation amplitude in the $\Lambda$CDM model. Specifically, we evolve the standard growing mode of $\Lambda$CDM dust density perturbations\footnote{
Here, and for all other example models we consider, we refer to the growing mode obtained for the $\Lambda$CDM model in the approximation of negligible radiation content---as the latter remains a subdominant component at such post-CMB times.
}~\cite{KodamaSasaki_1984} from the CMB epoch, with a typical amplitude at comoving wavelength $c/H_0$, $\Delta^{(d)} = 6 \times10^{-5}$, up to $z = 20$, giving $\Delta^{(d)}(\tinit) \approx 3.1 \times 10^{-3}$, and use these as ICs within our framework. For the expansion perturbation, the IC, $\mathcal{Z}(\tinit)$, is then derived directly from the time derivative of the $\Lambda$CDM dust density perturbation reached at that time, resulting in $\CZ(\tinit) \approx -0.16 H_0$. We note that this procedure is consistent with our approach, as the timescape model, as well as the other models that will be discussed in the following subsections, asymptotes to $\Lambda$CDM at early times.

We find that the comoving effective fluid framework produces faster structure growth in the dust at late times than the M\'esz\'aros approximation, owing to the coupling between the dust and effective fluid density perturbations within the first scenario. Indeed, we see that as the comoving effective fluid perturbation $\Delta^{(\mathrm{eff})}\com$ grows, the two predictions gradually diverge, while they coincide at early times when $\Delta^{(\mathrm{eff})}\com$ is still negligible.

\begin{figure}[htb!]
\begin{adjustwidth}{-2cm}{-1.5cm}
\centering
    \begin{minipage}{0.495\linewidth}
        \centering
        \includegraphics[width=\linewidth]{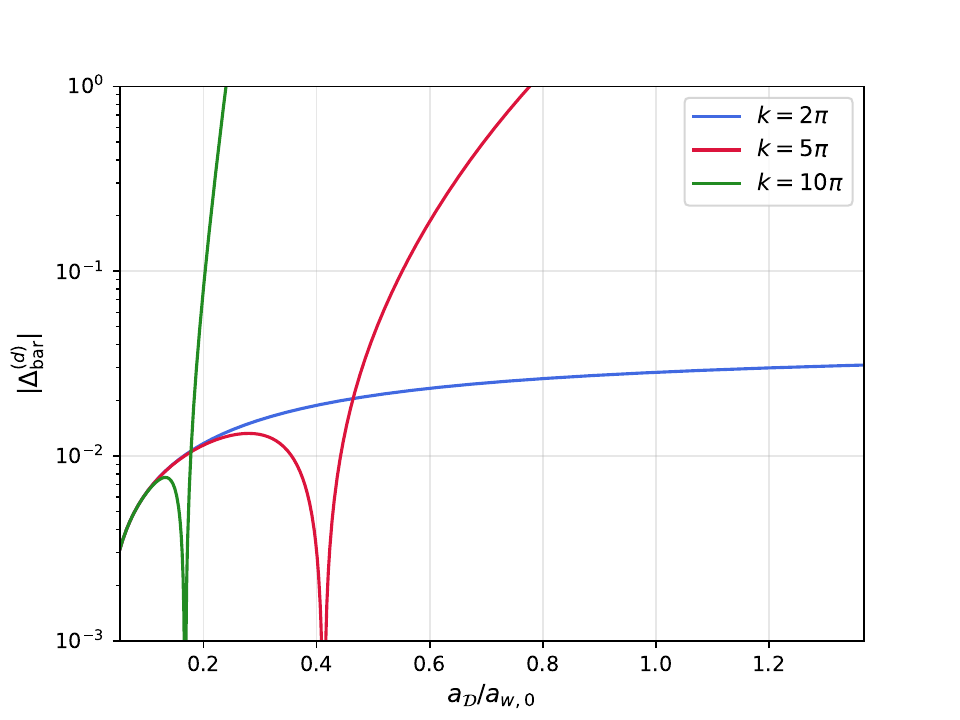}
        \centering
    \end{minipage}\hfill
    \begin{minipage}{0.495\linewidth}
        \centering
        \includegraphics[width=\linewidth]{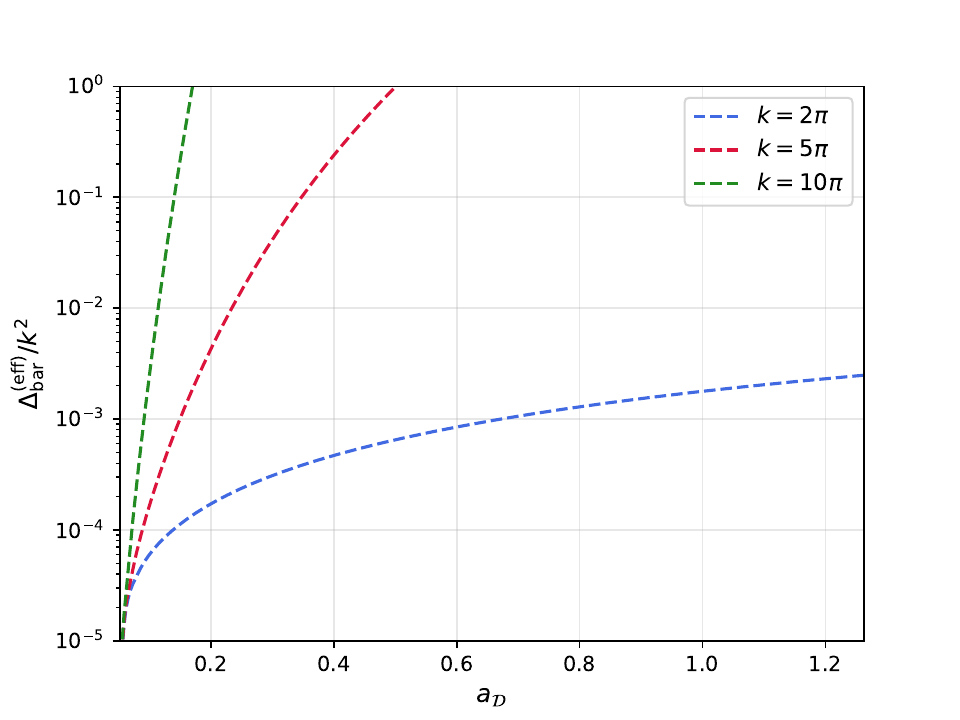}
        \centering
    \end{minipage}
\end{adjustwidth}
\caption{The growth of dust perturbations, $\Delta^{(d)}_{\mathrm{bar}}$, and effective fluid density perturbations, $\Delta^{(\mathrm{eff})}_{\mathrm{bar}}$, shown in the left and right panels respectively, for the timescape model, plotted as functions of the rescaled volume-average scale factor $a_\CD/a_{w,0}$, within the barotropic framework. We show the amplitude evolution for three different wavelengths on a spatially flat background ($K_\CD = 0$), at horizon ($k = 2\pi$) and sub-horizon scales. Here, the wavenumber $k$ is expressed in units of the present-day inverse Hubble length, $H_0/c$, as evaluated in the $\Lambda$CDM model following the best fit of~\cite{Planck_2018}. The ICs (at $z=20$) are seeded via the $\Lambda$CDM approximation at earlier times for $\Delta^{(d)}$ and $\CZ$, with the same values for all scales for direct comparison, and fixed to zero for the effective fluid perturbations.}
\label{fig:ts_3}
\end{figure}

In Fig.~\ref{fig:ts_3}, we show the behaviour of both the dust density perturbations (left panel) and effective fluid density perturbations (right panel), for the barotropic closure condition. We use the same initial conditions for $\Delta^{(\mathrm{eff})}$ (vanishing at $\tinit$), $\Delta^{(d)}$, and $\CZ$ as for the comoving picture discussed above, at the same initial time $\tinit$, and enforce again, now at initial time only, a comoving condition, $\tilde V\eff(\tinit) = 0$. This implies, in particular, that the dust perturbations keep the same initial amplitude across the scales we consider, thus enabling an easier comparison of the perturbations at different scales. Specifically, by doing so, we effectively reduce the initial density contrasts for smaller-scale perturbations, compensating for the intrinsic scale dependence of the $\Delta^{(i)}$ variables (as Laplacians of the associated densities).

As expected from the background conditions $w_{\mathrm{eff},\CD}<0$ and $c_{s,\mathrm{eff},\CD}^2<0$, we find that the perturbations exhibit instabilities in the sub-horizon regime, with a very fast growth of the effective fluid perturbations. Interestingly, while the latter remain monotonic, the dust density perturbations at sub-horizon scales undergo a reversal of their growth rate. Namely, initially growing overdensities transition into decaying modes, ultimately even turning into underdensities ($\Delta^{(d)}\baro < 0$). With the initial values we used here, this happens when the effective fluid perturbation has, in principle, already just left the linear regime; nevertheless, we stress that Fig.~\ref{fig:ts_3} shows a solution to the \emph{linear} equations, and hence the observed transition is indeed an additional, negative linear growth mode of the dust perturbation evolution sourced by the barotropic-picture effective fluid perturbation. The solution for $\{ \Delta^{(d)}\baro , \Delta\eff\baro \}$ can be arbitrarily rescaled by a constant factor, and a smaller initial dust perturbation would undergo the exact same growth dynamics while the system as a whole remains in the linear regime for longer. The observed reversal behaviour is highly unexpected and, we argue, ultimately unphysical. Local perturbations in the effective fluid---interpretable as manifestations of deviations in the spatial curvature and backreaction sourced by the matter perturbation---can hardly be expected to give rise to effects capable of reversing the clumping of matter on sub-horizon scales, as appearing in Fig.~\ref{fig:ts_3}. Such a bizarre phenomenon could produce (so far unobserved) signatures in cosmological tracers associated with the cosmic web, e.g., via an anomalous integrated Sachs-Wolfe effect~\cite{KofmanStarobinski_1985,Boughn_2004}.

\subsection{The Giani--von Marttens--Camilleri model}
\label{subsec:GMC}

This model (referred to hereafter as the GMC model) was recently introduced in~\cite{Giani_2025a}, where it was tested against combinations of cosmological data sets---CMB, BAO, Type~Ia supernovae, and independent $f\sigma_8$ measurements---and shown to outperform $\Lambda$CDM across this wide range of cosmological probes. Remarkably, the model may also offer a simultaneous resolution of both the Hubble and $\sigma_8$ tensions, by mimicking DDE effects through backreaction of inhomogeneities.

\subsubsection{GMC: theoretical framework}

In the GMC framework, the Universe is modeled via an FLRW background geometry in which the backreaction effects from non-virialised, nonlinearly evolving clustering regions, $(c)$, and expanding voids, $(v)$, are quantified by introducing two effective fluids, $\rho^{(c)}$ and $\rho^{(v)}$, which are dynamically coupled to a dust matter content, $\rho^{(m)}$, such that~\cite{Giani_2025a}
\begin{align}
&\dot{\rho}^{(m)} + \Theta\rho^{(m)} = - C(t) - V(t) \; ; \label{eq:GMC1}\\
&\dot{\rho}^{(c)} + \Theta\rho^{(c)}(1+w_{(c)}) = C(t) \; ; \label{eq:GMC2}\\
&\dot{\rho}^{(v)} + \Theta\rho^{(v)}(1+w_{(v)}) = V(t) \; .
\label{eq:GMC3}
\end{align}
Here, $w_{(c)}(t)$ and $w_{(v)}(t)$ are the effective equation of state parameters of the two effective fluids, $C(t)$ and $V(t)$ encode their respective couplings to the matter sector, and $\Theta$ denotes the background expansion rate which also accounts for the cosmological constant. We must stress that the interpretation of $\rho^{(m)}$ within the GMC framework is not that of the averaged dust density $\avg{\rholoc}$ introduced in the covariant averaging formalism, since it does not represent a conserved matter content. Rather, it should be understood as the average density of dust matter contained \emph{only} within virialised structures in the Universe. Within this framework, it is then of interest to introduce the total fluid source density---representing the conserved composite source driving the background expansion---as $\rhotot := \rho^{(m)} + \rho^{(c)} +\rho^{(v)}$, with EoS parameter $w_{\mathrm{tot}}(t) := (\rho^{(c)}/\rhotot )(t) \, w_{(c)}(t) + (\rho^{(v)}/\rhotot )(t) \, w_{(v)}(t)$.

We can then introduce the true averaged dust density, $\avg{\rholoc}(t) \propto \VD(t)^{-1}$, normalised to $\rho^{(m)}$ at an early initial time (e.g., $z = 20$ in~\cite{Giani_2025a} and in this work), and map the double-effective-fluid formulation introduced above---naturally arising from distinguishing between overdense and underdense non-virialised regions---to that of the Buchert averaging scheme discussed in Sec.~\ref{sec:buchert} (with $K_\CD = 0$) by setting~\cite{Giani_2025a}
\begin{align}
&\RD = -12\pi G \:\! \Delta\rho + 12\pi G \;\! \wtot \:\! \rhotot \; ; \\
&\QD = -4\pi G \:\! \Delta\rho - 12\pi G \;\! \wtot \:\! \rhotot \; ,
\end{align}
where $\Delta\rho := \rhotot-\avg{\rholoc}$. In addition, we directly obtain for the overall effective fluid encompassing backreaction and curvature effects, $\rho\eff_\CD = \Delta \rho$, and $p^{\mathrm{(eff)}}_\CD = \wtot \:\! \rhotot$. 

In order to close the system of evolution equations, Eqs.~\eqref{eq:GMC1}--\eqref{eq:GMC3}, an ansatz is introduced that relates the various energy densities through the non-virialised void and cluster fractions, $f_v(t)$ and $f_c(t)$, respectively. This links the energy densities as $\rho^{(c)} = f_c(t) \, \rhotot$, $\rho^{(v)} = f_v(t) \,\rhotot$, and $\rho^{(m)} = (1 - f_c(t) - f_v(t))\rhotot$~\cite{Giani_2025a}. The abundance of nonlinear, non-virialised structures and voids is then estimated using the Press–Schechter formalism and its extensions~\cite{Schechter_1974,Bond_1991,Sheth_2004,Jennings_2013,Verza_2024}. In the GMC model, this introduces two additional free parameters relative to $\Lambda$CDM, corresponding to the minimum comoving scale of collapsing and expanding regions contributing to backreaction effects, $\tilde{R}_v$ and $\tilde{R}_c$ respectively. The dynamics of the model are fully specified once an ansatz for $\wtot(t)$ is introduced, since the source terms in Eqs.~\eqref{eq:GMC1}–\eqref{eq:GMC3} can then be expressed directly in terms of $\rhotot$, $\wtot$, $f_v$, $f_c$, and their time derivatives. To this end, in~\cite{Giani_2025a} the authors employ the spherical collapse framework~\cite{Peebles_1967,Gunn_1972} to assign an effective internal energy to each non-virialised, nonlinear structure based on its expected spatial curvature and mass–radius relation. Integrating these internal energies and mass contributions (for both linear and nonlinear structures) over the Press–Schechter mass function---subject to minimal-scale cutoffs for voids and collapsing regions---yields evolving energy densities and pressures, whose ratio ultimately defines $\wtot$, thereby closing the system. It can then be solved numerically given $\Lambda$CDM-like initial conditions at $z = 20$. 

Here, we remark that in the GMC model---in accordance with the arguments presented in~\cite{Rasanen_2010,Rasanen_2012}---the average scale factor, $a_\CD$, normalised to unity at the present time ($a_{\CD}(t_0) = 1$), is assumed to be related to the cosmological redshift through the standard relation $a_\CD(z) = (1+z)^{-1}$. Interestingly, it is precisely this identification that allows the GMC framework to consistently employ the Press–Schechter formalism to determine the non-virialised void and cluster fractions.

\begin{figure}[htb!]
\begin{adjustwidth}{-2cm}{-1.5cm}
\centering
    \begin{minipage}{0.495\linewidth}
        \centering
        \includegraphics[width=\linewidth]{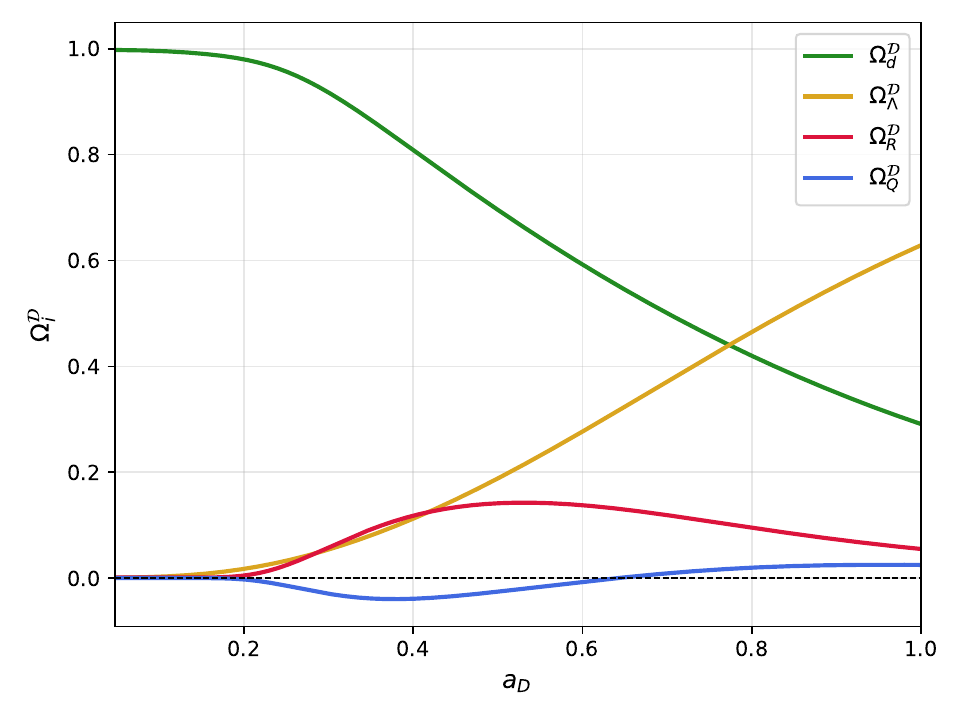}
        \centering
    \end{minipage}\hfill
    \begin{minipage}{0.495\linewidth}
        \centering        \includegraphics[width=\linewidth]{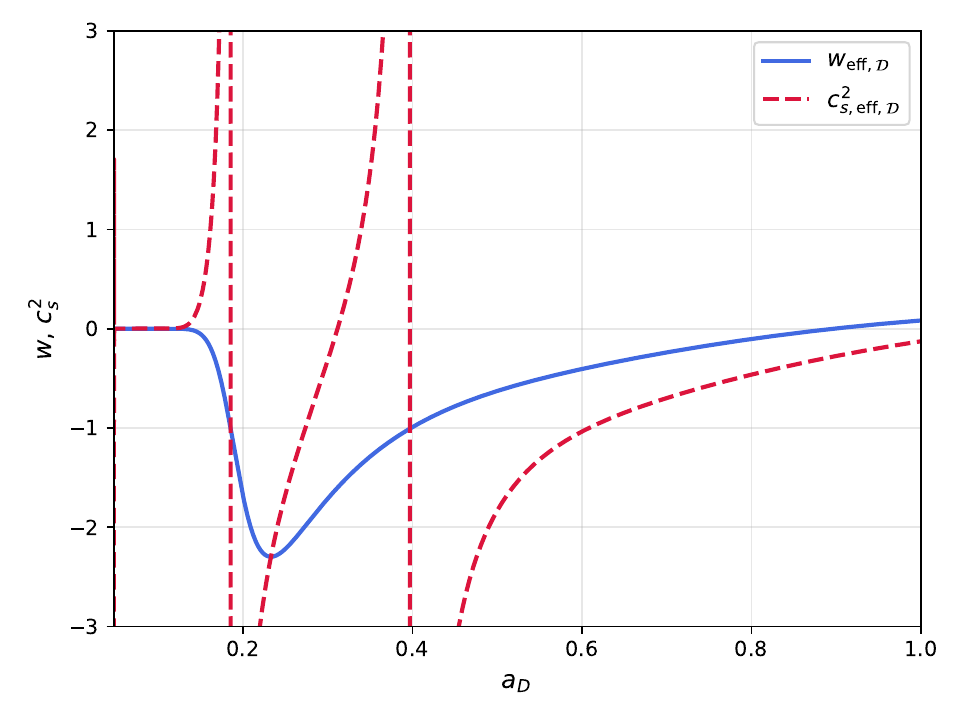}
        \centering
    \end{minipage}
\end{adjustwidth}
\caption{The $\Omega$ variables (left panel) and the effective fluid EoS parameter and squared sound speed (right panel), as functions of the effective scale factor $a_\CD$, in the GMC model with parameters $\tilde{R}_c =6.76$~Mpc, $\tilde{R}_v =4.57$~Mpc,  $H_{\CD,0} = 70.12$~km/(s$\cdot$Mpc), and $\Omega^{\CD,0}_d = 0.316$.}
\label{fig:gmc_1}
\end{figure}

We can then extract the $\Omega$ variables as well as the quantities characterising the effective fluid for the GMC model to be used as our two-fluid homogeneous background for CGI perturbations. In Fig.~\ref{fig:gmc_1} we plot for this background, as a function of the average scale factor $a_\CD(t)$ (starting from redshift $z = 20$), the $\Omega$ variables as defined in Eq.~\eqref{eq:omegas} (left panel), as well as the EoS parameter and squared sound speed of the effective fluid (right panel). Here, we set the parameters of the model to the best-fit values from~\cite{Giani_2025a}, i.e., $\tilde{R}_c =6.76$~Mpc, $\tilde{R}_v =4.57$~Mpc,  $H_{\CD, 0} = 70.12$~km/(s$\cdot$Mpc), and $\Omega^{\CD,0}_d = 0.316$.  

From the left panel of Fig.~\ref{fig:gmc_1} we see that, at early times, $\RD$ and $\QD$ are negligible, as expected, but become significant beyond $a_\CD \approx 0.2$. Interestingly, whilst the kinematical backreaction undergoes a sign change, the average spatial curvature remains negative ($\Omega^\CD_\mathcal{R} > 0$) throughout the Universe's expansion history. Moreover, in contrast to the timescape model (cf. Fig.~\ref{fig:ts_1}), the inclusion of a cosmological constant in the GMC model makes both the curvature and the kinematical backreaction remain subdominant, though non-negligible, relative to the other energy contributions. 

The right panel of Fig.~\ref{fig:gmc_1} shows that the EoS parameter for the effective fluid remains negative for most of the evolution of the GMC model, becoming slightly positive only at very late times. Moreover, the effective fluid exhibits two distinct formal phantom crossings, corresponding to divergences in its squared sound speed. These striking features highlight how dynamical effects arising from averaging procedures in cosmology can reproduce rich phenomenologies, even mimicking phantom DDE behaviours. Here, as in the timescape model, we note that, given the predominantly negative $w_{\mathrm{eff},\CD}$ and the pathological nature of $c^2_{s,\mathrm{eff},\CD}$, the M\'esz\'aros approximation is not safely applicable within the GMC framework. Indeed, we again expect the development of instabilities in the evolution of perturbations at sub-horizon scales, which would invalidate the approximation. 

\subsubsection{GMC model: CGI linear structure growth}
\label{subsubsec:GMC_CGI}

In~\cite{Giani_2025a}, a first study of the growth of linear perturbations in the GMC model was carried out to estimate the impact of backreaction from inhomogeneities on the $f\sigma_8$ cosmological probe. However, the authors (i) defined perturbations with respect to the virialised dust fluid $\rho^{(m)}$ rather than the true averaged dust density $\avg{\rholoc}$, and (ii) implemented the perturbative scheme within the M\'esz\'aros approximation.

Here, we argue that within an averaging framework in cosmology, perturbations should instead be considered with respect to $\avg{\rholoc}$, as only local deviations from a true average can properly be interpreted as such. Furthermore, although the use of the M\'esz\'aros approximation in~\cite{Giani_2025a} is motivated by the \emph{a priori} reasonable assumption that the impact of merging non-virialized clusters and voids can be neglected on sub-horizon scales, its applicability within a general-relativistic framework is not guaranteed. Indeed, as we see below, within the same GMC model, the approximation is not valid when considering the growth of perturbations of the averaged dust fluid. 

\begin{figure}[htb!]
    \centering
    \includegraphics[width=0.7\textwidth]{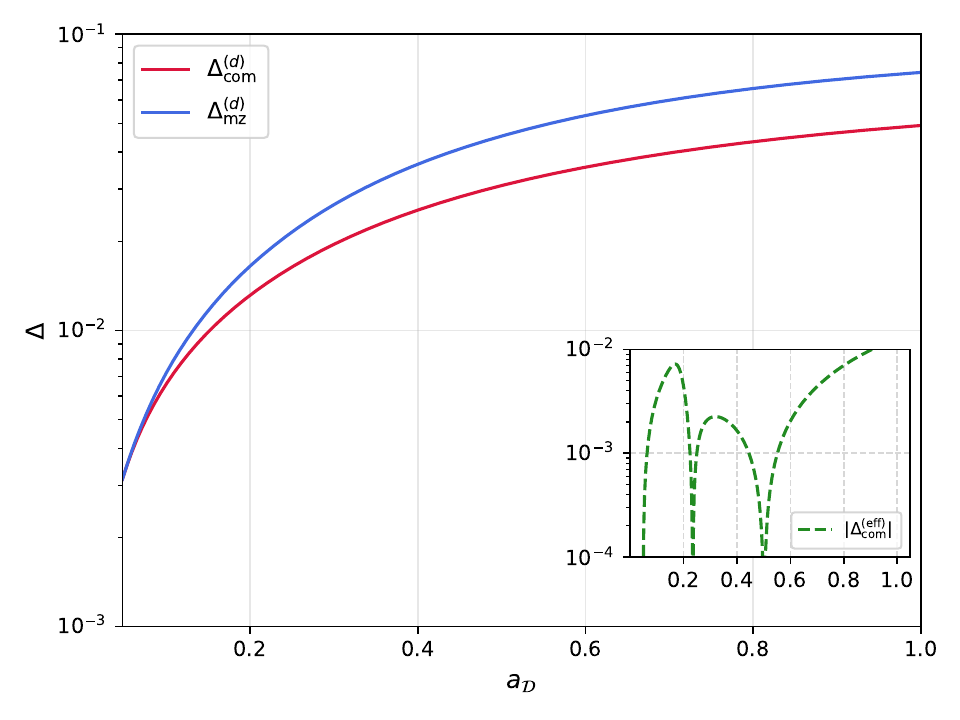} 
    \caption{Dust density perturbation growth within the comoving effective fluid framework, $\Delta^{(d)}_\mathrm{com}$, and the M\'esz\'aros approximation, $\Delta^{(d)}_\mathrm{mz}$, for a GMC background model, as functions of the average scale factor $a_\CD$. For the first scenario we also show the magnitude of the effective fluid density perturbations, $|\Delta^{(\mathrm{eff})}_\mathrm{com}|$, in the inset panel. The ICs for the dust density and expansion perturbations are set at $z = 20$, seeded using the $\Lambda$CDM-based approximation at earlier times as in Sec.~\ref{subsec:ts} above. The ICs for the effective fluid density perturbations are instead set to zero.}
    \label{fig:gmc_2}
\end{figure}

In Fig.~\ref{fig:gmc_2}, we show the growth of first-order scalar CGI dust perturbations with respect to $\avg{\rholoc}$ on a GMC background, as predicted by both the M\'esz\'aros approximation (here formally applied, through Eq.~\eqref{eq:original_Meszaros}) and the comoving effective fluid closure condition of Sec.~\ref{subsec:comoving}, where we have employed the same ICs as discussed in Sec.~\ref{subsubsec:ts_CGI}. We also plot the density perturbations of the effective fluid in the second scenario. We find that an early growth in amplitude of the effective fluid perturbations in the comoving framework leads to a growth history of the dust perturbations rapidly diverging from that given by the M\'esz\'aros approximation. In this case, the coupling to the (comoving-framework) effective fluid perturbations leads to a faster growth of the dust matter structures. This trend contrasts with that found in the timescape cosmology (see Fig.~\ref{fig:ts_2}), further highlighting the nontrivial interplay between the effective background and a chosen closure condition within the perturbative scheme, with important consequences for model fitting to cosmological probes of structure formation such as $f\sigma_8$.

\begin{figure}[htb!]
\begin{adjustwidth}{-2cm}{-1.5cm}
\centering
    \begin{minipage}{0.495\linewidth}
        \centering
        \includegraphics[width=\linewidth]{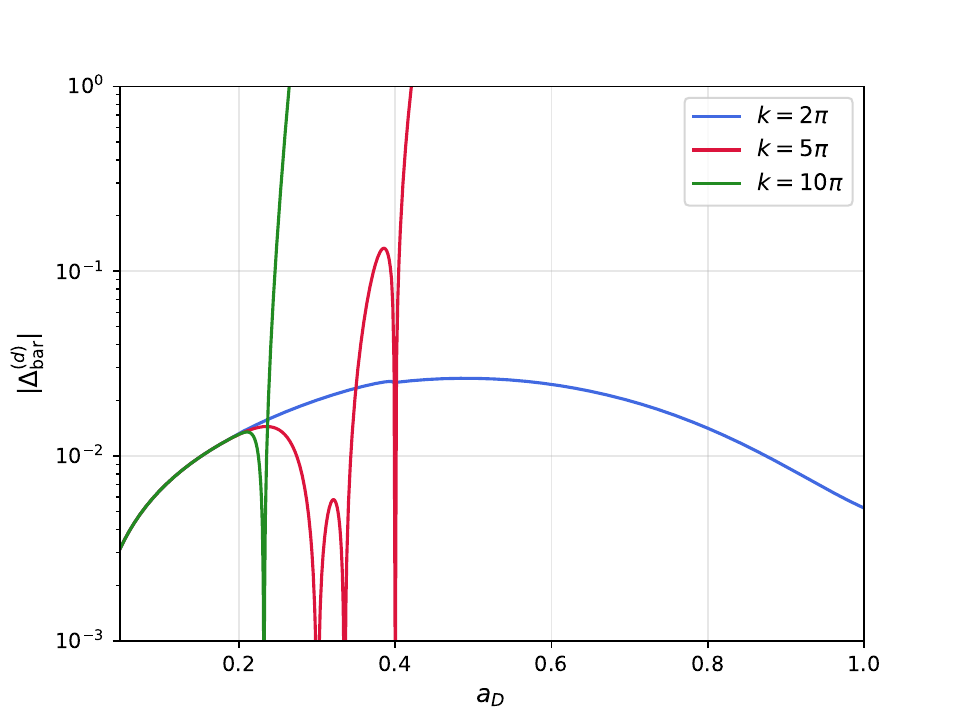}
        \centering
    \end{minipage}\hfill
    \begin{minipage}{0.495\linewidth}
        \centering
        \includegraphics[width=\linewidth]{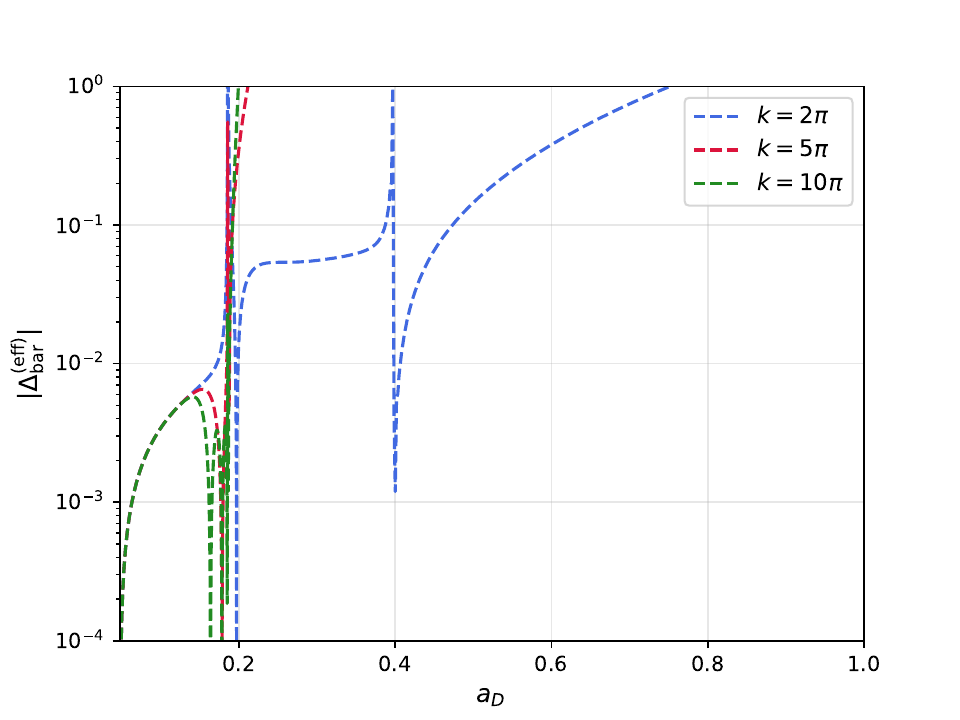}
        \centering
    \end{minipage}
\end{adjustwidth}
\caption{Horizon and sub-horizon dust (left panel) and effective fluid (right panel) density perturbation growth on a GMC background within the barotropic condition, as functions of the average scale factor $a_\CD$. We show the amplitude evolution for three different wavelengths on a spatially flat background ($K_\CD = 0$), at horizon ($k = 2\pi$) and sub-horizon scales. Here, the wavenumber $k$ is expressed in units of the present-day inverse Hubble length, $H_0/c$, as evaluated in the $\Lambda$CDM model following the best fit of~\cite{Planck_2018}. As in Sec.~\ref{subsec:ts} above, the ICs (at $z=20$) are seeded via the $\Lambda$CDM approximation at earlier times for $\Delta^{(d)}$ and $\CZ$, with the same values for all scales for direct comparison, and are set to zero for the effective fluid perturbations.}
\label{fig:gmc_3}
\end{figure}

In Fig.~\ref{fig:gmc_3}, we show the evolution of horizon- and sub-horizon-scale CGI perturbations for both the dust density (left panel) and the effective fluid density (right panel) under the barotropic closure condition. The initial conditions are again the same as those adopted in Sec.~\ref{subsec:ts}. As anticipated from the behaviour of $w_{\mathrm{eff},\CD}$ and $c_{s,\mathrm{eff},\CD}^2$ in Fig.~\ref{fig:gmc_1} (right panel), both sets of perturbations exhibit pathological evolutions. Specifically, we find that growing overdensities can transition into decaying modes, and vice versa, after the first phantom crossing, while the effective fluid perturbations even display clear divergences coinciding with the phantom crossings. We therefore conclude that, within a physically meaningful GMC cosmology---just like within the timescape model---the barotropic assumption cannot consistently describe perturbations of the effective fluid source.

\subsection{The Giani--von Marttens--Piattella model}
\label{subsec:GMP}
The above GMC model from~\cite{Giani_2025a} inspired another, simpler, recent phenomenological model proposed in~\cite{Giani_2025b}, hereafter referred to as GMP. It is designed to retain some qualitative features of the dynamics in the GMC model, still interpreted as arising from backreaction and evolving curvature, while framed under a simple form which does not make it an exact subcase of this previous model.
Like the GMC model above, the GMP phenomenology has been tested against cosmological observations from BAO, Type-Ia supernovae, and the CMB in~\cite{Giani_2025b}, constraining its parameters and resulting in a fit to the data better than $\Lambda$CDM and comparable to the usual DDE parametrisation.

\subsubsection{GMP: theoretical framework}

The GMP describes the averaged dynamics over some very large scale directly in terms of the total dynamical effect from the averaged matter density $\avg{\rholoc}$ and the effective energy density $\rho^\mathrm{eff}_\CD$ arising from the local inhomogeneity and geometry (from Eq.~\eqref{eq:def_rhoeff}), $\rhotot(t) := \avg{\rholoc}(t) + \rho^\mathrm{eff}_\CD(t)$. Backreaction effects are assumed to remain negligible in the early Universe, $\rhotot \approx \avg{\rholoc} \propto a_\CD^{-3}$, up to a certain transition time $t_T$ from where the formation of structures leads to a different evolution for $\rhotot$ --- with a sharp transition at $t_T$ for simplicity while keeping $\rhotot(t)$ continuous.

Specifically, the following ansatz is prescribed for $\rhotot$:
\begin{equation}
    \rhotot(t) = \left\{
    \begin{aligned}
        & \rhoinit  \frac{a_\CD(\tinit)^3}{a_\CD(t)^3} \; , \; \mathrm{for} \; t \leq t_T \; ; \\
        & \rhoinit \frac{a_\CD(\tinit)^3 a_\CD(t_T)^\epsilon}{a_\CD(t)^{3+\epsilon}} = \rho_T \frac{a_\CD(t_T)^{3+\epsilon}}{a_\CD(t)^{3+\epsilon}} \; , \; \mathrm{for} \; t > t_T \; ,
    \end{aligned}
    \right.
\end{equation}
for an arbitrary initial early time $\tinit$, a given transition time $t_T > \tinit$ and an exponent $\epsilon < 0$, with $\rhoinit := \rho(\tinit) = \rhotot(\tinit)$ and $\rho_T := \rho(t_T) = \rhotot(t_T)$.
This amounts to a vanishing effective-fluid contribution before $t_T$, and to an effective-fluid energy density post-transition given by
\begin{equation}
    \rho^\mathrm{eff}_\CD(t > t_T) := \rhoinit  \frac{a_\CD(\tinit)^3}{a_\CD(t)^3} \left(\frac{a_\CD(t_T)^\epsilon}{a_\CD(t)^{\epsilon}} - 1 \right) \; ,
     \label{eq:rhoeff_GMP}
\end{equation}
while the averaged dust density, by construction, keeps its inverse scaling with the volume at all times, yielding $\avg{\rholoc}(t) = \rhoinit \, a_\CD(\tinit)^3 / a_\CD(t)^3 \; \forall t$.
The effective fluid's pressure is then reconstructed from $\rho^\mathrm{eff}_\CD(t)$ and its derivative, from the expected effective energy conservation equation~\eqref{eq:intcond_eff}. This consistently sets $p^\mathrm{eff}_\CD(t) = \rho^\mathrm{eff}_\CD(t) = 0$ at $t < t_T$, and specifies a nonzero effective pressure
\begin{equation}
    p^\mathrm{eff}_\CD(t > t_T) := \frac{\epsilon}{3} \:\! \rho_T \:\! \frac{a_\CD(t_T)^{3+\epsilon}}{a_\CD(t)^{3+\epsilon}} < 0 \; ,
    \label{eq:peff_GMP}
\end{equation}
after the transition time.
This corresponds to an effective-fluid EoS parameter, at $t > t_T$, given by
\begin{equation}
   w_{\mathrm{eff},\CD}(t > t_T) = \frac{\epsilon}{3} \, \frac{a_\CD(t_T)^\epsilon \, }{a_\CD(t_T)^\epsilon - a_\CD(t)^{\epsilon}  }  \; .
\end{equation}
We note that, with $\epsilon < 0$, $w_{\mathrm{eff},\CD}$ remains nonpositive once well-defined (i.e. for $t > t_T$), and diverges to $- \infty$ at $t_T$, where $\rho^\mathrm{eff}_\CD(t_T) = 0$ while $\dot\rho^\mathrm{eff}_\CD(t_T) \neq 0$ (and thus $p^\mathrm{eff}_\CD(t_T) \neq 0$).
The above time dependences of the effective sources and EoS can also be straightforwardly expressed as functions of the effective scale factor $a_\CD(t)$, with the transition occurring at $a_T := a_\CD(t= t_T)$, a normalisation chosen so that $a_\CD(t_0) = 1$ at the present time $t_0$, and the initial time corresponding to a scale factor $\ainit := a_\CD(t=\tinit)$. As for the GMP model, the FLRW relation between the cosmological redshift $z$ and the effective scale factor $a_\CD$, is assumed to hold to a good approximation, $a_\CD \approx 1/(1+z)$.

The evolution of the kinematical backreaction and averaged curvature within this model can simply be recovered from the assumed $\rho^\mathrm{eff}_\CD(t)$ and $p^\mathrm{eff}_\CD(t)$ as given above by inverting Eqs.~\eqref{eq:def_rhoeff}--\eqref{eq:def_peff}, setting again a spatially flat background model, $K_\CD = 0$. This gives, at late times,
\begin{align}
    & \QD(a_\CD > a_T) = - 4 \pi G \left[ (1 + \epsilon) \left( \frac{a_T}{a_\CD} \right)^\epsilon - 1 \right] \frac{\rhoinit \, \ainit^3}{a_\CD^3}  \; ; \\
    & \RD(a_\CD > a_T) = - 12 \pi G \left[ \left(1 - \frac{\epsilon}{3} \right) \left( \frac{a_T}{a_\CD} \right)^\epsilon - 1 \right] \frac{\rhoinit \, \ainit^3}{a_\CD^3} \; ,
\end{align}
while $\QD$ and $\RD$ vanish by construction for $a_\CD < a_T$, i.e. $t < t_T$.

In the following we will set the free parameters of this model to their best-fit values obtained in~\cite{Giani_2025a}: $\epsilon = -0.58$; $a_T = 1/(1+1.14) \approx 0.467$; $ H_{\CD,0} = 68.94 \, \mathrm{ km/(s}\cdot \mathrm{Mpc)}$; $\Omega^{\CD,0}_d := 8 \pi G \avg{\rholoc}(t_0) \, / (3 H^2_{\CD,0}) = 8 \pi G \rhotot(t_0) \, a_T^{-\epsilon} / (3 H_{\CD,0}^2) = 0.33$.
The model includes a cosmological constant, whose present-day contribution can be evaluated directly from the averaged Hamiltonian constraint at $t_0$, $a_T^\epsilon \, \Omega^{\CD,0}_d + \Omega^{\CD,0}_\Lambda = 1$: $\Omega_\Lambda^{\CD,0} := \Lambda / (3 H_{\CD,0}^2) \approx 0.513$. 

\begin{figure}[htb!]
\begin{adjustwidth}{-2cm}{-1.5cm}
\centering
    \begin{minipage}{0.495\linewidth}
        \centering
        \includegraphics[width=\linewidth]{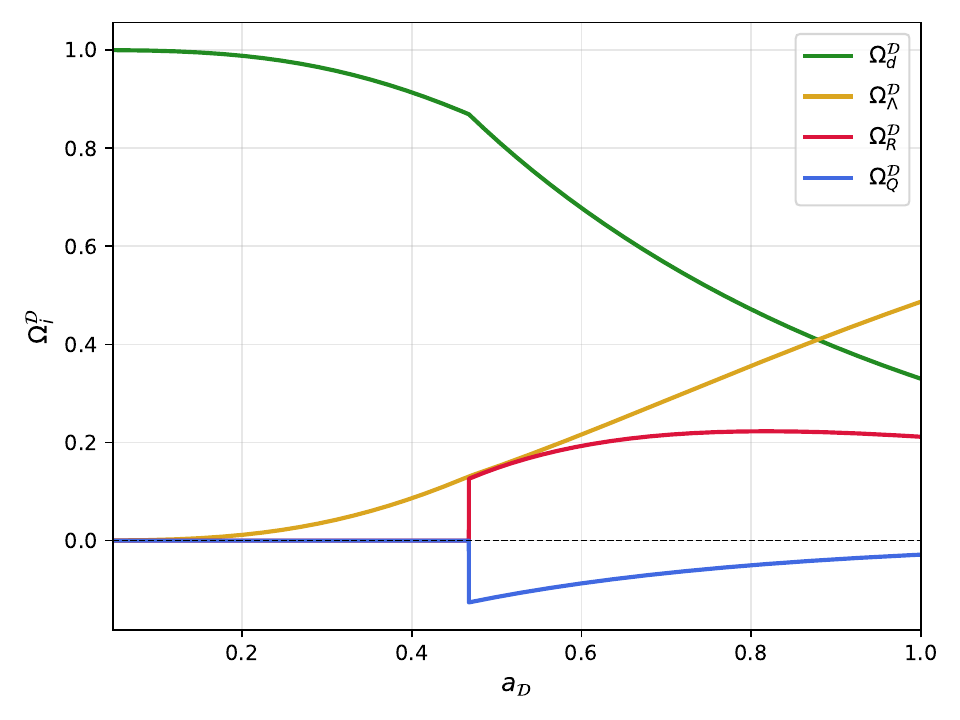}
    \end{minipage}\hfill
    \begin{minipage}{0.495\linewidth}
        \centering        \includegraphics[width=\linewidth]{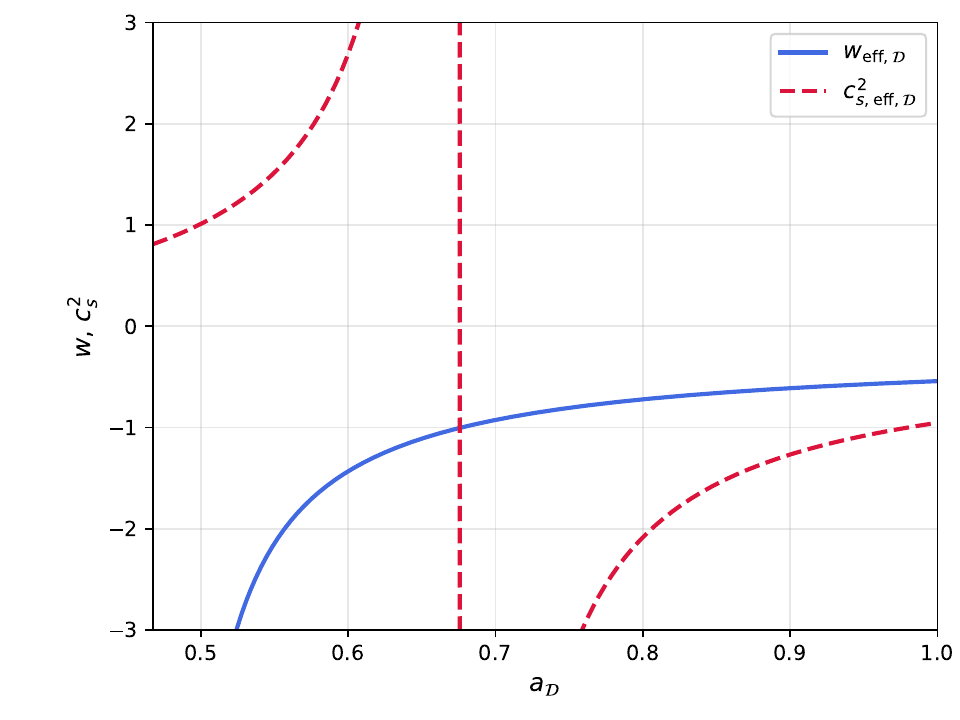}
    \end{minipage}
\end{adjustwidth}
\caption{Energy balance (left), and effective fluid EoS parameter and squared sound speed (right) as functions of the effective scale factor $a_\CD$ in the GMP model, with parameters $\epsilon = -0.58$; $a_T \approx 0.467$; $H_{\CD,0} = 68.94$~km/(s$\cdot$Mpc); $\Omega^{\CD,0}_d = 0.33$.}
\label{fig:gmp_1}
\end{figure}

On the left panel of Fig.~\ref{fig:gmp_1}, we show each contribution to the averaged Hamiltonian constraint in terms of the $\Omega$ variables of Eq.~\eqref{eq:omegas} as functions of $a_\CD$ after the transition time. From this plot, it is apparent that, after the transition occurs, the model has positive backreaction and negative average spatial curvature at all times; the relative contribution from $\QD$ to the energy budget slowly decays over time while that from $\RD$ remains around $20\%$, decreasing $\Omega_\Lambda^0$ with respect to its concordance $\Lambda$CDM value accordingly. The transition at $a_T$ with a nonzero pressure of the effective fluid at $a_\CD > a_T$ manifests itself as a discontinuity from zero to nonzero values in $\Omega^\CD_\mathcal{Q}$ and $\Omega^\CD_\mathcal{R}$. The combination $\Omega^\CD_\mathcal{Q} + \Omega^\CD_\mathcal{R}$ remains continuous with a change of slope at $a_T$ (with $\QD(a_T) = - \RD(a_T)$), and so do $\Omega^\CD_d$ and $\Omega^\CD_\Lambda$.

On the right panel, the effective fluid EoS and squared speed of sound parameters after the transition time are shown as functions of the effective scale factor. The divergence of $w_{\mathrm{eff},\CD}$ towards $a_\CD = a_T$ is apparent, while the effective sound speed also exhibits a divergence at some later time $t_1$ (with $a_\CD(t_1) \approx 0.676$), where $\dot\rho^\mathrm{eff}_\CD(t_1) = 0$, i.e. $w_{\mathrm{eff},\CD}(t_1) = -1$, while $\dot p^\mathrm{eff}_\CD$ never vanishes (see Eq.~\eqref{eq:peff_GMP}). The negative sign of $w_{\mathrm{eff},\CD}$ and this behaviour of $c^2_{s,\mathrm{eff},\CD}$ again point against the applicability of the M\'esz\'aros approximation at late times for this model. Nevertheless, it should be kept in mind that this approximation, but also its applicability criteria in terms of $w_{\mathrm{eff},\CD}$ and $c^2_{s,\mathrm{eff},\CD}$, rely on a barotropic picture for the non-dust (here, effective) fluid; we will see with this example that a different (e.g. comoving) representation of the effective fluid perturbation can sometimes lead to a dust perturbation evolution still rather compatible with the outcome of the M\'esz\'aros approximation---i.e., with the source-free evolution equation~\eqref{eq:original_Meszaros} for $\Delta^{(d)}$. 

\subsubsection{GMP model: CGI linear structure growth}

Similarly to the GMC model above, the study that introduced the GMP model~\cite{Giani_2025b} also discussed its first-order perturbation regime, making predictions for the $f \sigma_8$ parameter on this dynamical background. The same discussion as in Sec.~\ref{subsubsec:GMC_CGI} above for the GMC model perturbations applies in this case as well, as the linear growth in~\cite{Giani_2025b} was also described using the M\'esz\'aros approximation, and corresponded to a perturbation in the total dust plus effective energy density $\rhotot$ ---with $\rho^{(d)}$ replaced by $\rhotot$ in the coefficient of the last term of the evolution equation~\eqref{eq:original_Meszaros}.
Here, we will again consider the perturbations of the dust and effective fluids separately, and compare the dust structure growth in presence of $\Delta\eff$ to the predictions of the M\'esz\'aros approximation for the dust flow.

For both the comoving and barotropic representations of the perturbation evolution equations (as well as for the predictions of the M\'esz\'aros approximation for $\Delta^{(d)}$ used for comparison), we use the same initial conditions as in Sec.~\ref{subsubsec:ts_CGI} above at $z=20$. In this case, the $\Lambda$CDM evolution of the dust and expansion perturbations used between the CMB epoch and $z=20$ continues consistently as such until $a_\CD = a_T$, since the background is strictly $\Lambda$CDM until that point. We obtain accordingly $\Delta^{(d)}(a_T) \approx  3.0 \times 10^{-2}$, and $\mathcal{Z}(a_T) \approx -5.3  \times 10^{-2} H_0$, which we use as ICs for the subsequent evolution. Meanwhile, in both representations the effective fluid perturbations and tilt velocity are kept at zero until $a_\CD = a_T$ since their background energy sources strictly vanish until that time. However, as noted above, the background's effective-fluid pressure is discontinuous at the transition while $\rho_{\mathrm{eff},\CD}(a_T)$ is well defined and vanishing, leading to a formally infinite $w_\mathrm{eff,\CD}$ at $a_T$. For the evolution equations~\eqref{eq:dotdeltad_com}--\eqref{eq:dotZ_com} (comoving picture) or \eqref{eq:dotdeltad_bar}--\eqref{eq:dotV_bar} (barotropic picture) to remain well-defined, this prescribes a specific, nonzero value for $\Delta\eff$ immediately after the transition from the given expansion perturbation. From Eqs.~\eqref{eq:dotdeltaeff_com} and \eqref{eq:dotdeltaeff_bar}, this value is the same in both representations and at any wavelength, namely, $\Delta\eff\com(a_T) = \Delta\eff\baro(a_T) = \mathcal{Z}(a_T)/\Theta(a_T) \approx -9.2 \times 10^{-3}$. Note that the CGI density and pressure perturbation variables we use are defined with a normalisation by the corresponding fluid's background energy density, namely, here, $\Delta\eff \propto {}^{(3)}\!\Delta \rho\eff / \rho_{\mathrm{eff},\CD}$, $\tilde\CP\eff \propto {}^{(3)}\!\Delta p\eff / \rho_{\mathrm{eff},\CD}$. Hence, the density and pressure gradients and Laplacians themselves remain continuous and vanishing at $a_\CD = a_T$. Finally, in the barotropic picture, the tilt velocity $\tilde V\eff\baro(a_T)$ is also prescribed by Eq.~\eqref{eq:dotV_bar}, but the imposed value is in fact still zero (at any wavelength), thus keeping $\tilde V\eff\baro$ continuous across the transition. Once these values are set as ICs for the effective fluid density and tilt perturbations at $a_\CD = a_T$, there is no divergence in the system of linear perturbation evolution equations immediately after $a_\CD = a_T$, and their evolution can continue after the transition, with the background sources specified by Eqs.~\eqref{eq:rhoeff_GMP}--\eqref{eq:peff_GMP}.

\begin{figure}[htb!]
    \centering
    \includegraphics[width=0.7\textwidth]{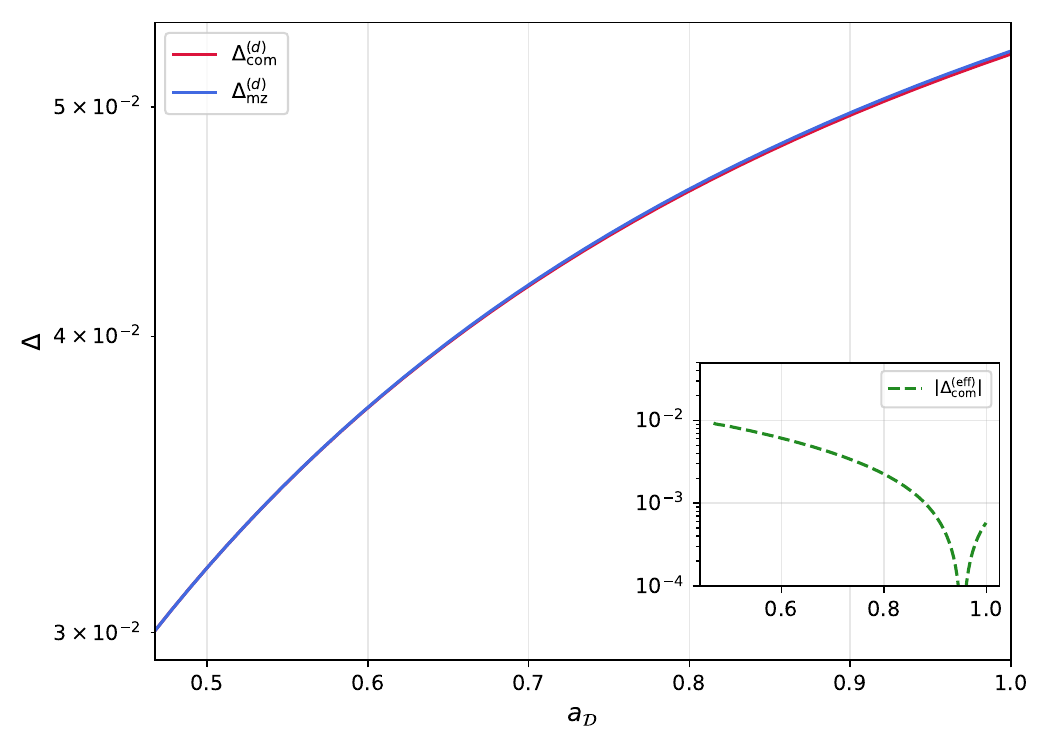} 
    \caption{Dust density perturbation growth within the comoving effective fluid framework, $\Delta^{(d)}_\mathrm{com}$, and the M\'esz\'aros approximation, $\Delta^{(d)}_\mathrm{mz}$, for a GMP background model, as functions of the average scale factor $a_\CD$ after the transition time. The inset panel shows the evolution of the magnitude of the effective fluid perturbations in the comoving framework, $|\Delta\eff_\mathrm{com}|$. The ICs for the dust density and expansion perturbations are seeded at the CMB epoch using the $\Lambda$CDM-based approximation as in Sec.~\ref{subsec:ts} above, and, in this case, evolved within $\Lambda$CDM up to $a_\CD = a_T$ since the background model remains $\Lambda$CDM until that point. The IC for the effective fluid density comoving perturbation at $a_\CD = a_T$ is set to the nonzero value $\Delta\eff\com(a_T) = \mathcal{Z}(a_T)/\Theta(a_T)$ imposed by the evolution equation~\eqref{eq:dotdeltaeff_com} (see main text), with the effective fluid energy density and pressure remaining zero overall until that time.}
    \label{fig:gmp_2}
\end{figure}

Fig.~\ref{fig:gmp_2} shows the evolution of the dust perturbation in the comoving framework for the GMP model as a function of its effective scale factor $a_\CD$ from the transition time onwards, as well as the prediction of the M\'esz\'aros approximation for comparison, using the initial conditions as detailed above. We also show as an inset the magnitude of the comoving effective fluid perturbation over the same scale factor range. Remarkably, for this model, the M\'esz\'aros approximation indeed appears to provide a very good description of the (scale-independent) dust perturbation growth obtained in the comoving frame. This is due to the effective fluid perturbations exhibiting a stable, decaying oscillatory behaviour with an overall small amplitude relative to the dust perturbation. This allows the source in the comoving-picture dust perturbation evolution equation~\eqref{eq:ddotdeltad_com} to be neglected, coinciding with the outcome of the M\'esz\'aros approximation (Eq.~\eqref{eq:original_Meszaros}) despite the latter being obtained from a barotropic framework.

\begin{figure}[htb!]
\begin{adjustwidth}{-2.cm}{-1.5cm}
\centering
    \begin{minipage}{0.495\linewidth}
        \centering
        \includegraphics[width=\linewidth]{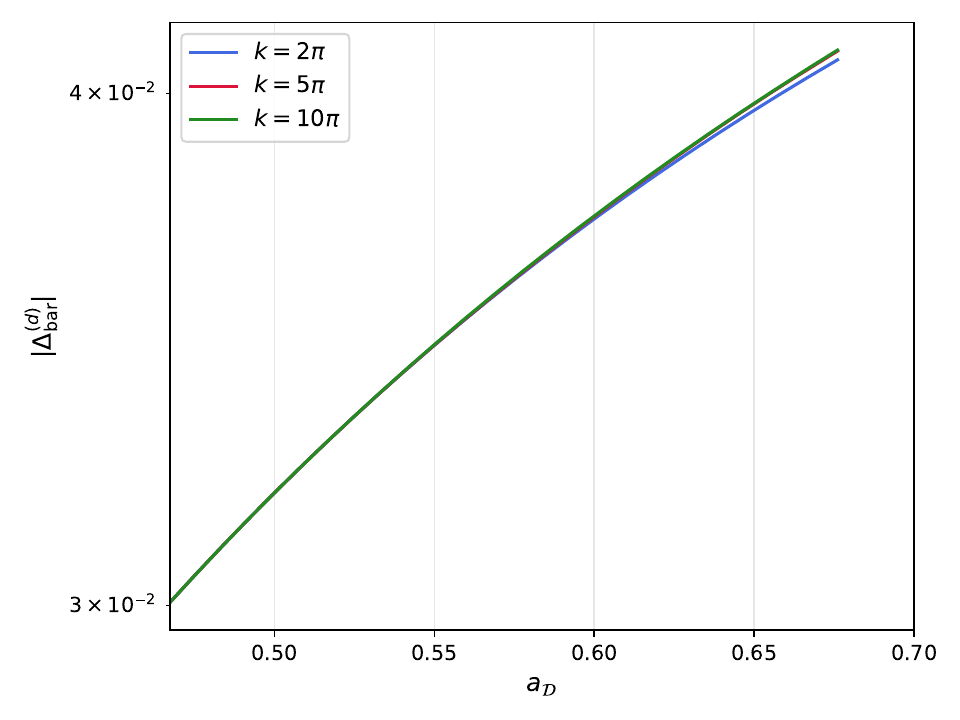}
    \end{minipage}\hfill
    \begin{minipage}{0.495\linewidth}
        \centering
        \includegraphics[width=\linewidth]{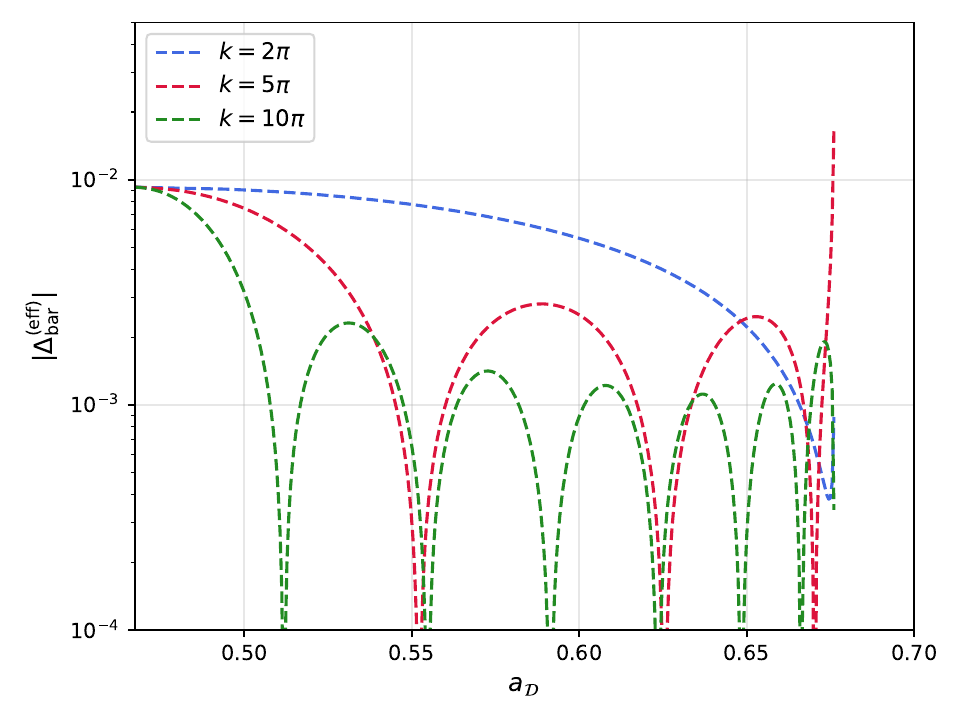}
    \end{minipage}
\end{adjustwidth}
\caption{Horizon and sub-horizon dust (left panel) and effective fluid (right panel) density perturbation growth on a GMP background within the barotropic condition. We show the amplitude evolution for three different wavelengths on a spatially flat background ($K_\CD = 0$), at horizon ($k = 2\pi$) and sub-horizon scales, as functions of the average scale factor $a_\CD$, between the transition time $t_T$ (where $a_\CD = a_T$) and the time $t_1$ where the evolution system breaks down (with $w_{\mathrm{eff},\CD}(t_1) = -1$). Here, the wavenumber $k$ is expressed in units of the present-day inverse Hubble length, $H_0/c$, as evaluated in the $\Lambda$CDM model following the best fit of~\cite{Planck_2018}. The ICs are seeded via the $\Lambda$CDM approximation for the dust density and expansion perturbation, as in Sec.~\ref{subsec:ts} above. The seed perturbations are then evolved till $a_\CD = a_T$ within $\Lambda$CDM, defining ICs for their $a_\CD \geq a_T$ evolution. The effective fluid source vanishes altogether for $a_\CD \leq a_T$. The values of its energy density and tilt perturbations at $a_\CD = a_T$ are then imposed by the evolution equation~\eqref{eq:dotdeltaeff_bar} to the scale-independent values $\Delta\eff\baro(a_T) = \mathcal{Z}(a_T)/\Theta(a_T)$, $\tilde V\eff\baro(a_T) = 0$ (see main text), and these values are thus set as ICs for the evolution after the transition.}
\label{fig:gmp_3}
\end{figure}

Within this model, at all scales, the description of barotropic-picture perturbations even breaks down altogether at the effective phantom crossing (at time $t_1$) as the effective fluid perturbations themselves diverge. We show in Fig.~\ref{fig:gmp_3} the short evolution of the dust (left panel) and effective fluid (right panel) energy density perturbations in that picture at horizon and sub-horizon scales as functions of the effective scale factor $a_\CD$, between $a_T$ and $a_\CD(t_1)$, using the (scale-independent) ICs discussed above at $a_\CD = a_T$. Until the framework breaks down at $t_1$, the dust perturbations undergo a steady growth with no notable features and little scale dependence, while the ultimately diverging effective fluid perturbations have an oscillatory behaviour in time with frequency increasing with $k$.

\subsection{The relativistic Zel'dovich approximation model}
\label{subsec:RZA}

The relativistic Zel'dovich approximation (RZA) is a framework for the description of structure formation in general-relativistic spacetimes, presented in~\cite{BuchertRZA_2012,BuchertRZA_2013,AllesRZA_2015} for irrotational dust and \cite{LiRZA_2018} for irrotational perfect fluids with pressure. It builds upon its well-known Newtonian counterpart by Zel'dovich~\cite{Zeldovich_1970} and, like the latter, allows for, e.g., nonlinear density variations with exact local mass conservation, while linearising or expanding in a deformation field (tied to the metric, in the relativistic case).

\subsubsection{RZA: general framework}

We will here restrict our attention to the first-order RZA scheme for dust as in~\cite{BuchertRZA_2012,BuchertRZA_2013}, and we refer the reader to these papers for more details.
In this framework, a model spacetime sourced by a single, inhomogeneous irrotational dust fluid (with $4-$velocity $u^\mu$) is again considered and described via a comoving and synchronous coordinate system $(t,X^i)$, so that the line element reads
\begin{equation}
    \mathrm{d}s^2 = - \mathrm{d}t^2 + g_{ij} \, \mathrm{d}X^i \mathrm{d}X^j \; ,
\end{equation}
where $t$ is a proper time for the dust flow whose level sets define flow-orthogonal hypersurfaces, and the spatial components $g_{ij}$ of the spacetime metric also correspond to the components of the induced spatial metric on the hypersurfaces.

This spatial metric is decomposed into a general basis of three spatial coframes $\mathbf{\eta}^a = \eta^a_{\; i} \, \mathrm{d}X^i$, $a = 1,2,3$:
\begin{equation}
    g_{ij} = G_{ab} \:\! \eta^a_{\;i} \:\! \eta^b_{\;j} \; ,
\end{equation}
with a time-independent Gram matrix $G_{ab}(X^k)$. This matrix can be chosen as the identity, $G_{ab} := \delta_{ab}$, so that the $\mathbf{\eta}^a$ define Cartan (i.e., orthonormal) spatial coframes, with an \emph{a priori} nontrivial form at any point in time, as was done in~\cite{BuchertRZA_2012}. Alternatively, selecting some initial slice $t = \tinit$, the components of the Gram matrix can be chosen to coincide with those of the spatial metric on that slice, $G_{ab}(X^k) := \delta_a^{\;i} \delta_b^{\;j}\, g_{ij}(\tinit,X^k)$, while initialising the coframes to $\mathbf{\eta}^a = \mathrm{d}X^a$, i.e., $\eta^a_{\;i}(\tinit, X^k) = \delta^a_{\;i}$. This latter option was used in~\cite{BuchertRZA_2013} and subsequent related papers, and we will also make this choice, for which the averaged quantities required to define our two-fluid background will not explicitly depend on the initial metric $G_{ab}$.

In order to define the approximation scheme itself, a reference dust FLRW metric is chosen, with scale factor $\aref(t)$ and density $\rhoref(t)$, and with its cosmic time mapped to the above $t$ coordinate. It includes the same cosmological constant as the inhomogeneous model universe, and may include a nonzero spatial curvature.
Local deviation to the reference homogeneous and isotropic dynamics is then encoded into the coframes, which play a role analogous to the gradient of the Lagrangian-frame displacement field in the Newtonian framework:
\begin{equation}
    \eta^a_{\;i} =: \frac{\aref(t)}{\arefi} \left( \delta^a_{\; i} + P^a_{\,\,i} \right) \quad ,  \quad \arefi := \aref(\tinit) \;\; ,
    \label{eq:RZAcoframesplit}
\end{equation}
with $P^a_{\,\,i}$ defining the small perturbation field, vanishing at the initial time $\tinit$, while the initial metric $G_{ab}$ is assumed to be everywhere close to the reference FLRW metric at $t=\tinit$ for consistency. The components $P^a_{\,\,i}$ of the perturbation field represent the dynamical variables, and the set of Einstein equations in $3+1$ form may be rewritten as evolution equations for these variables.
In the first-order RZA, these evolution equations are then linearised in $P^a_{\,\,i}$ and in the deviation of the initial metric from the reference. The perturbation field may then be split into its various time-evolution modes, which can be considered separately, and the leading modes can be assumed to take a separable form as
\begin{equation}
    P^a_{\,\,i}(t,X^k) = \xi(t) \SP^a_{\,i}(X^k) \; ,
    \label{eq:RZAseparable}
\end{equation}
with $\xi(\tinit) = 0$ and $\dot\xi(\tinit) = 1$.
Such a form is indeed compatible with the evolution equations, which reduce to ODEs and include a mode at linear order where the evolution of all components of $P^a_{\,\,i}$ follows that of its trace $\delta^{\;i}_{a} \:\! P^a_{\,\,i}$. This latter property translates in the Newtonian framework into the selection of the leading, longitudinal mode where the deformation and velocity fields remain proportional to their respective divergences. The separability assumption~\eqref{eq:RZAseparable} itself, through its first two derivatives, can be seen as a relativistic equivalent of the `slaving condition' enforcing the proportionality of the velocity and acceleration fields in the first-order Newtonian Zel'dovich approximation.

The linearised evolution of such a mode for $P^a_{\,\,i}$ is then determined by the scaling function $\xi(t)$ as $\xi(t) = (q(t) - q(\tinit)) / \dot q(\tinit)$ where $q(t)$ is a solution of
\begin{equation}
    \ddot q + 2 \, \frac{\dot \aref(t)}{\aref(t)}  \, \dot q - 4 \pi G \rhoref(t) \, q = 0 \; .
    \label{eq:ddotq_RZA}
\end{equation}
Accordingly, $\xi(t)$ satisfies 
\begin{equation}
    \ddot \xi + 2 \, \frac{\dot \aref(t)}{\aref(t)}  \, \dot \xi - 4 \pi G \rhoref(t) \left(\xi + \xi_1 \right) = 0 \; ,
    \label{eq:ddotxi_RZA}
\end{equation}
with $\xi_1 := q(t_i) / \dot q (t_i)$.
As in the standard Zel'dovich approximation, the decaying-mode solution to the above second-order differential equations is neglected, keeping only the growing mode---and resulting indeed in the separable form~\eqref{eq:RZAseparable} approximately describing the perturbation field as a whole, with the corresponding $\xi(t)$ solution. For the chosen reference FLRW model, with dust as the only fluid source, the growing mode for $q(t)$ can be characterised as the only solution of Eq.~\eqref{eq:ddotq_RZA} (up to normalisation) going to $0$ in the $\aref \rightarrow 0$ limit. This uniquely specifies the function $\xi(t)$ and the constant $\xi_1$ for each reference model of this class. In particular, for such reference models, the corresponding solution for $q(t)$ behaves as $q(t) \propto \aref(t)$ for small $\aref(t)$ in all cases, while this remains the exact solution at all times in the special case of an Einstein--de Sitter reference. Accordingly, $\xi_1$ will be close to the initial inverse Hubble constant, $\xi_1 \approx \aref(\tinit) / \dot \aref(\tinit)$ (and hence $\xi_1 > 0$), and equal to it in the Einstein--de Sitter case.

The final step of the RZA is to express all relevant observables as exact functions of the coframes, and thus of $P^a_{\,\,i}$, plus the initial conditions $G_{ab}(X^k)$ and $\rholoc(\tinit,X^k)$ where needed, and then evaluate these functionally for the approximate (linearised and leading-mode) solution obtained for $P^a_{\,\,i}$, without linearising the result. For instance, the spatial metric components in the $(X^i)$ coordinate basis will be evaluated as
\begin{equation}
    g_{ij} = G_{ab} \:\! \eta^a_{\;i} \eta^b_{\;j} =  \frac{\aref^2}{\arefi^2} \, G_{ab} \, (\delta^a_{\;i} + P^a_{\,\,i} ) \;\! (\delta^b_{\;j} + P^a_{\,\,j} ) \; ,
\end{equation}
for the above $P^a_{\,\,i}$ with given initial conditions, keeping the quadratic $G_{ab} \:\! P^a_{\,\,i} P^a_{\,\,j}$ term. Meanwhile, the local dust density $\rholoc$ is evaluated as
\begin{equation}
    \rholoc = \rholocinit \, \sqrt{\frac{\det(g_{ij})(\tinit)}{\det(g_{ij})}} = \frac{\rholocinit}{\det(\eta^a_{\;i})} = \frac{\rholocinit(X^k) \; \arefi^3}{\aref(t)^3 \;\! \det \! \big(\delta^a_{\;i} + \xi(t) \SP^a_{\;i}(X^k) \big)} \; ,
\end{equation}
solving the local mass conservation equation $\dot \rholoc +  \rholoc \, \nabla_\mu u^\mu  = 0$ exactly, with the initial rest-mass density $\rholocinit(X^k) := \rholoc(\tinit, X^k)$. This allows for nonlinear density contrasts, and, as expected and similarly to the corresponding Newtonian framework, describes local matter collapse up to arbitrarily large densities near shell-crossings, where the first-order scheme breaks down (with $\det(g_{ij}) \rightarrow 0$ in the flow-comoving coordinates $(X^i)$).

Many variables, such as $\rholoc / \rholocinit$ above, or the peculiar-expansion tensor $\theta^i_{\,j} := \nabla_j u^i - (\dot \aref / \aref) \:\! \delta^i_{\,j}$ (which determines the kinematical backreaction), can be expressed as functions of the coframes and their time derivatives only. Hence, in the first-order RZA, with the separable ansatz~\eqref{eq:RZAcoframesplit}--\eqref{eq:RZAseparable}, scalars constructed from such variables depend only on $\aref(t)$, $\xi(t)$, their derivatives, and the three principal invariants of the matrix $\SP^a_{\;i}(X^k)$ as initial conditions. It can be shown that the latter coincide with the principal invariants of the peculiar-expansion tensor on the initial slice,
\begin{equation}
     \Ii(X^k) := \theta^i_{\,i}(\tinit,X^k) \;\; ; \;\; \IIi(X^k) :=  \frac{1}{2} \left( \theta^i_{\,i} \:\! \theta^j_{\,j} - \theta^i_{\,j} \:\! \theta^j_{\,i} \right) (\tinit,X^k) \;\; ; \;\; \IIIi(X^k) :=  \det(\theta^i_{\,j})(\tinit,X^k) \; .
\end{equation}

\subsubsection{RZA: specific ansatz}

We define our inhomogeneous model spacetime as the result of a first-order dust RZA as above, with a flat $\Lambda$CDM reference $\aref(t)$, and with the second and third principal invariants of the peculiar-expansion tensor vanishing on average at the initial time over our spatial domain of interest $\CD$: $\avg{\IIi} (\tinit) = 0 = \avg{\IIIi}(\tinit)$. This latter simplifying assumption on quadratic and cubic terms in $\SP^a_{\;i}$ enables a consistent evaluation of terms like $\RD$ from several possible ways of expressing local variables, despite the specified coframes only satisfying the Einstein equations up to linear order in $\SP^a_{\;i}$ (see~\cite{BuchertRZA_2013} for a discussion of this property, and~\cite{VigneronBuchert_2019} for a detailed example of application of the assumption of vanishing second and third invariants within the RZA).
We denote by $\beta := \avg{\Ii}(t_i)  = \avg{\nabla_\mu u^\mu}(\tinit) - 3 \, (\dot\aref / \aref)(\tinit)$
the value of the remaining initial averaged peculiar-expansion invariant, i.e., the initial average of the peculiar-expansion scalar.
We will not need a full specification of the local values of the initial invariants $\Ii$, $\IIi$, $\IIIi$ as the terms ($\QD$, $\RD$, $\avg{\rholoc}$) specifying the dynamics of the effective scale factor $a_\CD(t)$ depend only on their averages on the initial slice in this scheme.

Within the above assumptions, the kinematical backreaction, averaged spatial curvature and effective scale factor for the domain considered are expressed as follows~\cite{BuchertRZA_2013}:
\begin{align}
   \label{eq:RZA_aD}   &a_\CD(t)  =  \frac{a_\CD (\tinit)}{\aref(\tinit)} \, \aref(t)\, \sqrt[3]{1 + \beta \, \xi(t)} \; ; \\
       &\QD(t)  =  - \frac{2}{3} \left( \frac{\beta \, \dot\xi(t)}{1 + \beta \, \xi(t)} \right)^2 \; ; \\
       \label{eq:RZA_RD}
  & \RD(t)  = - 4 \beta \, \frac{\ddot\xi(t) + 3 \, (\dot \aref / \aref)(t) \, \dot\xi(t)}{1 + \beta \, \xi(t)} = - 4 \beta \, \frac{(\dot\aref/\aref)(t) \, \dot\xi(t) + 4 \pi G \rhoref(t)  \left(\xi(t) + \xi_1 \right)}{1 + \beta \, \xi(t)} \; .
\end{align}
We note that this implies that $\QD(t) < 0$ and that $\RD$ has (at least near $\tinit$) an opposite sign to $\beta$.
It can be verified that, with these expressions, the integrability condition~\eqref{eq:intcond} coupling $\QD$ to $\RD$ is satisfied \emph{exactly} at all times.
Accordingly, the effective perfect-fluid source with energy density and pressure $\rho_\CD^\mathrm{eff}(t)$, $p_\CD^\mathrm{eff}(t)$ defined from $\QD(t)$ and $\RD(t)$ as per Eqs.~\eqref{eq:def_rhoeff}--\eqref{eq:def_peff} (choosing again $K_\CD = 0$) satisfy the effective energy conservation equation~\eqref{eq:intcond_eff} as expected.
We can use the freedom in the definition of $a_\CD(t)$ to set $a_\CD(\tinit) := \aref(\tinit)$ initially, simplifying the form of Eq.~\eqref{eq:RZA_aD}.

Finally, the averaged dust energy density is given as usual by the total mass conservation,
$\avg{\rholoc}(t) = \avg{\rholoc}(\tinit) \, (a_\CD(\tinit) / a_\CD(t))^3$, where, from the averaged Hamiltonian constraint at initial time, the initial averaged density is given by:
\begin{equation}
    \avg{\rholoc}(\tinit) = \rhoref(\tinit) \left( 1 - \beta \, \xi_1 \right) \; .
\end{equation}
If the reference FLRW model is considered as a best-fit description of the dynamics on the largest scales (e.g., the concordance $\Lambda$CDM model), the domain $\CD$ in this scenario may be chosen at a scale where structure is still visible, with an average density distinct from $\rhoref$ at all times. From the above relation, together with the definition of $\beta$ as the initial averaged peculiar expansion rate, and the sign of the averaged curvature from Eq.~\eqref{eq:RZA_RD} at initial time, a positive (resp. negative) $\beta$ corresponds to a domain chosen at $\tinit$ as an underdense void (resp. overdense wall), which expands faster (resp. slower) than the flat homogeneous reference, and has a negative (resp. positive) average spatial curvature. This interpretation of the averaged dynamics in this scenario as corresponding to a local domain, smaller than a typical statistical homogeneity scale, is indeed consistent with the fact that the assumption we use of vanishing averaged invariants $\avg\IIi(\tinit)$ and $\avg\IIIi(\tinit)$ is more relevant at such small scales~\cite{VigneronBuchert_2019}.

As in the previous subsections, the effective scale factor $a_\CD(t)$, average energy density $\avg{\rholoc}(t)$, and effective fluid energy density $\rho^\mathrm{eff}_\CD(t)$ and pressure $p^\mathrm{eff}_\CD(t)$, as determined above, define our effective two-fluid background for the CGI perturbation scheme. It can be verified that the above prescriptions for those quantities indeed satisfy exactly at all times the averaged Hamiltonian constraint and Raychaudhuri equation, Eqs.~\eqref{eq:avg_Hamilton_eff}--\eqref{eq:avg_Raych_eff}, in addition to the effective energy conservation equation---hence overall providing an exact solution to the Buchert equations, Eqs.~\eqref{eq:avg_Hamilton_eff}--\eqref{eq:intcond_eff}.

For our illustrative example, we consider an underdense region, with $\beta\xi_1 \approx 9.5\times10^{-3}$, i.e., $\beta \approx 0.51 H_0$. The underlying reference flat $\Lambda$CDM model is set to the Planck 2018 best-fit parameters~\cite{Planck_2018} and our initial time $\tinit$ is associated with a redshift of $z=20$ in that model. 

\begin{figure}[htb!]
\begin{adjustwidth}{-2cm}{-1.5cm}
\centering
    \begin{minipage}{0.495\linewidth}
        \centering
        \includegraphics[width=\linewidth]{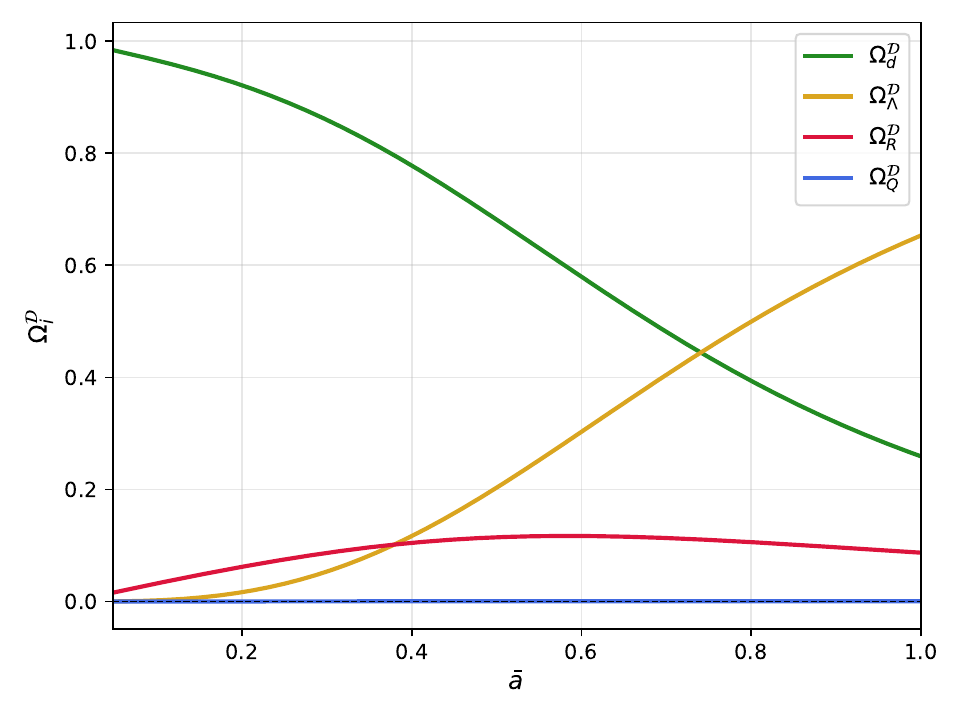}
    \end{minipage}\hfill
    \begin{minipage}{0.495\linewidth}
        \centering        \includegraphics[width=\linewidth]{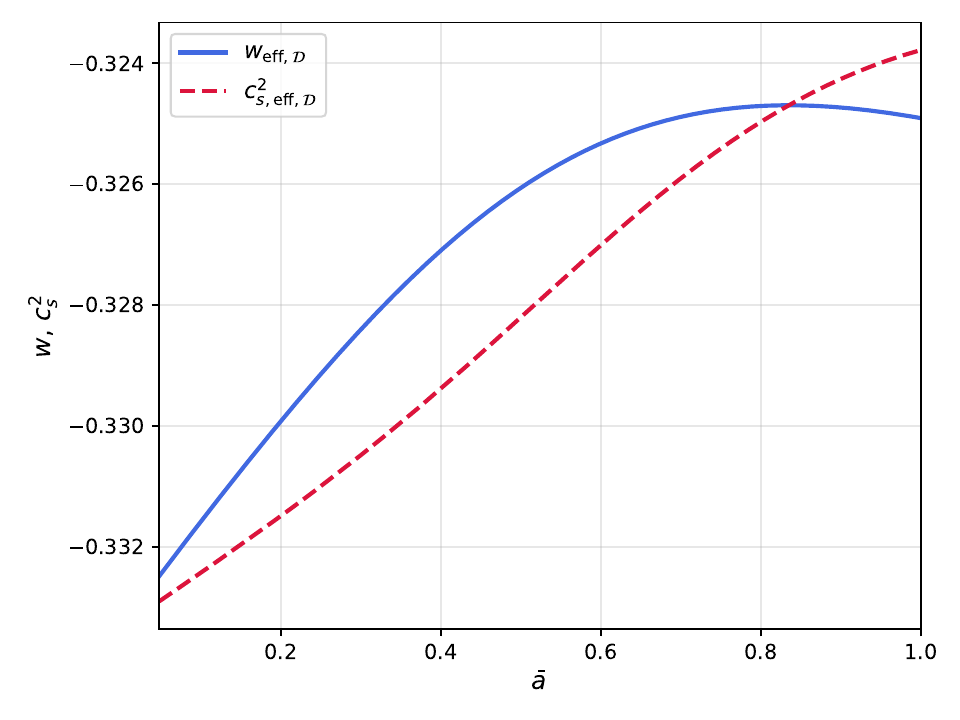}
    \end{minipage}
\end{adjustwidth}
\caption{Energy balance (left), and effective fluid EoS parameter and squared sound speed (right) as functions of the \emph{reference} $\Lambda$CDM scale factor $\aref$ in the RZA model we consider, with parameter $\beta\xi_1 \approx 9.5\times10^{-3}$ and Planck 2018~\cite{Planck_2018} parameters for the underlying reference flat $\Lambda$CDM model.}
\label{fig:rza_1}
\end{figure}

We show in Fig.~\ref{fig:rza_1} the contributions to the energy budget (from Eq.~\eqref{eq:omegas}) in this RZA model on the left panel, and the associated effective fluid EoS parameter and squared sound speed on the right panel. Here, the time dependence of these quantities is shown in terms of the scale factor of the reference $\Lambda$CDM model $\aref(t)$, which reaches $\aref(t_0) = 1$ at present time by definition. The effective scale factor of the averaging domain under consideration, $a_\CD(t)$, normalised to coincide with the reference scale factor on the initial slice, $a_\CD(\tinit) = \aref(\tinit) = 1/21$, reaches $a_\CD(t_0) = 1.047$ at present time, highlighting the faster growth of the underdense region $\CD$ compared to the homogeneous reference.
The kinematical backreaction, while constrained analytically to be nonpositive is this framework, remains negligible at all times compared to the other contributions to the energy budget, with an $\Omega^\CD_\CQ \, (>0)$ of order $10^{-4}$. The average curvature, on the other hand, starts off negative as expected for this underdensity and stays so at all times, and becomes a subdominant but non-negligible contribution at late times. It does massively dominate over $\QD$ among the effective sources however, and it does scale as $\RD \, a_\CD^{2} \approx \mathrm{const}$ to a good approximation in this particular model; this explains why the effective fluid EoS parameter and squared sound speed both remain close to (slightly above) those of a FLRW ``curvature'' equation of state, $w_{\mathrm{eff},\CD}(t) \approx -1/3 \approx c^2_{s,\mathrm{eff},\CD}(t)$ at all times. In particular, there is no phantom crossing nor divergence of the sound speed in this scenario. 

\textcolor{black}{Finally, we emphasise that the described behaviour of the $\Omega_i^\CD$ parameters is a direct consequence of the chosen $\Lambda$CDM reference model. Indeed, in the limit of a vanishing cosmological constant, the RZA average dynamics would instead more closely resemble those of timescape cosmology (see Fig.~\ref{fig:ts_1}), with structure formation alone sourcing an effective acceleration~\cite{BuchertRZA_2013}. However, we note that the physical interpretation of this acceleration–like behaviour would still differ fundamentally from that of the timescape scenario, since, in the RZA, this effect does not arise from any substantial distinction between regional observers nor from different calibrations for their associated clocks and rulers.}

\subsubsection{RZA: CGI linear structure growth}

We can now consider scalar CGI perturbations of the dust and effective fluid on the above background. We again set the ICs for these perturbations as in Sec.~\ref{subsubsec:ts_CGI}, with the same (typical) dust density perturbation at $\tinit$ (still associated with $\aref(\tinit) = 1/21$), and vanishing effective fluid perturbations at that initial time.

\begin{figure}[htb!]
    \centering
    \includegraphics[width=0.7\textwidth]{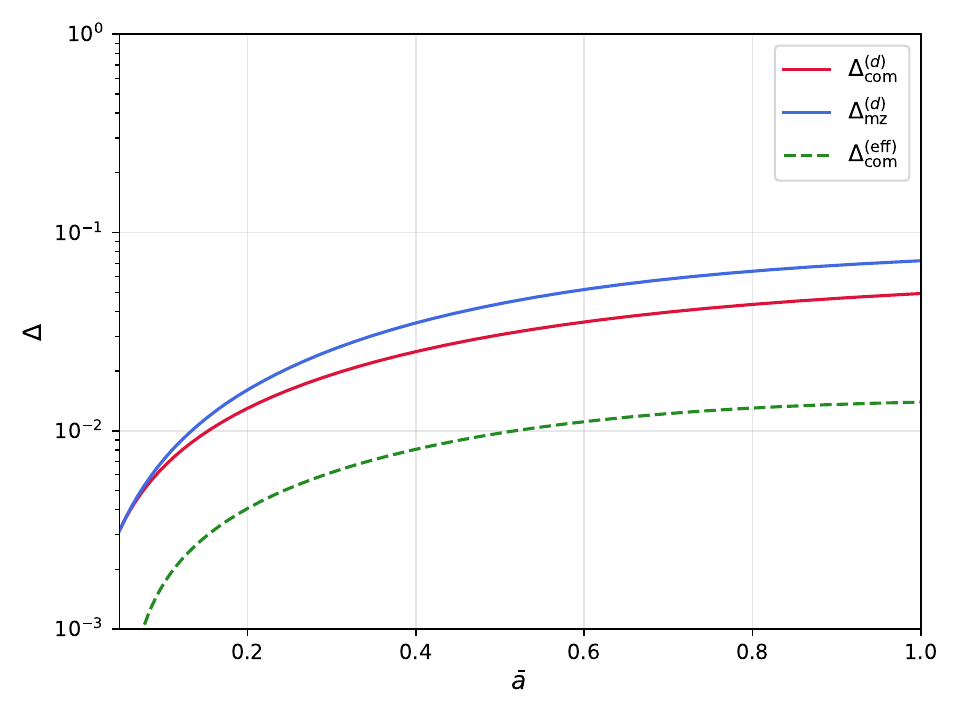} 
    \caption{Dust density perturbation growth within the comoving effective fluid framework, $\Delta^{(d)}_\mathrm{com}$, and the M\'esz\'aros approximation, $\Delta^{(d)}_\mathrm{mz}$, for the RZA background model presented above, as functions of the reference FLRW scale factor $\aref$. The ICs for the dust density and expansion perturbations are set at $z = 20$, seeded using the $\Lambda$CDM-based approximation as in Sec.~\ref{subsec:ts} above. The ICs for the effective fluid density perturbations are instead set to zero.}
    \label{fig:rza_2}
\end{figure}

In Fig.~\ref{fig:rza_2}, we show the growth of dust and effective fluid energy density perturbations in the comoving effective fluid picture, again using the reference FLRW scale factor $\aref$ as a time label. We also show the dust perturbation evolution predicted by the M\'esz\'aros approximation (i.e., using Eq.~\eqref{eq:original_Meszaros}) on this background for the same ICs, for comparison. Similarly to the case of the GMC model in Sec.~\ref{subsec:GMC} above, but unlike the other two examples, the growth of effective fluid perturbations in the comoving framework (in this case, steady and non-oscillatory) leads to a significantly slower evolution of the dust density perturbation. Interestingly, within the comoving picture, we also observe that the initially vanishing effective fluid energy density perturbation quickly settles into a constant ratio with the dust density perturbation, with $\Delta^{(d)}_\mathrm{com} / \Delta\eff_\mathrm{com} \approx \mathrm{const} \approx 3$ for $\aref \gtrsim 0.2$.

\begin{figure}[htb!]
\begin{adjustwidth}{-2cm}{-1.5cm}
\centering
    \begin{minipage}{0.495\linewidth}
        \centering
        \includegraphics[width=\linewidth]{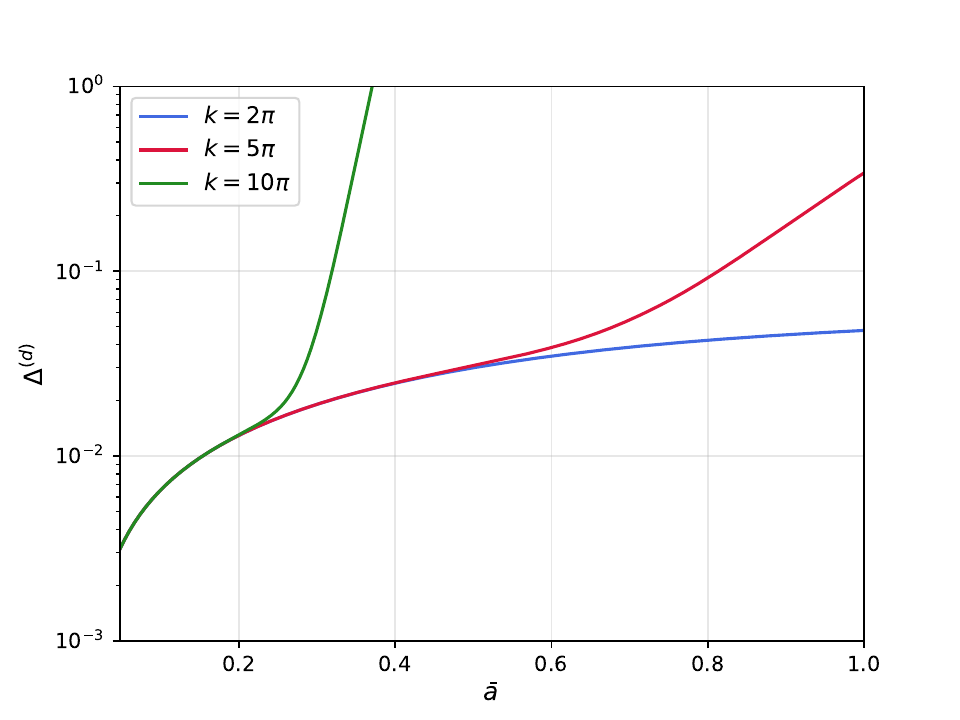}
    \end{minipage}\hfill
    \begin{minipage}{0.495\linewidth}
        \centering
        \includegraphics[width=\linewidth]{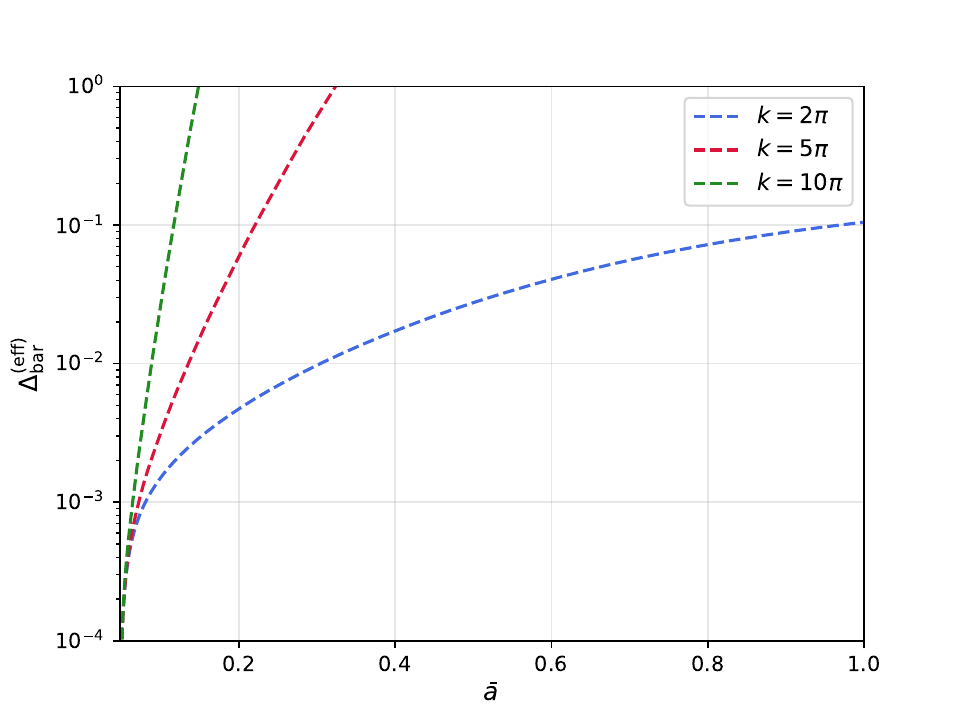}
    \end{minipage}
\end{adjustwidth}
\caption{Horizon and sub-horizon dust (left panel) and effective fluid (right panel) density perturbation growth on the RZA background presented above, as functions of the reference FLRW scale factor $\aref$ and within the barotropic condition. We show the amplitude evolution for three different wavelengths on a spatially flat background ($K_\CD = 0$), at horizon ($k = 2\pi$) and sub-horizon scales. Here, the wavenumber $k$ is expressed in units of the present-day inverse Hubble length, $H_0/c$, as evaluated in the $\Lambda$CDM model following the best fit of~\cite{Planck_2018}. The ICs at $z=20$ ($\aref = 1/21$) are seeded via the $\Lambda$CDM approximation for the dust density and expansion perturbation, and set to zero for the effective fluid perturbations, as in Sec.~\ref{subsec:ts} above.}
\label{fig:rza_3}
\end{figure}

We also show, in Fig.~\ref{fig:rza_3}, the growth of the dust (left panel) and effective fluid (right panel) energy density linear perturbations in the barotropic closure condition for the effective fluid, at horizon and sub-horizon scales. We note that these perturbations would typically only vary slightly in space across the averaging domain considered, which we interpret in the present case as a large underdense region, at most a few dozen megaparsecs in comoving diameter. As expected from the negative effective EoS parameters, the subhorizon effective fluid perturbations appear unstable. Specifically, as in the timescape background scenario (Sec.~\ref{subsec:ts}), the effective fluid perturbations grow rapidly, without oscillations, with smaller perturbation scales leading to faster growth. In this scenario, the dust perturbations maintain a moderate and scale-independent growth at first; at subhorizon scales, this is followed by a scale-dependent transition to a much faster growth. This occurs at very large values of $\Delta\eff\baro$ relative to $\Delta^{(d)}\baro$. Such a large amplitude ratio is required for the right-hand side of Eq.~\eqref{eq:ddotdeltad_bar} to have a significant impact on the dust perturbation evolution and make it effectively depart from its evolution equation within the M\'esz\'aros approximation, Eq.~\eqref{eq:original_Meszaros}, since the background value of $c^2_{s(\mathrm{eff})}$ stays very close to $-1/3$ in this model. With the initial values we used here, this also means that the effective fluid perturbation has, in principle, already left the linear regime by that point. Nevertheless, as discussed for the similar situation that occurs in the timescape example (Sec.~\ref{subsubsec:ts_CGI}), we are showing a linear solution, which could be arbitrarily rescaled in amplitude, and the faster growth in $\Delta^{(d)}$ indeed signals an additional linear growth mode due to the coupling to the effective fluid perturbations. In any case, the dynamics in this scenario implies that the sub-horizon backreaction and curvature perturbations (through the effective fluid) would at some stage become nonlinear while the dust perturbation itself still lies well within a linear regime. This is another seemingly unphysical consequence of the barotropic parametrisation choice, even though this example does not exhibit unexpected reversals of the growing trend for the dust perturbations.

\section{Conclusions}
\label{sec:conc}

In this work, we have analysed the linear growth of matter perturbations on top of a general background arising from the explicit spatial averaging of nonlinear local inhomogeneities over a given scale, in the matter-dominated era.

The dynamics of the background generally differ from those of a flat or constant-curvature, strictly homogeneous and isotropic counterpart, because the formation of nonlinear matter structures induces a \emph{backreaction} effect on the expansion and drives a nontrivial evolution of the averaged spatial curvature. We describe these effects using Buchert’s averaging scheme, specialised to inhomogeneous universes sourced by an irrotational dust fluid. The homogeneous--isotropic background we adopt---defined by the average expansion over a spatial domain comoving with the dust flow and by a chosen curvature parameter---encodes these effects as an effective second perfect-fluid component with pressure. This component influences the evolution of the scale factor alongside the averaged dust density, which retains the same scaling behaviour as in a strictly homogeneous dust model.

We then derived the evolution equations for linear deviations from this two-fluid effective background using the covariant and gauge-invariant multi-fluid perturbation formalism introduced by Ellis and Bruni, focusing on scalar perturbations and adopting the physical dust fluid as the reference frame. These partial differential equations---which reduce to ordinary differential equations in the space of eigenfunctions of the background’s spatial Laplacian---couple fluctuations in dust density to those in dust fluid expansion, effective fluid energy density and pressure, and relative (tilt) velocity between the two fluid components.

We discussed possible closure conditions for this system, which amount to specifying the nature of the effective fluid’s perturbations. This is not uniquely determined, since the effective fluid is only defined \emph{a priori} through its dynamical contributions to the background. Nonetheless, any perturbation in the dust density necessarily induces perturbations in the effective fluid via gravitational coupling. Therefore, within a general relativistic perturbation framework, the effective fluid cannot be assumed to remain homogeneous at all times.

We showed that, under specific conditions, the contributions of the effective fluid’s perturbations to the growth of dust perturbations can sometimes be neglected, so that the linear dust density perturbations formally satisfy the same source-free, scale-independent second-order differential equation as in a $\Lambda$CDM universe with linear perturbations of its single (dust) fluid component. This corresponds to the M\'esz\'aros approximation, originally formulated for perturbations in a mixture of dust and radiation in the early universe, here extended to a more general secondary perfect fluid. We emphasize that even when this approximation applies, the actual solution for the linear growth rate of dust structures differs from that in a $\Lambda$CDM background, because backreaction and dynamical curvature effects from nonlinear structure formation---accounted for in our effective background---modify the time dependence of the expansion rate and mass density, which enter as coefficients in the evolution equation. Moreover, this solution retains a dependence on the scale and location of the averaging domain used to construct the background if it is smaller than the statistical homogeneity scale.

When this approximation does not hold, a prescription must be provided for the nonzero pair of perturbations in the effective fluid’s energy density and pressure. We discuss in detail two possible choices, giving the corresponding closed set of evolution equations in each case, as well as the (higher-order)evolution equation that each choice implies for the dust density perturbations alone. 

The first option is to assume that the effective fluid retains, up to first order, the (non-constant) barotropic equation of state it obeys in the background. This is a natural choice for a physical fluid with a simple equation of state, such as radiation, for which it corresponds to adiabatic perturbations. However, when the energy density and pressure represent only effective sources, this choice is \emph{a priori} arbitrary and can lead to divergences in the coefficients of the perturbation evolution equations if the effective fluid undergoes a phantom crossing---associated with a formally infinite effective speed of sound---at certain times.

The second closure condition that we analysed is that of a comoving effective fluid, in which the relative tilt velocity is set to zero at all times, thereby binding the effective fluid’s $4$-velocity to the frame of the physical dust fluid. This choice extends to the first order the property of the FLRW background in which the $4$-velocities of all fluids define a single reference frame. Within our framework, it can be interpreted as implementing, at the level of perturbations, a feature inherited from the construction of the background effective fluid, which is defined via spatial averages in the rest frames of the dust flow in the original, fully inhomogeneous spacetime and evolved along that flow. The resulting evolution equations for the perturbations are relatively simple and free of divergent terms. Surprisingly, they also reduce directly to ordinary differential equations, fully independent of the scale of the perturbation itself---although their coefficients retain a dependence on the scale of the averaging domain used to construct the background---because the only spatial ($3$-Laplacian) operator in the general evolution equations acts on the relative tilt velocity.

We applied this formalism to four example models for the averaged dynamics of a spatial region, corresponding either to different representations of the local inhomogeneities or to a direct phenomenological assumption about the resulting effective fluid. We noticed that for all of these models, the M\'esz\'aros approximation should not be expected to hold, due to negative equation of state parameters for the effective fluid over at least part of the time range in each case, with effective phantom crossings occurring for two of the examples.

We then showed the growth of the dust and effective fluid perturbations over time for each of these models, from a typical small initial perturbation of the dust only, assuming either a barotropic or a comoving closure condition for the effective fluid. The corresponding evolution equations were solved numerically, choosing a vanishing spatial curvature for the effective two-fluid FLRW background. Along with the manifestly different equation sets, the results clearly highlight the dependence of the dust perturbation evolution on the choice of closure condition, i.e. on the assumed properties of the perturbed effective fluid, with even qualitatively very different behaviours in either case considered.

In the barotropic picture, we showed the growth of the perturbations at different scales. A  strong dependence on that scale is visible, around the horizon scale; the background (average) model also has a major impact on the dynamics of the perturbations. However, a common feature for all four models is a rapid growth of the amplitude of the effective fluid perturbations, especially at small scales, which typically excites an additional, rapidly evolving mode for the dust density perturbation compared to a perturbed $\Lambda$CDM scenario. At sub-horizon scales, the growth in amplitude of this matter perturbation appears too fast to be reasonably compatible with the observed structure formation in our Universe, and even more unphysical reversals in the density contrast sign over time occur in many cases. In several cases, the even faster growth of the effective fluid perturbations at sub-horizon scales would lead to the unexpected departure of the system from the linear regime while the dust perturbations are still very small. Additionally, for the two models where  phantom crossings of the effective fluid take place, those lead to sharp transitions in the perturbation dynamics at any scale, or even render the evolution across those points ill-defined altogether.

In the comoving picture, where the evolution of the perturbations is scale-independent, we compared the dust perturbation growth to what the M\'esz\'aros approximation would predict. As expected, the growth of the dust perturbation still depends on the background (average) model, albeit less strongly than in the barotropic scenario. In each case, a moderate growth is obtained at all times, remaining qualitatively similar overall to the M\'esz\'aros approximation solution. Nevertheless, in all but one example, the amplitude of the effective fluid density (and hence the kinematical backreaction and curvature) perturbation becomes large enough over time to have a significant impact on the dust density contrast, which grows slightly slower or faster than in the M\'esz\'aros approximation depending on the background model. No unphysical changes in trend nor divergences are observed. We consider this comoving choice to be a much more physically relevant parametrisation of the perturbations in the kinematical backreaction and average curvature, further motivated by the dust-frame construction of the averages which define our background from the inhomogeneous original spacetime.
Although its scale independence might seem unexpected, we stress that this characteristic, shared with perturbations of $\Lambda$CDM models in the matter-dominated era, is specific to the geodesic nature of the physical fluid defining the preferred frame. The presence of pressure gradients in that fluid would bring back a scale dependence, e.g., allowing for acoustic peaks in a dust and radiation mixture. Additionally, this approach remains fully independent of the arbitrary choice of spatial curvature for the homogeneous two-fluid background. This is another consequence of the disappearance of the $3-$Laplacian operator from the evolution equations, and thus of its curvature-dependent eigenvalues.

In this work, we focussed on scalar perturbation variables, which---as within the usual Bardeen perturbation formalism---are sufficient to capture the growth of structures at linear order. However, the covariant and gauge-invariant formalism used can also describe vector and tensor modes. We leave their more detailed investigation in dynamical, averaged backgrounds for future work. This would enable, for instance, the study of the development of vorticity and its contribution to anisotropic expansion and matter density gradients.

Another natural extension of this work would be the development of a consistent scheme for the growth of BAO, and possible impacts on other cosmological probes such as the $f\sigma_8$ parameter, within an averaged model universe that accounts for backreaction effects and curvature evolution from nonlinear structure formation at small scales. As inhomogeneities remain very small in the pre-recombination Universe, the perturbations in the dust/radiation mixture fluid could be treated in the standard approach, with a homogeneous background and no additional effective source. After the transition to the matter-dominated era, the acoustic peaks in the dust perturbations would grow in a scale-independent way, as in the $\Lambda$CDM framework, yet with a modified growth rate. This would come from averaged-over nonlinear effects, modelled as in this work via an effective fluid, with a comoving parametrisation of its perturbations.

Finally, the scheme presented here could be extended beyond the linear order in perturbations. Nonetheless, we stress that our approach already goes substantially beyond first-order standard perturbation theory on a strictly FLRW dust background, by relying on the averaged contributions from fully nonlinear structure formation below a certain scale to define its background. Accordingly, the interpretation of second-order perturbations or beyond on such a background would become more intricate, with such perturbations being, in principle, already part of the background. This is not a concern in a first-order framework, since shifting the background to include (or not) a given linear perturbation is a second-order effect on any other perturbations. However, it would impact the development of higher-order schemes, which we therefore leave for possible future work.

\acknowledgments

The authors wish to thank Marco Bruni, Thomas Buchert, Timothy Clifton, Leonardo Giani, Asta Heinesen, and David Wiltshire for their insights and our engaging conversations during the preparation of this paper. We also thank Francesco Bennetti, Christopher Harvey-Hawes, Morag Hills, Emma Johnson, Zachary Lane, Ryan Ridden-Harper, Shreyas Tiruvaskar for useful discussions. This work was supported by Marsden Fund Grant M1271 administered by the Royal Society of New Zealand, Te~Ap\=arangi.
\bibliography{main}
\newpage
\appendix
\section{Decoupled evolution equation for barotropic-picture dust perturbations}
\label{appendix:4thorder}

With a barotropic parametrisation of the effective fluid linear perturbations, these couple to the dust density perturbations in two second-order ordinary differential equations given by Eqs.~\eqref{eq:ddotdeltad_bar}--\eqref{eq:ddotdeltaeff_bar}. By taking two time derivatives of Eq.~\eqref{eq:ddotdeltad_bar}, combining it with Eq.~\eqref{eq:ddotdeltaeff_bar} to eliminate $\ddot\Delta\eff$, and using again Eq.~\eqref{eq:ddotdeltad_bar} and its first time derivative to eliminate $\Delta\eff$ and $\dot\Delta\eff$, one can obtain the full, source-free, fourth-order evolution equation for the linear growth of the dust density perturbation alone, taking the form
\begin{equation}
   \ddddot\Delta^{(d)} + \alpha(t) \;\! \dddot\Delta^{(d)} + \beta(t) \;\! \ddot{\Delta}^{(d)} + \frac{2}{3} \gamma(t) \;\! \dot{\Delta}^{(d)} - 4 \pi G \rho^{(d)} \zeta(t) \;\! {\Delta}^{(d)} = 0 \; .
\end{equation}
The coefficients of this equation are given by:
\begin{equation}
    \alpha(t) := -\frac{6 \, \csdot}{1 + 3 \:\! c_{s(\mathrm{eff})}^2}+\Theta \left(\frac{10}{3} - c_{s(\mathrm{eff})}^2 \right) \; ; 
\end{equation}
\begin{align}
    \beta(t) := {} & \frac{k^2}{a^2} c_{s(\mathrm{eff})}^2 - \frac{13 \, \csdot}{1+3\:\!c_{s(\mathrm{eff})}^2}\,\Theta +18 \left[ \frac{\csdot}{1 + 3 \:\! c^2_{s(\mathrm{eff})}} \right]^2\! - \frac{3 \, \csddot}{1 + 3 \:\! c_{s(\mathrm{eff})}^2} + 2 \:\! \Theta^2\left(\frac{4}{3} - c_{s(\mathrm{eff})}^2\right) \nonumber \\ 
    & {\!}  -4\pi G  \left[\left(\frac{10}{3}-c_{s(\mathrm{eff})}^2 \right)\rho^{(d)}+\left( \frac{10}{3} +8 \:\!w_{(\mathrm{eff})} +2 \:\! c_{s(\mathrm{eff})}^2 \right)\rho\eff \right] +\Lambda \left[ \frac{7}{3} - c^2_{s(\mathrm{eff})} \right] \; ; 
\end{align}
\begin{align}
    \gamma(t) := {} & \Theta \, \frac{k^2}{a^2} \, c_{s(\mathrm{eff})}^2 + \frac{\csdot}{1+3\:\!c_{s(\mathrm{eff})}^2} \left( - 7 \:\! \Theta^2 + 12\pi G  \left[5 \rho^{(d)}+2 \left(1+3 \:\!w_{(\mathrm{eff})} \right) \;\! \rho\eff \right] - 6 \:\! \Lambda \right)  \nonumber \\
    & {\!} +18 \:\! \Theta \left[ \frac{\csdot}{1+3 \:\! c_{s(\mathrm{eff})}^2} \right]^2  - \frac{3 \, \csddot}{1+3\:\!c_{s(\mathrm{eff})}^2} \, \Theta + \Theta^3  \left(\frac{2}{3} -  c_{s(\mathrm{eff})}^2 \right) \nonumber \\
    & {\!}  + \Lambda \;\! \Theta \left( 3 - 2 \:\! c^2_{s(\mathrm{eff})} \right) -2\pi G \;\! \Theta \;\! \rho^{(d)} \left(6-7\:\!c_{s(\mathrm{eff})}^2 \right) \nonumber \\ 
    & {\!}   -2\pi G \;\! \Theta \;\! \rho\eff \left[ \, 6 \left(1+3\:\! w_{(\mathrm{eff})} \right) - c_{s(\mathrm{eff})}^2 \left(7+15 \:\! w_{(\mathrm{eff})}+ 9 \:\! c_{s(\mathrm{eff})}^2 (1+w_{(\mathrm{eff})})\right) \right] \vphantom{\frac{\csdot}{1 + 3 \:\! c^2_{s(\mathrm{eff})}}}  \; ; 
\end{align}
\begin{align}
    \zeta(t) := {} &  c_{s,\mathrm{eff}}^2 \left( \frac{k^2}{a^2}  - \frac{\Theta^2}{3} + 4\pi G \left[\rho^{(d)} + \left(1+3 \:\! w_{(\mathrm{eff})}  \right) \rho\eff \right] - \Lambda \right)  \nonumber \\
    &{\!} -3 \, \frac{\csddot + \Theta \csdot }{1+3\:\!c_{s(\mathrm{eff})}^2}  + 18 \left( \frac{\csdot}{1+3\:\!c_{s(\mathrm{eff})}^2} \right)^2  \; . 
\end{align}

\section{Homogeneous-density effective fluid parametrisations}
\label{appendix:mc}

As an alternative to the barotropic and comoving parametrisations introduced in Sec.~\ref{subsec:barotropic}--\ref{subsec:comoving}, one may want to close the scalar perturbation evolution equations~\eqref{eq:system4_dust}--\eqref{eq:system4_V} by imposing that the perturbed effective fluid remains homogeneous in its own local rest frames up to first order---that is, requiring the spatial gradient in those frames $\left( g_{\mu\nu} + u_\mu\eff u_\nu\eff \right) \nabla^\nu$ to vanish at first order for $\rho\eff$ and $p\eff$. This is a natural choice if the effective fluid is to be interpreted as a minimally coupled scalar field $\Phi_\CD$ (with $\nabla_\mu(\Phi_\CD) \propto u_\mu\eff$), even when perturbations are included.
However, the perturbations in the effective fluid energy density and pressure cannot be simultaneously assumed to vanish identically in the effective fluid frame, even only for their scalar parts, for the same reason as for the corresponding perturbations $\Delta\eff, \CP\eff$ in the dust fluid's frame---such an assumption in Eqs.~\eqref{eq:system4_dust}--\eqref{eq:system4_V} would set all scalar perturbations to zero at all times for both fluids.

Instead, one can close that system by only setting, for instance, the effective \emph{energy density} source perturbation in the effective fluid's local rest frames to zero:
\begin{equation}
    0 = \left( g_{\mu\nu} + u_\mu\eff u_\nu\eff \right) \nabla^\nu \rho\eff \FO D_\mu \rho\eff - \Theta \rho\eff \left( 1 + w_{\mathrm{eff})} \right) V_\mu\eff \: .
\end{equation}
For the scalar CGI variables, this implies
\begin{equation}
    \Delta\eff \FO a \;\! \Theta \tilde V\eff \; .
\end{equation}

Through this relation, the tilt velocity $\tilde V\eff$ is expressed in terms of the dust-frame effective fluid density perturbation $\Delta\eff$, and by substituting $\tilde V\eff$ accordingly in Eq.~\eqref{eq:system4_V}, the dust-frame effective fluid pressure perturbation $\tilde\CP\eff$ can also be expressed in terms of $\Delta\eff$ and $\dot\Delta\eff$.
We are then left with the remaining three evolution equations for the density and expansion perturbation variables $\Delta^{(d)},\Delta\eff,\CZ$:
\begin{align}
    & \dot{\Delta}^{(d)} + \CZ \FO 0 \; ; \label{eq:dotdeltad_mc} \\
    & \dot{\Delta}^{(\mathrm{eff})} + \frac{1}{2 \Theta} \left[4 \pi G \rho^{(d)} + 4 \pi G \rho\eff \, ( 1 + 3 \:\!w_{(\mathrm{eff})}) - \Lambda + \left(\frac{1}{3} - 2 \:\! w_{(\mathrm{eff})} \right) \Theta^2 - \frac{k^2}{a^2} \right] \Delta^{(\mathrm{eff})} \nonumber \\& \qquad \qquad \qquad {} + \frac{1}{2} (1 + w_{(\mathrm{eff})}) \, \CZ \FO 0 \; ; \label{eq:dotdeltaeff_mc} \\
    & \dot \CZ + \left[ \frac{2}{3} + \frac{ 6 \pi G \rho\eff}{\Theta^2} (1 + w_{(\mathrm{eff})})  \right] \Theta \CZ  + 4 \pi G \rho^{(d)} \Delta^{(d)} \nonumber \\
    &\qquad \qquad \qquad - \frac{6 \pi G \rho\eff}{\Theta^2} \left[ 4 \pi G \rho^{(d)} + 4 \pi G \rho\eff \, (1 + 3 \:\! w_{(\mathrm{eff})}) - \Lambda - \frac{\Theta^2}{3} + \frac{k^2}{a^2} \right] \Delta^{(\mathrm{eff})} \FO 0 \; . \label{eq:dotZ_mc}
\end{align}
We note that this system remains dependent on the scale $k$ of the perturbation, as in the barotropic framework (Sec.~\ref{subsec:barotropic}), but like the comoving framework (Sec.~\ref{subsec:comoving}) it does not depend directly on the background squared sound speed of the effective fluid $c^2_{s,\mathrm{eff},\CD}$ and thus should remain stable when crossings of an effective phantom EoS occur.

As for the other closure conditions discussed in Sec.~\ref{sec:Us}, substituting $-\dot\Delta^{(d)}$ for $\CZ$ (from Eq.~\eqref{eq:dotdeltad_mc}) straightforwardly transforms Eq.~\eqref{eq:dotZ_mc} into a second-order differential equation for the dust density perturbation $\Delta^{(d)}$, sourced by the dust-frame effective fluid density perturbation $\Delta\eff$.
That perturbation is generally nonvanishing due to the change of frame (through the nonzero $V\eff$), and is here determined via the evolution equation~\eqref{eq:dotdeltaeff_mc}. The latter can either be considered directly as a first order ODE sourced by both $\Delta^{(d)}$ and $\dot{\Delta}^{(d)}$, or be converted into a second-order ODE sourced by $\Delta^{(d)}$, similarly to Eq.~\eqref{eq:ddotdeltaeff_com} in the comoving representation, by taking another time derivative and replacing $\dot\CZ$ through Eq.~\eqref{eq:dotZ_mc}.
As in the barotropic and comoving representations, the three evolution equations~\eqref{eq:dotdeltad_mc}--\eqref{eq:dotZ_mc} can also be combined into a source-free differential equation for the dust perturbation, only up to third-order time derivatives.

The additional requirement of a vanishing scalar effective pressure perturbation in the effective fluid frame would amount to
\begin{equation}
    \tilde \CP\eff \FO a \;\! \Theta \tilde V\eff c_{s(\mathrm{eff})}^2 \FO c_{s(\mathrm{eff})}^2 \Delta\eff \; ,
\end{equation}
and thus would additionally impose a barotropic relation between the dust-frame effective fluid perturbations. Yet, as can be inferred by comparing Eq.~\eqref{eq:dotZ_mc} above and Eq.~\eqref{eq:dotZ_bar}, the above system of evolution equations is not compatible in general with that of the barotropic representation (Eqs.~\eqref{eq:dotdeltad_bar}--\eqref{eq:dotV_bar}), preventing this additional condition on the effective pressure perturbation. As requiring homogeneity of only the energy density part of the effective fluid source in its own rest frames is not as clearly interpreted physically, we do not include the choice of this closure condition in our investigation of linear structure growth in various averaged cosmological models in Sec.~\ref{sec:examples}.

\end{document}